\documentclass[smallextended]{svjour3_mh}       % onecolumn (second format)
\smartqed  % flush right qed marks, e.g. at end of proof

\usepackage{fix-cm}

\usepackage{graphicx}
\usepackage{amssymb,amsmath}
\usepackage{mathptmx}
\usepackage{bm}
\usepackage{enumerate}
\usepackage{epstopdf}

\usepackage{color}
\usepackage[utf8x]{inputenc}
\usepackage[T1]{fontenc}

\definecolor{K}{rgb}{0,0,0}
\definecolor{R}{rgb}{1,0,0}
\definecolor{G}{rgb}{0,1,0}
\definecolor{B}{rgb}{0,0,1}
\definecolor{C}{rgb}{0,1,1}
\definecolor{M}{rgb}{1,0,1}
\definecolor{Y}{rgb}{1,1,0}

\definecolor{KR}{rgb}{0.5,0  ,0  }
\definecolor{KG}{rgb}{0  ,0.5,0  }
\definecolor{KB}{rgb}{0  ,0  ,0.5}
\definecolor{KC}{rgb}{0  ,0.5,0.5}
\definecolor{KM}{rgb}{0.5,0  ,0.5}
\definecolor{KY}{rgb}{0.5,0.5,0  }

\definecolor{redyellow}{rgb}{1,.5,0}
\definecolor{yellowgreen}{rgb}{.5,1,0}
\definecolor{greencyan}{rgb}{0,1,.5}
\definecolor{cyanblue}{rgb}{0,.5,1}
\definecolor{bluemagenta}{rgb}{.5,0,1}
\definecolor{magentared}{rgb}{1,0,.5}

\definecolor{black}{rgb}{0,0,0}
\definecolor{darkgray}{rgb}{.25,.25,.25}
\definecolor{gray}{rgb}{.5,.5,.5}
\definecolor{lightgray}{rgb}{.75,.75,.75}
\definecolor{white}{rgb}{1,1,1}

\def\textK#1#2{\textcolor{K}{\hbox to #1{#2\hfill}}}
\def\textR#1#2{\textcolor{R}{\hbox to #1{#2\hfill}}}
\def\textG#1#2{\textcolor{G}{\hbox to #1{#2\hfill}}}
\def\textB#1#2{\textcolor{B}{\hbox to #1{#2\hfill}}}

\def\textRY#1#2{\textcolor{redyellow}{\hbox to #1{#2\hfill}}}
\def\textMR#1#2{\textcolor{magentared}{\hbox to #1{#2\hfill}}}
\def\textBM#1#2{\textcolor{bluemagenta}{\hbox to #1{#2\hfill}}}
\def\textCB#1#2{\textcolor{cyanblue}{\hbox to #1{#2\hfill}}}
\def\textGC#1#2{\textcolor{greencyan}{\hbox to #1{#2\hfill}}}
\def\textYG#1#2{\textcolor{yellowgreen}{\hbox to #1{#2\hfill}}}

\def\textKR#1{\textcolor{KR}{#1}}
\def\textKY#1{\textcolor{KY}{#1}}
\def\textKG#1{\textcolor{KG}{#1}}
\def\textKC#1{\textcolor{KC}{#1}}
\def\textKB#1{\textcolor{KB}{#1}}
\def\textKM#1{\textcolor{KM}{#1}}

\def\geoK{\textK{24pt}{}}
\def\geoR{\textR{24pt}{\hbox to 12pt{\hfill $\!\bigcirc\!$\hfill}}}
\def\geoB{\textB{24pt}{\hbox to 12pt{\hfill $|\,|        $\hfill}}}
\def\geoG{\textG{24pt}{\hbox to 12pt{\hfill $)(          $\hfill}}}

\def\sR{{\rm sin}_{\tiny\hbox to 4pt{\hfill $\!\bigcirc\!$\hfill}}\!}
\def\sB{{\rm sin}_{\tiny\hbox to 4pt{\hfill $|\,|        $\hfill}}\!}
\def\sG{{\rm sin}_{\tiny\hbox to 4pt{\hfill $)(          $\hfill}}\!}

\def\cR{{\rm cos}_{\tiny\hbox to 4pt{\hfill $\!\bigcirc\!$\hfill}}\!}
\def\cB{{\rm cos}_{\tiny\hbox to 4pt{\hfill $|\,|        $\hfill}}\!}
\def\cG{{\rm cos}_{\tiny\hbox to 4pt{\hfill $)(          $\hfill}}\!}

\def\<{\langle}
\def\>{\rangle}
\renewcommand{\b}[1]{\mathbf{#1}}
\newcommand{\eps}{\epsilon}

\definecolor{jgGreen}{rgb}{0.0, 0.5, 0.0}
 \journalname{myjournal}

\begin{document}

\title{Random sequential adsorption of discs on constant-curvature surfaces: plane, sphere, hyperboloid, and projective plane}

\titlerunning{Random sequential adsorption on constant-curvature surfaces}       

% Authors: full names plus addresses.
\author{
  Elizabeth R. Chen
  \and
  Miranda Holmes-Cerfon
}


\institute{E. Chen \at
              \email{bethchen@seas.harvard.edu}           %  \\
           \and
           M. Holmes-Cerfon \at
              \email{holmes@cims.nyu.edu} 
}

\date{}

\maketitle

\begin{abstract}
We present an algorithm to simulate random sequential adsorption (random ``parking'') of discs on constant-curvature surfaces: the plane, sphere, hyperboloid, and projective plane, all embedded in three-dimensional space. 
We simulate complete parkings by explicitly calculating the boundary of the available area in which discs can park and concentrating new points in this area. 
This makes our algorithm efficient and also provides a diagnostic to determine when each parking is complete, so there is no need to extrapolate data from incomplete parkings to study questions of physical interest.
 
We use our algorithm to study the number distribution and density of discs parked in each space, where for the plane and hyperboloid we consider two different periodic tilings each. 
We make several notable observations: (i) On the sphere, there is a critical disc radius such the number of discs parked is always exactly four: the random parking is actually deterministic. 
We prove this statement rigorously, and also show that random parking on the surface of a $d$-dimensional sphere would have deterministic behaviour at the same critical radius. 
(ii) The average number of parked discs does not always monotonically increase as the disc radius decreases: on the plane (square with periodic boundary conditions), there is an interval of decreasing radius over which the average \emph{decreases}. We give a heuristic explanation for this counterintuitive finding. 
(iii) As the disc radius shrinks to zero, the density (average fraction of area covered by parked discs) appears to converge to the same constant for all spaces, though it is always slightly larger for a sphere and slightly smaller for a hyperboloid. Therefore, for parkings on a general curved surface we would expect higher local densities in regions of positive curvature and lower local densities in regions of negative curvature.

\end{abstract}

\section{Introduction}

Many physical processes involve objects that stick randomly and irreversibly to other objects. 
Proteins, viruses, bacteria, colloids, and macromolecules adsorb to surfaces or liquid-solid interfaces \cite{Evans:1993hx,Talbot:2000wj}.
Small colloids, resembling spheres, may stick to the surface of a bigger colloid, creating building blocks for new materials with novel shapes \cite{Schade:2013ee}. 
The structure of granular materials such as sand can be studied by considering the irreversible placement of grains around a central grain \cite{Newhall:2011ca}.

Such processes are often modelled using random sequential adsorption (RSA): objects (such as discs) stick one-by-one to a surface,
with locations chosen uniformly at random from the surface area that leads to no objects overlapping.
Eventually, there is no more available surface area so the process terminates. 
One is then interested in statistical quantities like the number distribution of attached objects,
the average density of area covered, or the spatial distribution of objects. 
Because the one-dimensional model resembles parking a car on the curb in a busy city, this process is often called random \emph{parking};
note that it is distinct from random \emph{packing}. 
Variants of RSA can include non-uniform disc sizes or shapes \cite{Newhall:2011ca,Ciesla:2016jn},
modelling the motion of objects before they find a parking spot \cite{Mansfield:1996wu,Erban:2007hl}, or non-uniform sticking distributions,
as in cooperative sequential adsorption in which the sticking probability depends on the locations of the objects already attached \cite{Talbot:2000wj}. 

Analytic formulas for various statistical quantities are available in limited situations, such as parkings on one-dimensional space \cite{Renyi:1958uy,Evans:1993hx},
but for the realistic situation of dimension at least two one must obtain these statistical quantities by simulation.
Simulations have provided insight into a great number of random parking applications,
yet almost all simulations to date have considered the case of objects attaching to a plane, or higher-dimensional Euclidean space. 
Spheres as substrates have been considered in a small number of studies \cite{Rosen:1986ue,Mansfield:1996wu,Schade:2013ee}.
However, there are virtually no studies on spaces other than these, %nor on planes with other kinds of boundary conditions, 
despite the fact that real substrates are often curved.
In addition, no simulation on the plane has used any boundary condition other than periodic on a rectangular lattice, to our knowledge. 

The goal of this paper is twofold. First, we present an algorithm to simulate a complete random parking of discs on spaces of constant curvature: the sphere, plane, hyperboloid, and projective plane.
By ``complete,'' we mean the algorithm generates samples with exactly zero space to park another disc.
On the plane and hyperboloid, the algorithm can handle parkings on any translation-invariant tiling with periodic boundary conditions. 

Second, we use this algorithm to investigate the number distributions of parked discs as a function of the disc radius.
The number distribution for large disc radius in particular has received little attention in theory and simulation,
but is of interest in materials science, for example, where one can use irreversible adsorbtion to design particular kinds of clusters \cite{Mansfield:1996wu,Schade:2013ee,Phillips:2012jo}.
On the sphere, we find empirically there is a critical radius at which all random parkings produce exactly four discs.
We ask whether this is a feature peculiar to the surface of a three-dimesional sphere,
and prove theoretically that a critical radius occurs for parkings on the surface of spheres in all dimensions $d$, at which all parkings contain $d{+}1$ discs. 
Several other spaces contain such critical radii or intervals of radii.
We also point out a peculiar property of the plane with periodic boundary conditions: the average number of discs parked is a non-monotonic function of disc radius.
The average sometimes decreases as the radius decreases, a counterintuitive finding that can be traced to the geometric complexity induced by the boundary conditions.  

Finally, we consider the density, i.e. the average fraction of surface area covered by a parking, as the disc radius goes to zero,
and show that while this appear to converge to the same constant on all spaces,
the density is always slightly higher than the constant for spaces of positive curvature and slightly slower for spaces of negative curvature.
Therefore, for parkings on an arbitrary curved surface, one would expect higher local area coverage in areas of positive curvature and conversely for negative curvature. 

Our algorithm has two major advantages over others that have been proposed. First is its ability to handle spaces with curvature and more complex boundary conditions.
Second is that it produces complete parkings.
To see why this second property is important, imagine a simple algorithm to produce a non-complete parking:
repeatedly choose a point, representing the center of the disc,
uniformly from the entire surface, and if it is too close to the points already parked then reject it.  
When enough points are parked this process becomes extremely slow because there is little available surface area,
so a naive algorithm would terminate each sample after a given time or consecutive number of failures. 
This naive algorithm does not detect when it has reached a complete parking, so not only is it inefficient,
wasting time trying to fit in new points when there is no more space,
but it is also inaccurate, because it often stops before a complete parking is reached.
It simulates a time-dependent version of a random parking, which approaches a complete parking as the length of each simulation increases,
but this process is slow: theoretical and numerical estimates show that the average density $\phi(t)$ of discs on a plane at time $t$ (the number of attempted placements)
approaches the limiting value $\phi_\infty$ as $\phi_\infty-\phi(t)\sim t^{{-}1/2}$ \cite{Feder:1980jw,Pomeau:1980,Swendsen:1981cd}. 
Nevertheless, with this formula in hand one can extrapolate data from a time-dependent simulation to estimate statistical quantities for the complete parking,
a strategy followed by a large number of researchers (with small modifications), such as \cite{Rosen:1986ue,Mansfield:1996wu,Feder:1980jw,Hinrichsen:1986}.

The naive algorithm can be improved by estimating the area available to park new points, and concentrating new points near this area.  
Several studies have considered methods to do this for discs parking on a plane with periodic boundary conditions.
One method is to divide the plane into small boxes and determine which boxes have space to park another disc.
One can then concentrate new points in these boxes, and stop when there are no more boxes \cite{Doge:2001gt,Torquato:2006eh}.
Even this improved method can be slow, because the available space in each box can be much smaller than the size of the box itself,
so one may need to stop before a complete packing is reached and extrapolate data anyways \cite{Torquato:2006eh}.
By continually refining the boxes one may reach a complete parking \cite{Zhang:2013cf},
but even so this method is not exact because it estimates whether a given box contains available space or not by sampling, so there is a small chance for errors.
Another method involves determining the available area from a Delaunay triangulation of the parked points,
and surrounding this by polygons in which new points are concentrated \cite{Lotwick:2007ej}.
This method can proceed until completion, but it is computationally demanding to calculate the areas for all of the small triangles,
and it is not clear how easy it is to adapt the method to objects of different sizes or parking on other spaces. 
Other methods have introduced heuristics to estimate where the available space is, without computing it exactly \cite{Feder:1980jw,Hinrichsen:1986},
or have introduced methods that compute the available space but do not sample it with the correct density \cite{Tanemura:1979vs}. 

Our strategy to produce a complete parking is to calculate the boundaries of all the regions where a new disc can be inserted.
After enough attempts in the naive algorithm, these regions are almost always isolated polygons with curved boundaries.
We concentrate new points in the polygons by sampling uniformly from the circumdiscs surrounding the polygons.
When there is no more boundary left then there is no more available space left, so we have a complete parking.
Calculating the boundaries is conceptually the same for all surfaces which have analytic formulas available for the geodesic distance,
allowing our code to be applied equally to all surfaces of constant curvature. 

The outline of the paper is as follows. In section \ref{sec:overview} we give a high-level overview of the algorithm,
introducing the necessary vocabulary and the basic ideas upon which it is based. 
In section \ref{sec:numerical} we analyze the statistical predictions of the algorithm,
both for large disc radius where certain ``critical radii'' lead to deterministic behaviour in various spaces,
as well as the limit where the radius shrinks to zero.
We also comment on the efficiency and accuracy of the algorithm.
In section \ref{sec:critical} we rigorously prove that the critical radius observed numerically on the sphere actually is a critical radius,
and also show it would be a critical radius for parkings on the surface of a $d$-dimensional sphere for any $d>1$.
In section \ref{sec:algorithm} we give the mathematical details necessary to implement the algorithm.
Section \ref{sec:conclusion} is a brief discussion and conclusion. 

%\todo[inline]{MHC:insert more description here, of point vs line. See emails June 25, morning.}
%\begin{verbatim}
%June 25, 10:18am: 
%point space vs line space are two complementary surfaces
%(you have their equations, so you know what they are)
%
%oh darn it!! i forgot to mention the norms!!
%
%usually, i map points to column vectors, and lines to row vectors
%so when you take the product of point and line, this is just the usual dot product
%(product of row matrix and column matrix)
%
%when you take the product of two points, you use the point-dot
%and for two lines, you use the line-dot
%(weve been using them, without calling them point-dot \& line-dot
%they are the funny things going on with the minus signs)
%
%and conversely, when you convert between points \& lines, you might need to add minus sign
%
%cross product of two points gives line, cross product of two lines gives point
%
%June 25, 10:45am:
%point vectors on the surface (point space) are really lines (rays thru (0,0,0)) in 3D
%similarly, the line curve on the surface (point space) is really surface (thru the two points & (0,0,0)$) in 3D
%so the line vector is really the normal to this surface
%
%June 25, 10:54am:
%the various dot products correspond to sin/cos (sinh/cosh) of certain quantities (length, angle)
%\end{verbatim}

\section{Parking algorithm}\label{sec:overview}

We consider parkings on constant curvature surfaces: the sphere, plane, hyperboloid, and projective plane. 
The three geometries $\{\textcolor{R}{\bigcirc},\textcolor{B}{||},\textcolor{G}{)(}\} = \{\textcolor{R}{\rm elliptic},\textcolor{B}{\rm parabolic},\textcolor{G}{\rm hyperbolic}\}$
(elliptic includes the sphere and projective plane), can be described either by their point space,
which is the collection of points that form a particular surface, or by the line space, which is the dual space to point space.\footnote{
While it is more conventional to think in point space and most of the discussion to follow will refer to this space,
we find some calculations are more conveniently implemented in line space so keep both representations for flexibility.
A summary of the relevant formulas to convert from one to the other is in the appendix, section \ref{app:geometry}.}
We have point vectors $\<x,y,z\>$ that live in point space, and line vectors $\<u,v,w\>$ that live in line space, defined by the following equations:
\begin{equation*}\begin{array}{l}
\geoR\textR{90pt}{$z^2 = 1-x^2-y^2$}\textR{90pt}{$u^2+v^2 = 1-w^2$}\textR{144pt}{\textcolor{redyellow}{sphere} • \textcolor{magentared}{projective·plane}}\\
\geoB\textB{90pt}{$z^2 = 1      $}\textB{90pt}{$u^2+v^2 = 1   $}\textB{144pt}{\textcolor{bluemagenta}{plane} • \textcolor{cyanblue}{plane}}\\
\geoG\textG{90pt}{$z^2 = 1+x^2+y^2$}\textG{90pt}{$u^2+v^2 = 1+w^2$}\textG{144pt}{\textcolor{greencyan}{hyperboloid} • \textcolor{yellowgreen}{hyperboloid}}
\end{array}\end{equation*}

The last column indicates which of the spaces the equations apply to, with color denoting the particular tiling (see below).  
The projective plane can be thought of as a sphere where diametrically opposite points are identified. For the plane and hyperboloid, we choose the positive-$z$ branch of the surface, or equivalently identify points with opposite $z$-values.

For convenience, we also define the following notation:
\begin{equation*}\begin{array}{l}
\geoR\textR{90pt}{$\sR\chi = \sin\chi $}\textR{90pt}{$\cR\chi = \cos\chi $}\textR{144pt}{\textcolor{redyellow}{sphere} • \textcolor{magentared}{projective·plane}}\\
\geoB\textB{90pt}{$\sB\chi = \chi     $}\textB{90pt}{$\cB\chi = 1        $}\textB{144pt}{\textcolor{bluemagenta}{plane} • \textcolor{cyanblue}{plane}}\\
\geoG\textG{90pt}{$\sG\chi = \sinh\chi$}\textG{90pt}{$\cG\chi = \cosh\chi$}\textG{144pt}{\textcolor{greencyan}{hyperboloid} • \textcolor{yellowgreen}{hyperboloid}}
\end{array}\end{equation*}

We represent discs by their centers, which are point vectors, and use the words ``disc'' and ``point'' interchangeably. 
We suppose the discs have radius $\varrho$, measured along geodesics on the surface, and require that they do not overlap with each other. 
This means that for any two points $\bm\xi$, $\bm\zeta$ on the surface, we 
require $\text{dist}(\bm\xi,\bm\zeta) ≥ 2\rho$, where $\text{dist}(\cdot,\cdot)$ is the geodesic distance.
Given a collection of points on a surface, we call it a \emph{parking} if no discs overlap, and the parking is \emph{complete} if there is no space to add another point without violating a distance constraint.

The plane and hyperboloid are unbounded so we impose boundary conditions by considering periodic parkings on regular polygona that tile the entire space by translation. 
Each single polygon is a tile, and for each point that is parked on one tile, there is a point parked on all the other tiles at the same relative location. 
This approach also works for the projective plane, which can be treated with as a sphere with a particular tiling, as well as the sphere which is a single tile. 

For any given space we consider tilings $[p,q]$ consisting of $p$-polygons, meeting $q$ at each vertex.
In our examples and numerical simulations we consider the following tilings: plane: $[4,4]$, $[6,3]$  (tiled by squares, hexagons respectively), hyperboloid: $[8,8]$, $[10,5]$ (tiled by octagons, decagons respectively), projective plane: $[p,2]$, sphere: $[2,1]$. 
The tilings for the projective plane consist of two half-spheres, with $p$ vertices along the equator ($p$ must be even); all values of $p$ are equivalent. The tiling for the sphere is simply the whole sphere, and it is written $[2,1]$ because 
the sphere can be cut along a meridian between the poles to form a 2-gon which covers the entire sphere. 
We find it convenient to label these tilings in colour, so will denote them $\{\textcolor{yellowgreen}{[8,8]},\textcolor{greencyan}{[10,5]},\textcolor{cyanblue}{[4,4]},\textcolor{bluemagenta}{[6,3]},\textcolor{magentared}{[p,2]},\textcolor{redyellow}{[2,1]}\}$.
Other tilings are also possible and we give a list of the families in the appendix section \ref{app:tiles}.

An important property of the tile is its inradius $\epsilon$, the distance from the center to the nearest edge midpoint. This sets an upper bound for the disc radius $\rho ≤ \eps$, for otherwise a disc would overlap with a translated copy of itself on a neighbouring tile. 
Therefore, $\eps$ gives a natural scale with which to measure the radius $\rho$. 
For the sphere, we set $\epsilon$ equal to half the circumference. 
Formulas defining the tiles and $\epsilon$ are given in the appendix, section \ref{app:coords}.

The algorithm proceeds in two stages: a coarse stage, which chooses points uniformly from the entire surface, 
and a fine stage, which  calculates the regions where new points can park, and then concentrates new points near these.
We give an overview of each of these stages, leaving the mathematical details to section \ref{sec:algorithm}.

\subsection{Coarse stage}\label{sec:coarse}
In the first stage we choose points uniformly at random from the entire surface, and reject them if they are too close to other parked points. 

On the sphere, we can sample uniformly from the entire surface, since this can be thought of as a single tile. For the other surfaces, we identify  
 a disc enclosing a single tile, the \emph{coarse disc}, and sample uniformly from this, resampling if the point falls outside the tile. 
 Once we have a sample point, we check the distances to all other points, and reject it if it is within a distance of $2\varrho$ from any other parked point. For the sphere this is straightforward, and for the tiled surfaces we must also consider the translated copies of the tile. 
 
 If the point is accepted (not rejected), then we calculate the boundaries of the regions where new points can be added. This is done so we know when to stop: if there are no more boundaries, then there is no more space to add points.
 
\begin{figure*}
\begin{center}
\includegraphics[width=4.5in]{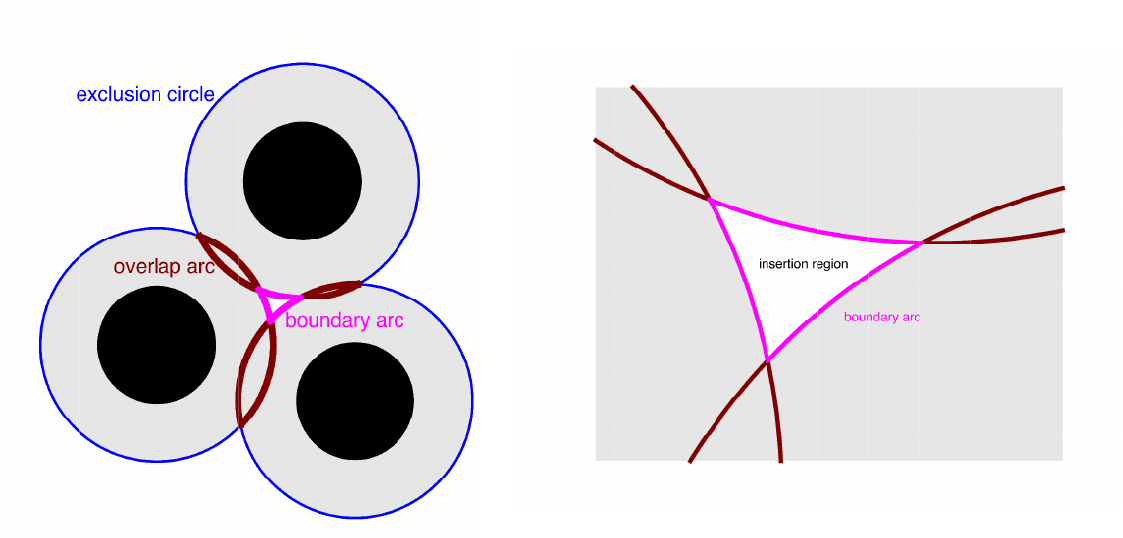}\\
\end{center}
\caption{Left: example of three parked discs (black circles), along with their exclusion circles (boundary of grey regions; union of blue, red, pink arcs), overlap arcs (union of red, pink arcs) and boundary arcs (pink). 
Right: zoomed-in image showing the small, polygonal insertion region bounded by boundary arcs.}\label{fig:exclusioncircle}
\end{figure*}

 The key observation behind this calculation 
is that each parked point defines a disc of radius $2\rho$ in which no other point can be parked. We call the boundary of this disc the  \emph{exclusion circle.}
Some parts of each exclusion circle will overlap with other exclusion circles;  we call these covered arcs  \emph{overlap arcs.}
If a piece of an exclusion circle does not overlap with any other exclusion circle, then it forms part of the boundary of a region where we can insert points. We call such an arc a \emph{boundary arc}, and the region it surrounds an \emph{insertion region}. 
See Figure  \ref{fig:exclusioncircle} for an illustration of the terminology. 
Typically, the boundary arcs can be pieced together to form closed loops, so the insertion regions are curved polygons bounded by circular arcs, as in  Figure \ref{fig:exclusioncircle} (left). 
This is not always the case; for example the insertion regions could be annuli, or have a more complicated topology, an issue we discuss in section \ref{sec:errors}.

We terminate coarse stage, and move to fine stage, once there are a fixed number of rejections $n_{\text{coarse}}$ in a row. In addition, if there are no more boundary arcs, then we terminate the sample altogether, and re-start the algorithm with a new sample.

Note that other termination criteria are possible, such as  a fixed number of trial points, or a geometrical criterion that detects whether all regions of available area are polygons. The latter would be the best, but we have found it difficult to formulate this criterion mathematically and computationally. 
Many RSA algorithms used in practice end at this stage, without testing whether there is any available area.

\subsection{Fine stage}
In this stage we explicitly compute the insertion regions where a point can be parked without overlapping another. New points are then concentrated near these regions. 

We first calculate the list of boundary arcs, as in section \ref{sec:coarse}. 
Then, we sew them together to find the boundaries of the insertion regions 
using our table of boundary arcs, which lists their neighbours on either side. 
We end up with a list of loops of boundary arcs, where each loop by assumption forms the boundaries of a small, polygonal insertion region.  

To sample from the insertion regions, we enclose each polygon in a \emph{fine disc} centered at the centroid of the region, 
whose radius is the distance to the farthest vertex of each polygon. 
Then, we sample points uniformly from the fine discs, where each disc is chosen in proportion to its area. If the point also lies in the polygon inside the corresponding disc, then it is tentatively parked; otherwise, the point is rejected. This ensures the point is sampled uniformly from the insertion regions. Once a point is tentatively parked, we check the distances with other parked points, and reject if it is too close. If we ultimately accept the point, then we calculate the boundary arcs again. We stop if there are none, but otherwise repeat the above steps by computing the insertion regions.

\subsection{Examples of random parkings}\label{sec:visual}

\begin{figure*}
\center\includegraphics[width=4.5in,height=6in]{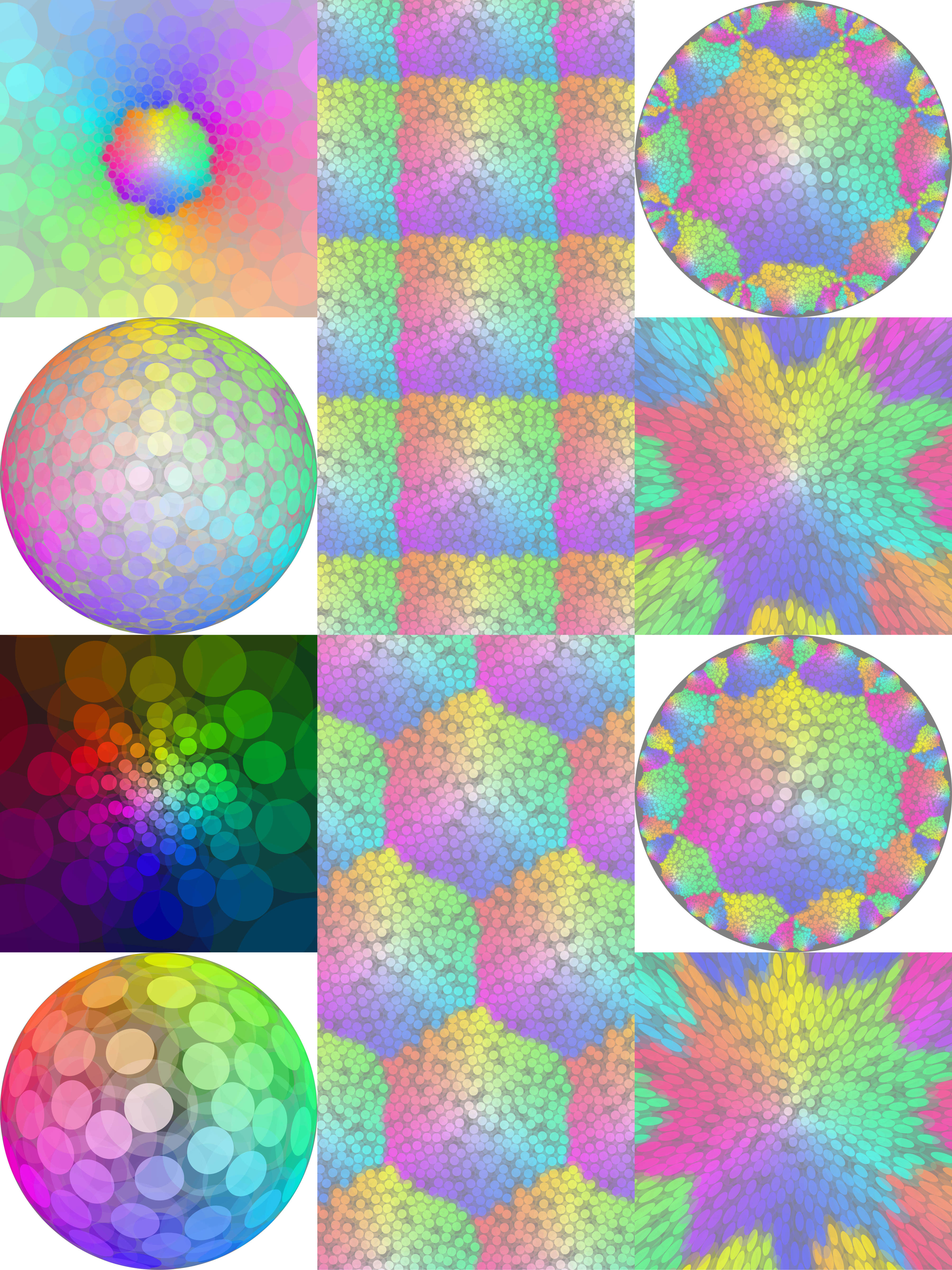}
\caption{Examples of complete random parkings on each of the spaces we considered, for  $\rho = \eps e^{{-}3}$. Discs are colored and exclusion circles are transluscent circles surrounding them. 
Each parking on each space is shown in two projections: Stereographic (A) and Orthogonal (B). Top row: projective plane, plane [4,4], hyperboloid [8,8], all in projection A. Second row: same as top, but projection B. Third row: sphere, plane [6,3], hyperboloid [10,5], projection A. Last row: same as third row, but projection B. }
\label{fig:parkings}
\end{figure*}

Figure \ref{fig:parkings} shows a typical output of the algorithm, one complete parking for each space using a disc radius such that $\ln\tfrac{\eps}{\rho} = 3$. Discs are plotted in less-transparent colours and exclusion circles are the same colour as the corresponding disc, but more transparent. One can verify by visual inspection that the exclusion circles completely cover each space, so these are complete parkings.

\begin{figure*}
\center\includegraphics[width=4.5in,height=6in]{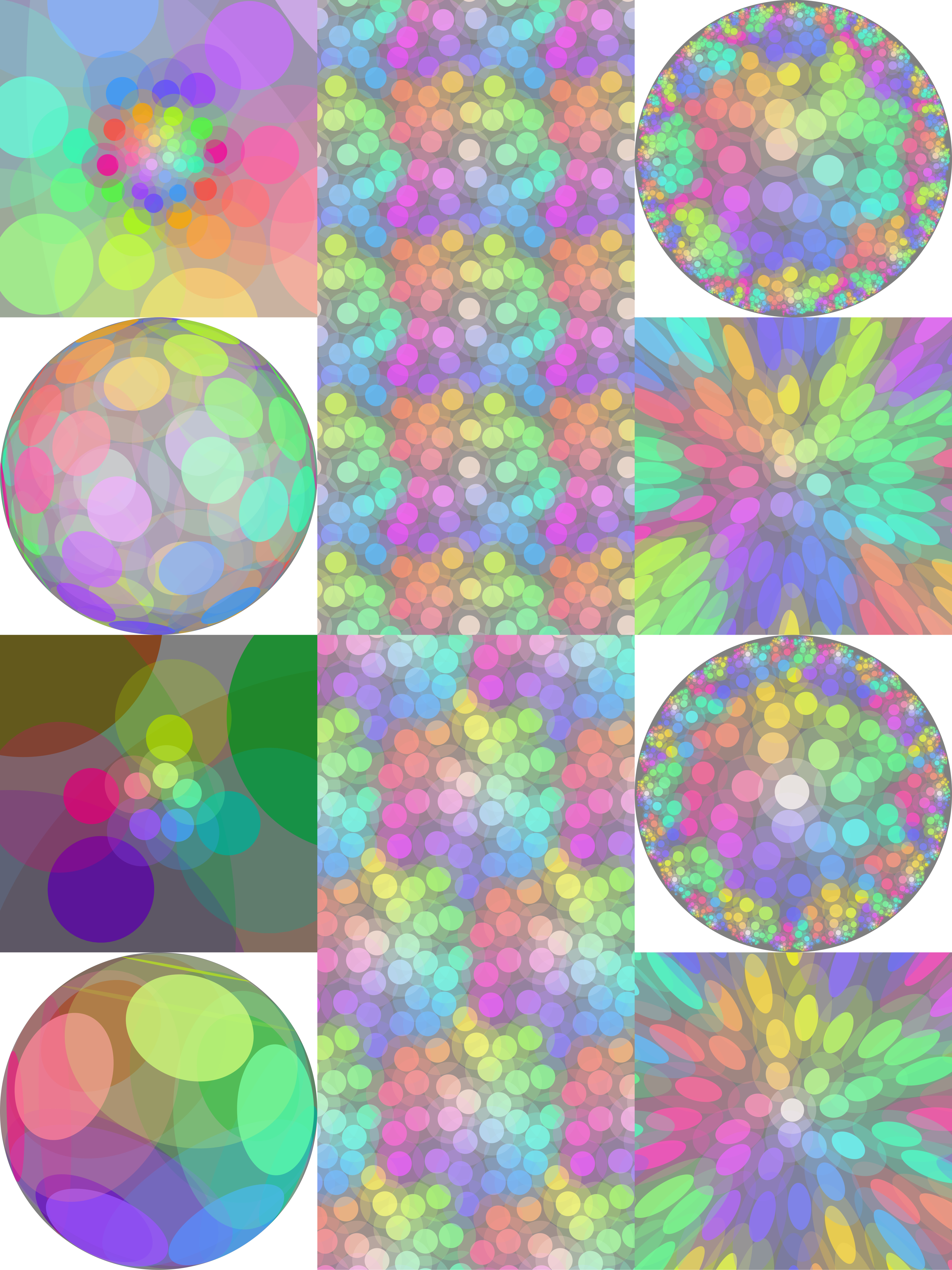}
\caption{
Examples of incomplete random parkings on each of the spaces we considered, for $\rho = \eps e^{{-}2}$, obtained by stopping the program just before switching to fine mode. The transluscent exclusion circles no longer cover the entire space: there are small holes, the insertion regions, where a new disc could fit. 
The order of the spaces is the same as in Figure \ref{fig:parkings}.
}
\label{fig:holes}
\end{figure*}

Figure \ref{fig:holes} shows examples of incomplete parkings which were terminated just before moving to fine stage, also using $\ln\tfrac{\eps}{\rho} = 3$. 
We purposely selected samples with a large number of holes, to illustrate the shape of typical insertion regions. Now the exclusion circles do not cover the entire area of the tile, but rather leave uncovered a few tiny, polygonal regions where one can add a new disc. Clearly sampling uniformly from the entire coarse disc would be extremely inefficient. 

\begin{figure*}
\center\includegraphics[width=4.5in,height=6in]{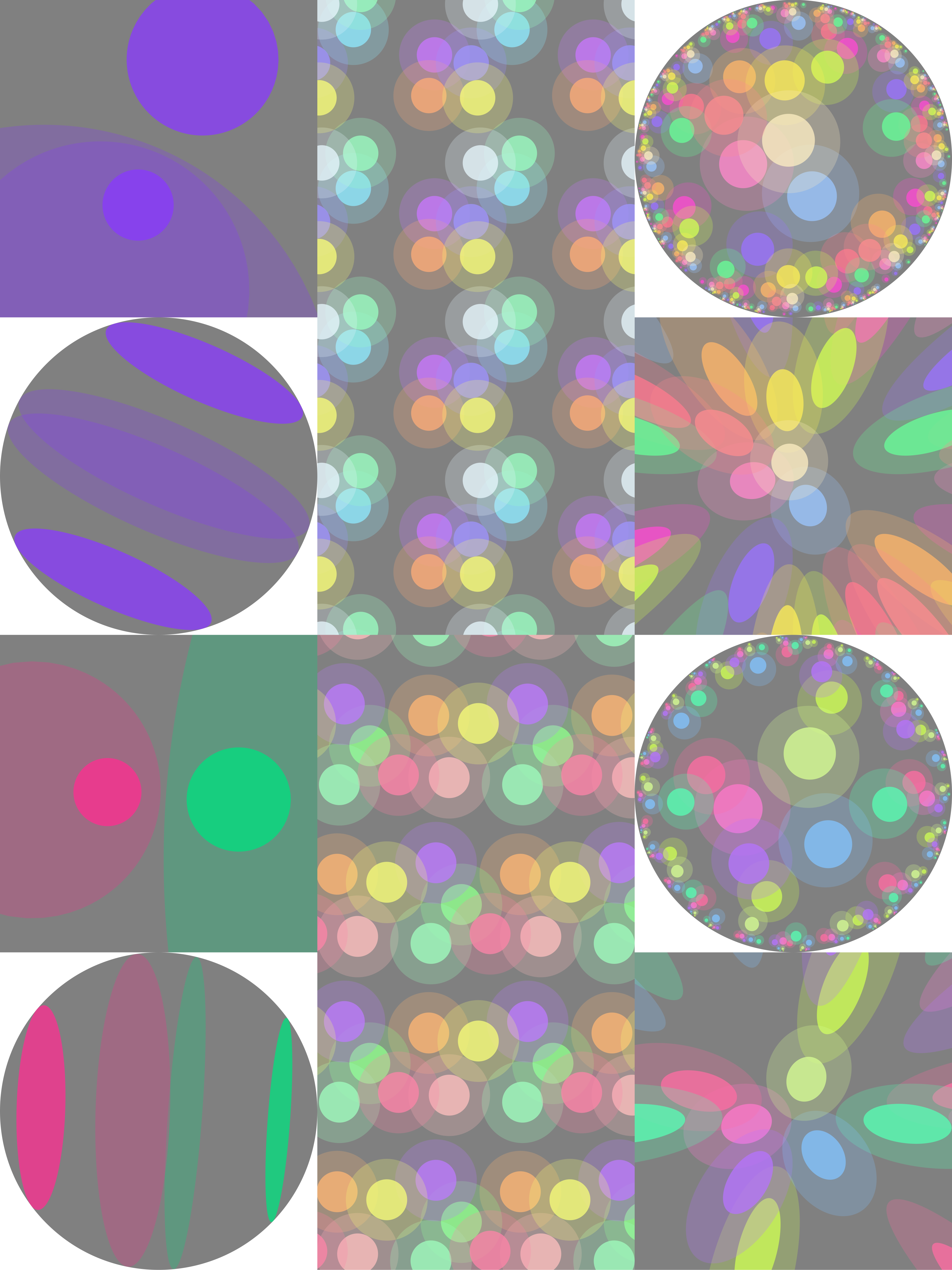}
\caption{Examples of non-polygonal insertion regions, $\rho = \eps e^{{-}3/2}$. The space not covered by the transluscent exclusion circles  stretches across the boundary of each tile and connects to a translated version of itself. 
The ordering of the pictures is the same as in Figures \ref{fig:parkings}, \ref{fig:holes}.}
\label{fig:strips}
\end{figure*}

Figure \ref{fig:strips} shows examples of insertion regions that are not polygonal, a problem that will be discussed in more detail in section \ref{sec:errors}. 

\clearpage

\clearpage

\clearpage

\clearpage

\subsection{Summary of the algorithm}

Here is an outline of the algorithm. Details for each point are given in section \ref{sec:algorithm}. 
\begin{enumerate}[(A)]
\item Coarse stage 
  \begin{enumerate}[(1)]
    \item Sample a point uniformly from the coarse disc
    \item Check distances with other parked points, and either accept or reject
    \item Calculate boundary arcs 
    \item Check exit criteria: if there are no boundary arcs, then stop algorithm, or if $n_{coarse}$ points in a row were rejected, move to fine mode. 
  \end{enumerate}
\item Fine stage 
  \begin{enumerate}[(1)]
    \item Compute the polygonal insertion regions 
    \item Construct the fine discs enclosing the polygons
    \item Sample a point uniformly from the insertion regions
    \item Check distances with other parked points, and either accept or reject
    \item Calculate boundary arcs 
    \item Check exit criterion: if there are no boundary arcs, then stop algorithm. 
  \end{enumerate}
\end{enumerate}

%%%%%%%%%%%%%%%%%%%%%
%%% %%   Numerical Results     %%%%%
%%%%%%%%%%%%%%%%%%%%%

\section{Numerical results}\label{sec:numerical}

\begin{figure*}
\center\includegraphics[width=4.5in,height=6in]{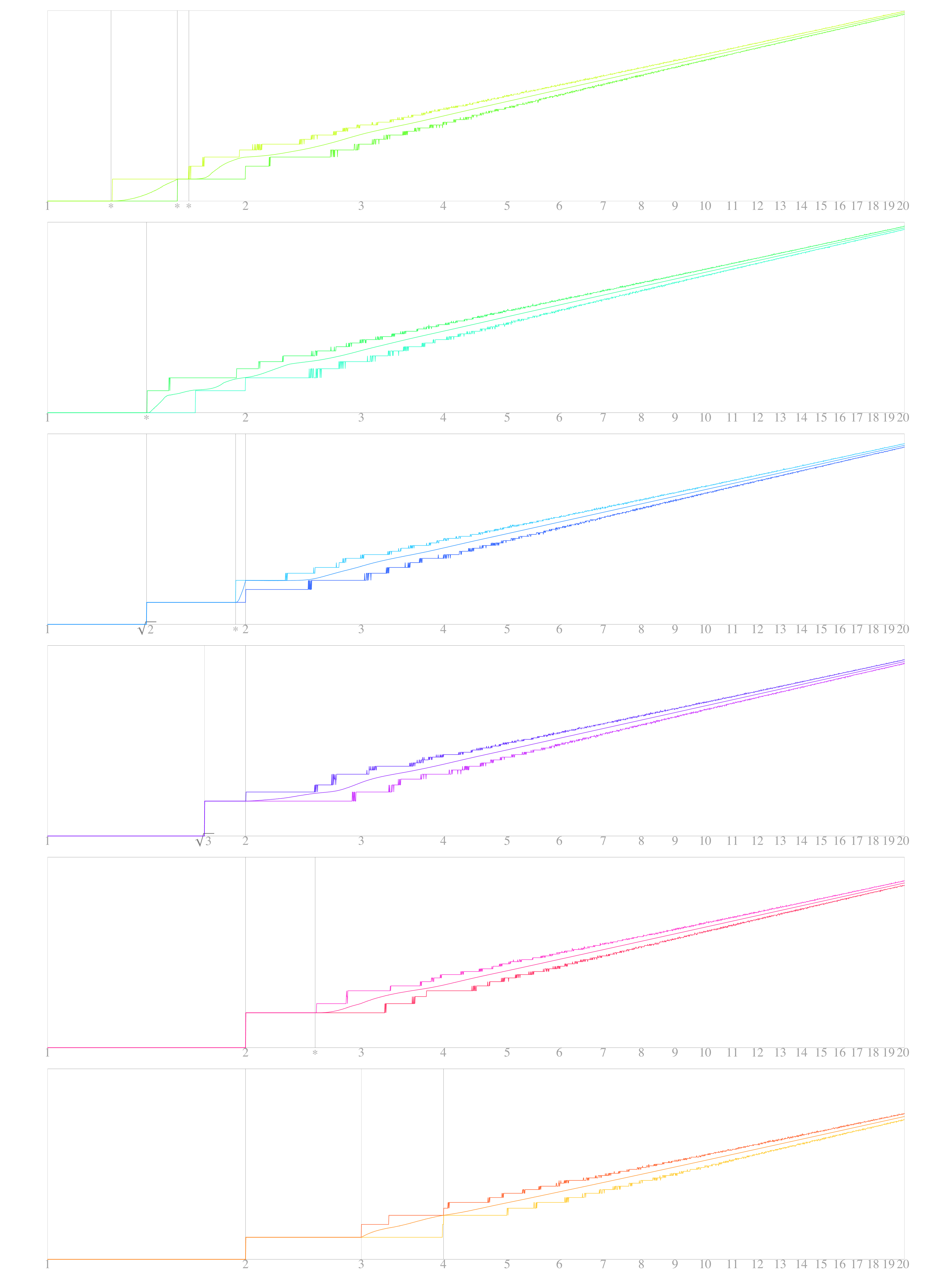}
\caption{Average number of discs parked (middle lines), and maximum and minimum numbers (top, bottom curves), as a function of $\ln\tfrac{\eps}{\rho}$ (horizontal axis, labels are $\tfrac{\eps}{\rho}$). Vertical axis is the logarithm of all three quantities and ranges from 0 to 6. From top to bottom: $\{\textcolor{yellowgreen}{[8,8]},\textcolor{greencyan}{[10,5]},\textcolor{cyanblue}{[4,4]},\textcolor{bluemagenta}{[6,3]},\textcolor{magentared}{[p,2]},\textcolor{redyellow}{[2,1]}\}$.}
%\caption{vertical: $[0 ≤ \ln(\{{\rm min},{\rm mean},{\rm max}\}\,{\rm cluster{\cdot}size}) ≤ 6]$\hbox to 5in{}
%horizontal: $[0 ≤ \ln({\rm max{\cdot}radius/radius}) ≤ 3]$ • tiles: $\{\textcolor{yellowgreen}{[8,8]},\textcolor{greencyan}{[10,5]},\textcolor{cyanblue}{[4,4]},\textcolor{bluemagenta}{[6,3]},\textcolor{magentared}{[p,2]},\textcolor{redyellow}{[2,1]}\}$}
\label{fig:dens1}
\end{figure*}

\begin{figure*}
\center\includegraphics[width=4.5in,height=6in]{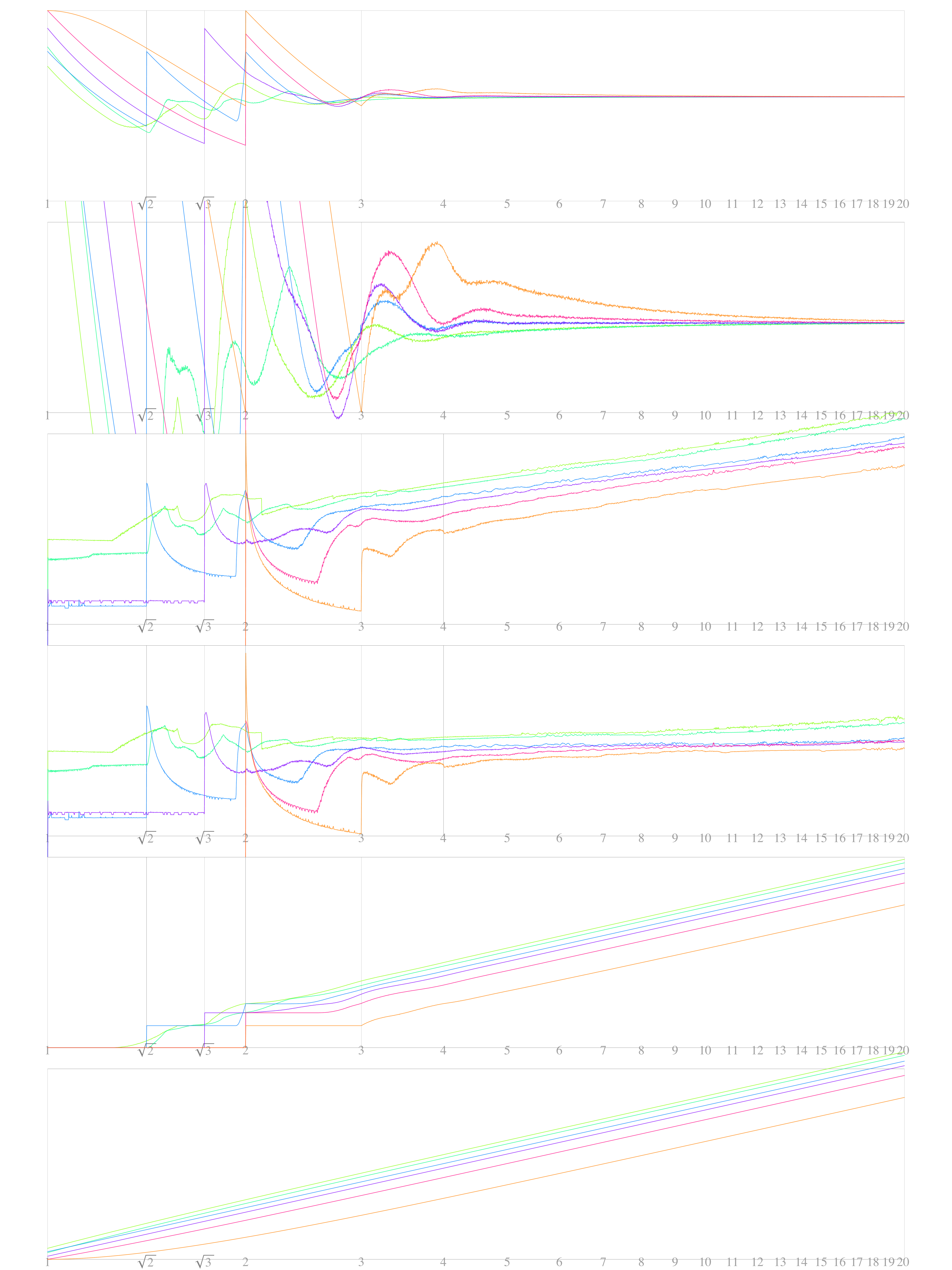}
\caption{Averages of various measurements as a function of $\ln\tfrac{\eps}{\rho}$ (horizontal axis, labels are $\tfrac{\eps}{\rho}$). From top to bottom: (1) Density of coverage, vertical axis $[0,1]$. (2) Density of coverage, vertical axis $[0.5,0.6]$. 
(3) Logarithm of cpu time in second (vertical axis [{-}6,6]). 
(4) Logarithm of (cpu time / average number of discs parked) (vertical axis [{-}6,6]). 
(5) Logarithm of (tile area / disc area) (vertical axis [0,6]). 
(6) Logarithm of average number of discs parked (vertical axis [0,6]).
Curve colors correspond to the following tiles: $\{\textcolor{yellowgreen}{[8,8]},\textcolor{greencyan}{[10,5]},\textcolor{cyanblue}{[4,4]},\textcolor{bluemagenta}{[6,3]},\textcolor{magentared}{[p,2]},\textcolor{redyellow}{[2,1]}\}$
%\tobeth{I wish we could get rid of the bottom two plots.}
}
\label{fig:dens2}
\end{figure*}

We used our numerical algorithm to investigate and compare the statistical properties of random parkings on different spaces. 
%We consider the following tilings: plane: $[4,4]$, $[6,3]$  (squares, hexagons), hyperboloid: $[8,8]$, $[10,5]$ (octagons, decagons), projective plane: $[p,2]$, sphere: $[2,1]$, written $\{\textcolor{yellowgreen}{[8,8]},\textcolor{greencyan}{[10,5]},\textcolor{cyanblue}{[4,4]},\textcolor{bluemagenta}{[6,3]},\textcolor{magentared}{[p,2]},\textcolor{redyellow}{[2,1]}\}$.
We generated random parkings on each of the six different spaces $\{\textcolor{yellowgreen}{[8,8]},\textcolor{greencyan}{[10,5]},\textcolor{cyanblue}{[4,4]},\textcolor{bluemagenta}{[6,3]},\textcolor{magentared}{[p,2]},\textcolor{redyellow}{[2,1]}\}$ for values of radius equally spaced at intervals of $0.001$ in log-space, so the the range is $\rho\in\{e^0,e^{0.001},\cdots,e^{2.999},e^3\}$. For each value of $\rho$ we generated $10^4$ samples (complete parkings) and kept track of the number of points parked in each sample. We used a coarse-to-fine switching parameter of $n_{\text{coarse}} = 10^4$. We now analyze the statistical properties of the parkings both for large $\rho$ (section \ref{sec:largep}), and for small $\rho$ in the limit that the $\rho\to 0$ (section \ref{sec:density}), and comment on the efficiency and accuracy of the algorithm (section \ref{sec:alganalysis}). 
The results are shown in Figures \ref{fig:hist}, \ref{fig:dens1}, \ref{fig:dens2}, and are discussed momentarily. 
Note that in all figures referenced, the horizontal axis is units of $\tfrac{\eps}{\rho}$ so the radius decreases as we move to the right.

\subsection{Large discs and critical radii}\label{sec:largep}

\begin{figure*}
\center\includegraphics[width=4.5in,height=6in]{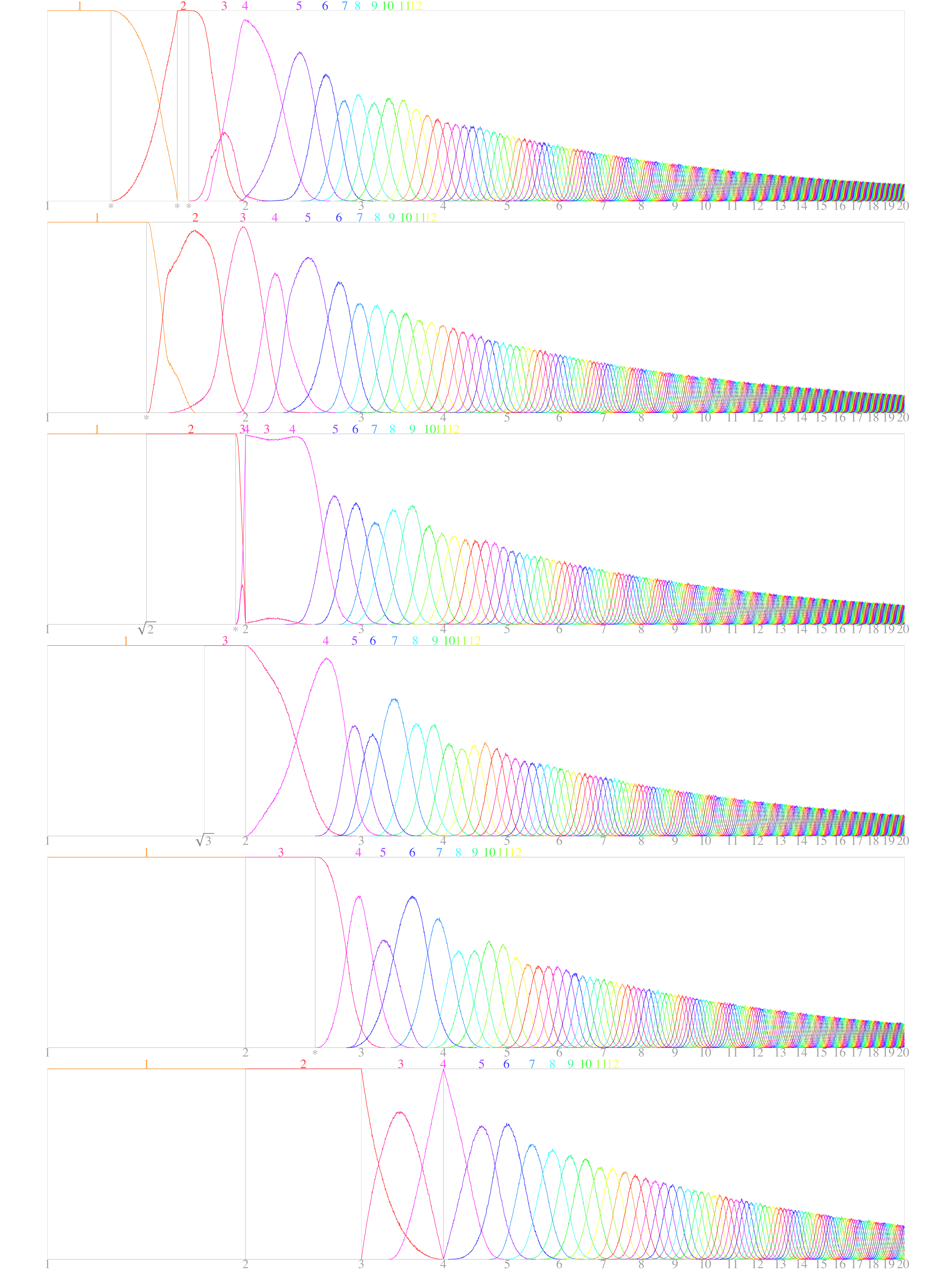}
\caption{Curves showing the number distribution of parked discs in a complete parking as a function of $\ln\tfrac{\eps}{\rho}$ (horizontal axis; labels are $\tfrac{\eps}{\rho}$). 
Each curve labelled $k$ (top) shows the fraction of parkings with $k$ discs parked. 
Vertical axis ranges from 0 to 1. From top to bottom: $\{\textcolor{yellowgreen}{[8,8]},\textcolor{greencyan}{[10,5]},\textcolor{cyanblue}{[4,4]},\textcolor{bluemagenta}{[6,3]},\textcolor{magentared}{[p,2]},\textcolor{redyellow}{[2,1]}\}$.}
%hyperboloid [8,8], hyperboloid [10,5], plane [4,4], plane [6,3], projective plane [p,2], sphere [2,1]. }
%\caption{vertical: $[0 ≤ {\rm cluster{\cdot}size} ≤ 1]$ distribution\hbox to 5in{}
%horizontal: $[0 ≤ \ln({\rm max{\cdot}radius/radius}) ≤ 3]$ • tiles: $\{\textcolor{yellowgreen}{[8,8]},\textcolor{greencyan}{[10,5]},\textcolor{cyanblue}{[4,4]},\textcolor{bluemagenta}{[6,3]},\textcolor{magentared}{[p,2]},\textcolor{redyellow}{[2,1]}\}$}
\label{fig:hist}
\end{figure*}

Figure \ref{fig:hist} shows the histograms of the number of points parked as a function of disc $\rho$, for the six different spaces. In each plot there are many curves, each showing the fraction, or probability, of parking a given number of discs as $\rho$ varies. We call a parking a \emph{$k$-parking} when it has exactly $k$ discs parked, and call the curve measuring the fraction of $k$-parkings the \emph{$k$-curve}.  At any particular value of $\rho$ the sum of the values of the $k$-curves therefore equals 1.

\clearpage
\clearpage

\clearpage
\clearpage

For very large $\rho$ (left on the horizontal axis) all parkings are 1-parkings: they yield exactly 1 disc. As $\rho$ decreases, other numbers become possible and typically we see a distribution of $k$-values: some samples yield more parked discs than others. Even though all samples are complete parkings, it is possible for them to contain different numbers of discs because depending on how the discs are arranged, which is random, they may block more or less of the space for other discs to park. 

Interestingly, for some spaces there are particular values of $\rho$ for which all samples appear to give exactly the same number $k>1$ of discs. 
On the sphere there is a particular value of $\rho$ at which all samples are 4-parkings.   On the projective plane, there are an interval of values for which all samples are 3-parkings. 
When such a value of $\rho$ is isolated we call it a \emph{critical radius} and when it occurs over an interval we call it a \emph{critical interval}. The only true critical radius we have found is on the sphere, although on the plane [4,4] there is a radius which is almost critical, where the fraction of 4-parkings reaches 99.3\%.
Critical intervals appear to occur for the hyperboloid [8,8] (2-parkings), plane [4,4] (2-parkings), plane [6,3] (3-parkings), projective plane (3-parkings), and sphere (2-parkings). 
Except for the hyperboloid [8,8], these intervals typically occur directly after the transition from 1-parkings. 

The plane [4,4] is interesting because the histograms for 3-parkings, 4-parkings each have two local minima, and therefore the average number of points parked does not increase monotonically as the radius decreases. 

Why do these critical radii and intervals occur? We briefly discuss two cases: the critical radius on the sphere and the critical interval on the projection plane. Then  we discuss the peculiar properties of parkings on the plane [4,4].

\subsubsection{Critical radius on the sphere}

The critical radius on the sphere occurs at $\rho = \tfrac{\eps}{4}$, where there are always exactly 4 discs parked. For a unit sphere this corresponds to a geodesic radius of $\tfrac{\pi}{4}$, a Euclidean space radius of $\sqrt{\tfrac{1}{2}}$, or a constraint that disc centers subtend an angle of at least $\tfrac{\pi}{2}$. 
Why does this critical radius to exist? 

A heuristic explanation is as follows. First, there must always be at least four points parked. To see why, suppose there are already three points parked. These lie on a plane which can be rotated to lie either entirely above or entirely below the equator of the sphere. Then, another point can fit on the opposite pole. Next, we must argue there cannot be more than four points parked. Actually we can fit six points, by putting two points on the north and south poles, and four around the equator. But, to fit in five or more points requires at least three to be exactly on the equator, an event with Lebesgue measure zero, hence probability zero. %(see figure \ref{fig:2}, box \note{is it somewhere in this figure?}). 

One might wonder how this critical radius behaves in different dimensions -- is it special to parkings in dimension 3? We prove in section \ref{sec:critical} that for random parkings on the surface of a $d$-dimensional sphere, the same critical radius always yields exactly $d{+}1$ parked discs. 
This also rigorously proves the heuristic argument above, for $d=3$. 

We can use this theoretical result as a benchmark for our algorithm: 
we ran the algorithm using the critical radius $\rho = \tfrac{\eps}{4}$ (accurate to floating-point precision) and always parked exactly 4 discs -- we never saw any other sizes.
This partly verifies that the program is working correctly, i.e. that it is simulating complete random parkings. 

Other studies have noticed this critical radius without being able to simulate it exactly. For example, Mansfield et al. \cite{Mansfield:1996wu} considered small spheres attaching to the surface of a larger sphere, a problem which is equivalent to discs parking on the surface of a sphere. The authors noticed that at a particular ratio of sphere radii, the parking problem produced nearly 100\% yield of clusters with 4 small spheres parked. Because they ran simulations in the equivalent of our coarse mode only, and did not compute the geometrical stopping criterion, they were not able to observe that the yield is in fact exactly 100\%. 
Later, in \cite{Schade:2013ee} we used the algorithm described here to verify that the yield is 100\%, and then used this fact to experimentally create clusters  of colloidal particles consisting of 4 small spheres bound to a central sphere with high yield.

\subsubsection{Critical interval on the projective plane}

A critical interval occurs on the projective plane, where for $\rho\in(\rho_c,\tfrac{\eps}{2})$, there is always exactly 3 discs parked, where $\rho_c\approx \tfrac{\eps}{2.56}$ is  estimated numerically. Recall that for the projective plane  $\tfrac{\eps}{2} = \tfrac{\pi}{4}$, which is the same geodesic distance as the critical radius on the sphere. 

The critical interval occurs here for a similar reason as on the sphere. We can think of the projective plane as a sphere such that whenever one point is parked, another must be parked at the opposite pole.
For $\rho > \tfrac{\eps}{2}$ any single point parked on the projective plane blocks all other points, for the same reason that two points on opposite poles on the sphere block all other points (Theorem \ref{thm:bounds}, (i) in  section \ref{sec:critical}). 
Now consider $\rho < \tfrac{\eps}{2}$. Without loss of generality, we can assume the first point is parked at the north pole. Then there is a small annulus around the equator where another point can fit, and no matter where it is put, there are two small intervals to put a third (and its polar opposite). Therefore we can always fit at least 3 discs on the projective plane. We won't be able to fit 4, until we are able to fit at least 8 on the sphere, so this behaviour persists over an interval.

\subsubsection{Not-quite critical radius on the plane [4,4]}

The plane [4,4] looks like it might have a critical radius at $\rho = \tfrac{\eps}{2}$, where from far away the 4-curve appears to reach 1. 
However, zooming in on the figure (or carefully analyzing the neighbouring data points) shows that, numerically at least, the fraction of 4-parkings never reaches exactly 1 -- the maximum we have ever obtained numerically is 99.3\%. 

Is this a real critical radius, with 100\% yield, (in which case our code would be making detectable errors)? 
Or is there actually a small probability of obtaining a 3-parking? 
We argue that this effect is real -- that the yield of 4-parkings never reaches exactly 100\%, but rather  there is a very tiny probability of obtaining 3-parkings even at $\rho = \tfrac{\eps}{2}$, a probability our code has correctly detected. 

\begin{figure*}
\begin{center}
\includegraphics[width=4.5in]{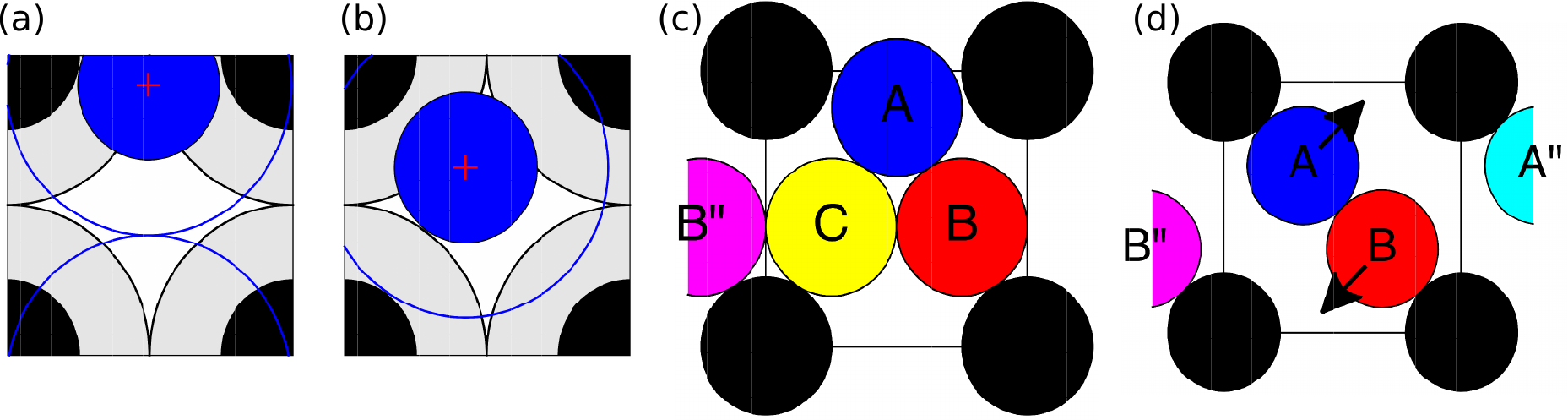}
\end{center}
\caption{(a,b,c) Illustration of the near-critical behaviour on the plane [4,4]. A disc is parked near the corner of the tile (black filled circle), and another (blue filled circle) attempts to park in the remaining space. Each discs' exclusion circle is black or blue respectively and the insertion region is white. 
If the blue disc parks near the cusp (a) or middle (b) of the insertion region, there is room for two more discs. 
Only a very special configuration obtained from (c) as described in the text will allow exactly 3 discs, blocking a fourth. 
(d) Illustration of the non-monotonic change in average number of discs with radius. 
When the discs are sheared as illustrated, they form a configuration where there is no space for a fourth disc. As the discs become larger, the size of the region they can access by shearing shrinks, so the size of the region where they can block a fourth disc shrinks also.}
\label{fig:44}
\end{figure*}

Here is a heuristic explanation. 
Let the tile be a square with side length 2 centered at the origin, and the discs have radius $\tfrac{1}{2}$, so the exclusion circles have radius 1. 
Suppose, without loss of generality, that the first disc P is parked at the corner $({\pm}1,{\pm}1)$.
The insertion region is a concave diamond-shaped region in the middle of the square, with cusp-like corners at the top, bottom, left, and right; see Figure \ref{fig:44} (a,b). 
The only way to cover the whole area of the diamond is to put a point exactly in the middle, but this is an event with probability zero. Therefore, we can assume the second point is somewhere in the interior of the north-west corner of the diamond, and we will always be able to fit a third point. 

If the second point is near a corner (say the top one), then there is a wiggly strip that crosses the diamond (say horizontally) where new points can be added.
The length of this strip is 2, and it has a finite width except at a single point, so no matter where a third point parks, there will be space for a fourth one (Figure \ref{fig:44} (a)).
If the second point is near the middle of the diamond (say in the top-left quadrant), then there will be two neighbouring corners (the bottom and right corners) with a small amount of space to park. Since these corners are a distance of $\sqrt{2}$ apart, we can usually put one point in each of them (Figure \ref{fig:44} (b)).
Therefore, we expect to park four points with high probability. 

To show there is an arrangement where only three points can park, consider the setup in Figure \ref{fig:44} (c). Here discs A,B,C are exactly tangent, forming an equilateral triangle with center aligned with the center of the tile. Disc A is at $\<0,\sqrt{3}{-}1\>$, disc B is at $\<\tfrac{1}{2},\tfrac{1}{2}\!\sqrt{3}{-}1\>$, disc C is at $\<{-}\tfrac{1}{2},\tfrac{1}{2}\!\sqrt{3}{-}1\>$. Note that disc C is also tangent to B'', the translated copy of B in the neighbouring tile to the left. 
We will show how to move A, B a small amount, to exclude any space for C. 
Suppose disc A moves a small amount to the left, by rolling along B. It can also move a small amount up or down without hitting P or B, so it can move around in an open neighbourhood without overlapping B. Disc B can now move a small amount sideways (with a bit of wiggle room up and down) without overlapping A or P. However, disc C now has no room to park, since if it does not move, it overlaps with A, and if it moves sideways with A, it overlaps with B''. 
Therefore, we obtain a configuration A, B, P, where we cannot fit a fourth point, and each of A,B can move a little bit within this configuration, so the set of such configurations has non-zero probability. 
The probability of such a configuration should be very small, since it depends on the size of the gap between A and P, which in turn is very small.

\subsubsection{Non-monotonic density for the plane [4,4]}

The near-critical radius on the plane [4,4] is related to a peculiar property: the average number of parked discs does not increase monotonically as $\rho$ decreases; there is a small interval of radii over which the average number actually decreases. This is evident from the histograms, which show that 
 as $\rho$ decreases beyond the near-critical value, the number of 4-parkings drops slightly while the number of 3-parkings increases (the number of 4-parkings eventually rises again before dropping to zero). The number of 5-parkings is zero over most of this range, so the average number of discs parked (and hence the density of discs) actually decreases over an interval. 
This is counterintuitive, because one would expect that as the discs get smaller, we should on average be able to fit more of them in a given space. 

We give a heuristic explanation for this non-monotonic behaviour. The basic idea is that as the radius decreases, space to put in new points opens up in just the right place so that a disc parked there blocks space for any other discs. 
Suppose the discs are small enough that exactly 3 fit along the diagonal, as in Figure \ref{fig:44} (d).
This configuration blocks all the space, leaving only 3 points parked. There is room to wiggle within this configuration, by shearing the interior discs: moving A to the north-east, and B to the south-west. This still blocks all the space, until we shear them too much, in which case space will open up on the left-hand side of the square. 

As the discs grow larger, we can still use this set of configurations to achieve a 3-parking as long as the discs have been sheared from their diagonal positions.
When the discs are too big, however (at the near-critical-radius size), the interior discs cannot fit anywhere near the diagonal, and there is only a very small region where this shearing adjustment can make them fit. Therefore, the set of configurations where they can block space for a fourth disc is smaller, so it is more likely to be able to put in a fourth discs. For this space and boundary conditions, smaller discs can more effectively block a fourth one, so are more likely to lead to a 3-parking.   This argument shows the  boundary conditions can have complex effects, because they are nonlocal.

\subsection{Small-disc parkings: toward a limiting density}\label{sec:density}

We now compare parkings in the different spaces in the limit as $\rho\to 0$. 
Figure \ref{fig:dens1} shows  the average number of discs parked as a function of $\rho$, as well as the maximum number and minimum numbers found numerically. All curves appear to asymptote to a line with a slope of 2 in log-log space, 
which is as expected since the surfaces are two-dimensional. 
%Figure \ref{fig:dens2} (2nd from top) shows the average number for each space, on the same set of axes, where the same asymptotic slope is clear in comparison.  
%Figure \ref{fig:dens2} (top) shows the area ratios: (Area of entire tile) / (Area of one disc), for all six surfaces. These curves asymptote to a slope of 1, as expected.\note{check asymptotic slope}

A problem of interest in applications is the limiting density of discs, i.e. the average fraction of area covered by the discs (area covered by discs / area of tile), in the limit when$\rho\to 0$. 
Figure \ref{fig:dens2} (top) shows the density for all spaces as a function of $\rho$, and Figure \ref{fig:dens2} (2nd from top) shows the same density with a zoomed-in vertical axis. 
For large $\rho$, the density has significant variations both among and within the different spaces, at some points even appearing to jump discontinuously. % -- at $\tfrac{\eps}{\rho} = \sqrt{2}$ for the plane [4,4], at $\tfrac{\eps}{\rho} = \sqrt{3}$ for the plane [6,3], and at $\tfrac{\eps}{\rho} = 2$ for the sphere, projective plane, and plane [4,4]. These are exactly the critical values where the number of discs parked jumps from being entirely 1 to entirely 2 or 3, so it is natural that the density should be discontinuous. 
However, as $\rho$ decreases the densities converge to a limiting constant, which appears to be the same constant for all spaces. 
We estimate the limiting densities by computing the average of the data points for the smallest 100 radii ($18.17 ≤ \tfrac{\eps}{\rho} ≤ 20.1$), and show the results in  Table \ref{tbl:density}. 
For both tilings on the plane we find a density of $0.5471 \pm 0.0002$.  
This is consistent with the values found in the literature (most of which were obtained by extrapolating data from incomplete parkings): 
0.5473  $\pm$ 0.0009   \cite{Tanemura:1979vs},
0.5591  $\pm$ 0.0010   \cite{Lotwick:2007ej},
0.547   $\pm$ 0.002    \cite{Feder:1980jw}, 
0.5471  $\pm$ 0.0051   \cite{Hinrichsen:1986},
0.54700 $\pm$ 0.000063 \cite{Torquato:2006eh}. 
%Note that most of these techniques involved some kind of extrapolation, except \cite{Tanemura:1979vs} which did not sample uniformly from the available space, and \cite{Lotwick:2007ej}, which is far off the others for a reason we do not understand. \todo{check extrapolation comment}

The limiting densities for the sphere, hyperboloid, and projective plane appear to approach that of the plane as $\rho\to 0$, although for the smallest values for which we computed the density they are still detectably different (0.05\% lower for the hyperboloid, and 0.2\% higher for the sphere).  
It is notable that, for $\rho$ small enough that the density is monotonic, the density on the sphere and projective plane is always  strictly larger than that on the plane, and the density on the hyperboloid is always strictly smaller. This has implications for random parking on a more general curved surface: 
in regions of positive curvature, one would expect a greater local density, and in regions of negative curvature, a smaller density. 
The difference will be more noticeable for discs that are larger compared to the local radius of curvature of the surface, and it appears to be more pronounced for positively-curved surfaces. 

It is notable that for $\tfrac{\eps}{\rho}\gtrapprox 5$, there is no discernible difference in the densities between the two different tilings for each of the plane and the hyperboloid. Therefore, this is the value beyond which the the boundary conditions in each space do not matter. 

The fact that all six spaces give the same density is interesting when compared to systems where discs interact with a potential that has a decay scale, such as in the Thomson problem \cite{bowick}. There, the density of optimal configurations does depend on the curvature of the space, which causes optimal configurations to contain defects, i.e. deviations from perfect lattices, and the area ratio of defects does not disappear as the discs become very small. Therefore, the densest packing on a sphere is different from that on a plane or hyperboloid \cite{Clare:1991wq,Bausch:2003gq}, and this could potentially be the case for a random parking of discs that can interact. 

\clearpage

\begin{table}
\begin{center}\begin{tabular}{l}
\textK {30pt}{                  }\textK {60pt}{           }\textK {100pt}{density                       }\textK {30pt}{cpu }\textK {60pt}{cpu/disc}\\
\textRY{30pt}{$[\phantom{0}2,1]$}\textRY{60pt}{sphere     }\textRY{100pt}{$0.54832\quad\pm\quad 0.00032$}\textRY{30pt}{2.42}\textRY{60pt}{0.46}\\
\textMR{30pt}{$[\phantom{0}p,2]$}\textMR{60pt}{proj plane }\textMR{100pt}{$0.54740\quad\pm\quad 0.00020$}\textMR{30pt}{2.53}\textMR{60pt}{0.53}\\
\textBM{30pt}{$[\phantom{0}6,3]$}\textBM{60pt}{plane      }\textBM{100pt}{$0.54712\quad\pm\quad 0.00016$}\textBM{30pt}{2.22}\textBM{60pt}{0.22}\\
\textCB{30pt}{$[\phantom{0}4,4]$}\textCB{60pt}{plane      }\textCB{100pt}{$0.54710\quad\pm\quad 0.00018$}\textCB{30pt}{2.23}\textCB{60pt}{0.23}\\
\textGC{30pt}{$[          10,5]$}\textGC{60pt}{Hyperboloid}\textGC{100pt}{$0.54681\quad\pm\quad 0.00016$}\textGC{30pt}{2.61}\textGC{60pt}{0.60}\\
\textYG{30pt}{$[\phantom{0}8,8]$}\textYG{60pt}{Hyperboloid}\textYG{100pt}{$0.54683\quad\pm\quad 0.00013$}\textYG{30pt}{2.79}\textYG{60pt}{0.78}
\end{tabular}\end{center}
\caption{Left: Density for small $\rho$, computed using the average of the last 100 data points ($18.17 ≤ \tfrac{\eps}{\rho} ≤ 20.1$). Error bars are 95\% confidence intervals, assuming the distribution is normal. Extra decimal points than are significant are shown to give a sense of the discrepancy between tilings; the discrepancy between identical spaces with different tilings is consistent with the size of the error bars. 
Right: Slope $m$ of best-fit line for $\log(\texttt{cpu}) = m\log\tfrac{\eps}{\rho} + b$, where \texttt{cpu} is either the raw cpu-time, or the normalized cpu-time (cpu-time divided by mean number of discs).
The best-fit line was computed for data such that $\tfrac{\eps}{\rho} > 5$.
}%}\label{tbl:cpu}
\label{tbl:density}
\end{table}

\subsection{Efficiency and accuracy of the algorithm}\label{sec:alganalysis}

In this section we comment on the practical aspects of the algorithm: its efficiency, and the possibility for errors.  

\subsubsection{Efficiency}

Figure \ref{fig:dens2} (3, 4th from top) shows the CPU time of the simulations performed to make Figures \ref{fig:hist},\ref{fig:dens1},\ref{fig:dens2}, in seconds. The simulations were done on a 2.8 GHz core 2 duo Macbook Pro. The plot third from top shows the raw cpu time, and the plot 4th from top shows the normalized cpu time, namely the raw cpu time divided by the mean number of discs parked. %Both vertical axes are a log scale. 

For large $\rho$, both the raw and normalized cpu time jumps discontinuously at the critical radii for each space, where the average number parked also jumps discontinuously. 
At these radii the discs are tightly packed, so the last disc (or discs) doesn't have much space to park, so one must typically run coarse mode to its completion and park the last disc in fine mode. At slightly larger radii the cpu time decreases sharply, because the discs are  loosely packed so coarse mode can typically park all the discs before switching to fine mode.
The cpu time reaches local minima at radii roughly corresponding to local minima in density. 
 
As $\rho\to 0$, the raw cpu time approaches a power law $\propto\rho^{-\alpha}$ where the power is $\alpha\approx 2.2$ (planes), $\alpha\approx 2.4$ (sphere), $\alpha\approx2.6-2.8$ (hyperboloids). The normalized cpu time approaches very nearly the same power minus 2 (Table \ref{tbl:density}.)  
Since $\rho \propto n^{{-}2}$ for large $n$, where $n$ is the average number of discs parked, the computational complexity of the algorithm is approximately $O(n^\alpha)$. For all spaces this is slightly larger than the $O(n^2)$ complexity that would be expected from checking all pairs of discs for overlap (a step which could be made slightly more efficient by using neighbour lists, for example \cite{Frenkel:2001uy}).
The algorithm is most efficient for planes, and least efficient for hyperboloids.

\subsubsection{Geometrical errors}\label{sec:errors}

Our algorithm assumes the insertion regions are polygonal, or at the very least are contained in the interior of their circumdiscs. 
When the topology of the insertion regions is more complicated, the algorithm may get stuck in an infinite loop or produce incorrect results. 

Figure \ref{fig:strips} gives several examples of non-polygonal insertion regions that occurred in actual random parkings (the examples were created by an earlier version of our program with a bug that created such regions with unusually high probability). 
The first column, rows 3,4 shows a region on the sphere, where two large discs are parked nearly at opposite poles. The insertion region is then a thin annulus near the equator, with a boundary formed from two disconnected arcs. 
An annulus can also occur on the projective plane (first column, rows 1,2), except when drawn as a sphere, the points are always exactly diametrically opposed, and the insertion region has the topology of a M\"obius strip.
Columns 3, 4, show examples of strips on the spaces with periodic tilings, where an insertion region stretches across a tile and connects to a translated version of itself on the other side. We call this an infinite strip.

We can sometimes detect infinite strips because when we sew the boundary arcs together we may return to a translated version of our original arc. In addition, if we don't return after a given (large) number of boundary arcs, we may have an infinite strip or a floating-point roundoff error that prevents us from detecting it, so we abort the sample. In both cases we return an error. 

The probability of such errors is controlled by the parameters $n_{coarse}$, and the number of translated copies of each tile we consider to check for overlap between discs (see section \ref{sec:tiles}). Increasing both of these reduces the number of errors due to non-polygonal regions but increases the computational cost. 

In case error detection doesn't work and we get stuck in an infinite loop, we include a parameter in the code, $n_{fine}$, which aborts fine mode if there are more than this many rejections in a row, and records an error. With $n_{coarse} = n_{fine} = 10^4$, for all the data generated for each space (a total of $3\times 10^7$ samples per space), we detected 1 error each on the plane [6,3] and hyperboloid [8,8],   3 on the hyperboloid [10,5], and none on the other spaces. These errors are insignificant compared to the statistical error due to having a finite number of samples.

We have tried to find a better criterion than using a fixed number of failures to determine when to switch to fine mode. For example, we tried criteria based on the total arc length of the boundary arcs, the total uncovered area, etc. These work reasonably well for large $\rho$, but did not produce significant gains for small $\rho$ since the extra calculation time grew too quickly with the number of discs parked.

%%%%%%%%%%%%%%%%%%%%%
%%% %%   Critical size ratio     %%%%%
%%%%%%%%%%%%%%%%%%%%%

\clearpage

\section{Critical radius on the $d$-dimensional sphere}\label{sec:critical}

%\note{maybe change $d$ to $d{+}1$ (i.e. use $d$ to mean the dimension of the surface is $d$)}

%\comment{leaving out detailed calculation of $M_d$ for $r>r_c$. could add this -- it's not that hard. (1/2 page?)}
%\bigskip

In this section, we prove that there is a critical radius at which 
all complete random parkings on the sphere are 4-parkings. 
This result is used to benchmark our numerical algorithm, but is of theoretical interest in itself. 
It follows from our derivation of the more general case of a random parking on the surface of a $d$-dimensional sphere, $d ≥ 2$, where we show that at geodesic radius $\rho = \tfrac{\eps}{4}$, all complete random parkings contain exactly $d{+}1$ discs with probability 1 (Theorem \ref{thm:critical}).
The general result is proved inductively, starting from the base case of random parking on a 2-dimensional sphere, which is a one-dimensional circle, or a line with periodic boundary conditions.  

Consider the surface of a unit sphere $S^{d{-}1} \in \mathbb{R}^d$, centered at the origin. 
Given two points $\b{x},\b{y}\in S^{d{-}1}$,
let $\phi_{\b{x}\b{y}}\in[0,\pi]$ be the angle subtended by the vectors associated with each point, calculated as $\cos\phi_{\b{x}\b{y}} = \b{x}\cdot\b{y}$.
A collection of points $\b{x}_1,\b{x}_2,\cdots,\b{x}_n\in S^{d{-}1}$ is a parking if all pairs satisfy a distance constraint (for this section all distances are calculated in Euclidean space), or equivalently a constraint on their angles:
\begin{equation}\label{eq:con}
|\b{x}_i-\b{x}_j| ≥ r\quad\Longleftrightarrow\quad 1{-}\tfrac{1}{2}{r^2} ≥ \cos\phi_{\b{x}_i\b{x}_j} \qquad \forall\: i,j ≤ n.
\end{equation}
The constraint depends on a constant $r$ which 
we call the radius in this section, even though it is technically a diameter. 
We will show that the critical radius is $r = r_c = \sqrt{2}$. At this radius 
the angles must satisfy $\phi_{\b{x}\b{y}} ≥ \tfrac{\pi}{2}$, or equivalently 
all dot products must be non-positive: $\b{x}_i\cdot\b{x}_j ≤ 0$. This is the key property that makes this radius critical, and shows that Theorem \ref{thm:critical} is equivalent to the statement that if one chooses points at random in $\mathbb{R}^d$ according to some continuous density, and requires the pairwise dot products to be non-positive, then this process must always terminate (with probability 1) after exactly $d{+}1$ points are chosen; in other words, the distance of the chosen points from the origin is irrelevant for observing this deterministic behaviour. 

Let us define the following: let $M_d(r)$ be the maximum number of points that can be parked for a given $r$, and let $m_d(r)$ be the minimum number.
By minimum, we mean the minimum number $l$ such that there exists a configuration with $l$ points, but where it is impossible to fit in another point without violating a distance constraint. 
Our strategy will be to derive the values or bounds for $M_d(r)$, $m_d(r)$ for certain values of $r$. %These numbers are important because for a particular $r$,our numerical algorithm must always terminate after a number of points $n$ such that $m_d(r) ≤ n ≤ M_d(r)$. It does, and this is one major difference between our algorithm and previous ones. 
The main results are encapsulated in the following two theorems.

\begin{theorem}\label{thm:critical}
Let $r_c = \sqrt{2}$. With probability 1, a complete random parking of discs of radius $r = r_c$ on $S^{d{-}1}$ contains exactly $d{+}1$ points.
\end{theorem}

\begin{theorem}\label{thm:bounds}
Consider a random parking of $n$ points on $S^{d{-}1}$. 
%collection of collection of $n$ points parked on the surface of a unit sphere in $d$ dimensions, such that any two parked points $\b{x},\b{y}$ satisfy $|\b{x}-\b{y}|≥ r$.
%Let $r_c = \sqrt{2}$ be the ``critical'' radius. 
%For $\sqrt{2} = r_c ≤ r < 2$, we have:
The maximum and minimum satisfy the following bounds:  
\begin{enumerate}[(i)]
\item $m_d(r) = 2$\quad for $r_c < r < 2$
\item $m_d(r_c) = d{+}1$
\item $M_d(r) ≤ d{+}1$\quad for $r_c < r < 2$
\item $M_d(r_c) = 2d$.
\end{enumerate}
\end{theorem}

The proofs rely on the following lemma:
\begin{lemma}\label{lem:dot}
When $r = r_c$, if the number $n$ of points in a random parking on $S^{d{-}1}$ satisfies $n > d{+}1$,
then there is at least one pair of points $\b{x}$, $\b{y}$ such that $\b{x}\cdot\b{y} = 0$. 
\end{lemma}

\subsection{Proofs}

\begin{proof}[Theorem \ref{thm:critical}]
By Theorem \ref{thm:bounds} (ii), there will always be at least $d{+}1$ points parked.
By Lemma \ref{lem:dot}, if there are more than $d{+}1$ parked then there is at least one pair whose dot product is zero.
But the probability of this is zero, since the set of points $\b{b}$ such that $|\b{b}| = 1$, $\b{a}\cdot\b{b} = 0$ (for fixed $\b{a}$) is $(d{-}2)$-dimensional,
so it has zero Lebesgue measure on a $(d{-}1)$-dimensional surface. Therefore with probability 1, there are exactly $d{+}1$ points parked. \qed
\end{proof}

Note that this result also follows from continuity: the minimum at $r_c$ is $d{+}1$, and the maximum jumps from $d{+}1$ when $r$ is infinitesimally larger, to $2d$ at exactly $r_c$.
If the average number of points parked changes continuously with $r$, then we can park only $n = d{+}1$ points at $r_c$.  

\bigskip

Next consider Theorem \ref{thm:bounds}. 
To show that $M_d(r) = k$, we must (a) demonstrate a parking of $k$ points, and (b) show there does not exist a parking of $k{+}1$ points without violating the distance constraints.
To show that $m_d(r) = l$, we must show that (c) there is a configuration of $l$ points such that it is impossible to add another,
and (d) given $l{-}1$ points in any configuration, we can always add at least one more. We refer to these steps as (a), (b), (c), (d) below. 

We write the coordinates of a point $\b{x} = \<x_1,\cdots,x_d\>$, and write $\b{e}_i = \<0,\cdots,1,\cdots,0\>$ for the unit vector with a 1 in the $i$th position, and zeros elsewhere.

\begin{proof}[Theorem \ref{thm:bounds}, (i)]
Choose two points at opposite poles, say $\b{x} = \b{e}_1$, $\b{y} = {-}\b{e}_1$. 
Any other point $\b{z} = \<z_1,\cdots,z_d\>$ will have $\phi_{\b{x}\b{z}} ≤ \tfrac{\pi}{2}$ if $z_1 ≥ 0$, and $\phi_{\b{y}\b{z}} ≤ \tfrac{\pi}{2}$ if $z_1 ≤ 0$.
Since when $r > r_c$ the constraint \eqref{eq:con} is $\phi_{\cdot,\bm z} > \tfrac{\pi}{2}$, it is not possible to add another point. This shows (a). 

Clearly if there is one point parked, we can always put another at the opposite pole provided $r < 2$. This shows (b). \qed
\end{proof}

\begin{proof}[Theorem \ref{thm:bounds}, (ii)]
First we show (d), i.e. that $m_d(r_c) ≥ d{+}1$. 
Suppose there are $n < d{+}1$ points parked. These points lie in an $(n{-}1)$-dimensional plane. 
Rotate the coordinate system so the plane is $x_d = u$, where $u ≤ 0$.
Then we can put another point at $\b{e}_d$, as this will make an angle of at least $\tfrac{\pi}{2}$ with all points in the plane. 

Next we show (c), i.e. that $m_d(r_c) ≤ d{+}1$, by demonstrating a configuration of $d{+}1$ points such that it is impossible to fit another point. 
Consider the regular $d$-dimensional simplex whose $d{+}1$ vertices lie on the unit sphere. 
The points on the sphere that are farthest from the vertices of the simplex are those obtained by projecting the centroid of each face onto the surface of the sphere.
We will show, by direct calculation, that the distance between a face's vertex and its projected centroid is less than $r_c$. 

The vertices of the simplex are given by
\begin{equation}
\b{x}_i = \b{s}+\b{r}_i\quad(i = 1,\cdots,n{-}1),\qquad
\b{x}_n = \b{t},
\end{equation}
where 
\begin{equation}
\b{s} = \big(\sqrt{n^{{-}3}}{-}\sqrt{n^{{-}2}{+}n^{{-}3}}\big)\;\b{1},\qquad
\b{r_i} = \sqrt{1{+}n^{{-}1}}\;\b{e}_i,\qquad
\b{t} = {-}\sqrt{n^{{-}1}}\;\b{1},
\end{equation}
with $\b{1} = \<1,\cdots,1\>$. It can be checked that $|\b{x}_i| = 1$, and that $\b{x}_i\cdot \b{x}_j = {-}n^{{-}1}$ for all $i,j$.
The projected centroid of the face bounded by vertices $\{\b{x}_1,\cdots,\b{x}_{n{-}1}\}$ is ${-}\b{t}$.
The distance from a vertex on this face to the centroid is $|\b{x}_i+\b{t}| = \sqrt{2}\;\sqrt{1{-}n^{{-}1}} ≤ \sqrt{2} = r_c$.
By symmetry, this holds for all faces. Therefore, there is no room to park another point in this configuration.
\qed
\end{proof}

\begin{proof}[Theorem \ref{thm:bounds}, (iii)]
Note that by \eqref{eq:con}, (iii) is equivalent to showing that the maximum number of vectors in $\mathbb{R}^d$ with a pairwise negative dot product is $d+1$. 
We show this by induction on $d$. For $d = 2$ it is straightforward to show that $M_d(r) ≤ 3$ when $r > r_c$, by considering equally-spaced points on a circle (they can be normalized to lie on a circle).

Suppose this holds for $d{-}1$, and consider a collection of $n$ points in $\mathbb{R}^d$.
Rotate the coordinate system so that one point lies along the positive branch of the axis defined by $\b{e}_d$.
The other points all satisfy $x_d < u<0$, where $x_d$ is the $d$th coordinate and $u$ is a constant.
Project these  points onto the plane $x_d = u$, by setting the $d$th coordinate to $u$:
$P\b{x} = \<x_1,\cdots,x_{d{-}1},u\>$. The points (there are $n{-}1$ of them) now lie in a $(d{-}1)$-dimensional plane. % that is isometric to $\mathbb{R}^{d{-}1}$.
If $n > d{+}1$, then by the induction hypothesis there must be a pair of projected points $P\b{a},P\b{b}$ with non-negative dot product:
$a_1b_1+\cdots+a_{d{-}1}b_{d{-}1}+u^2 ≥ 0$. Since $a_d,b_d < 0$ and $|a_d|,|b_d| > 0$, we have $a_db_d > u^2$, so $\b{a}\cdot\b{b} > P\b{a}\cdot P\b{b} ≥ 0$, so the points (prior to the projection) do not have pairwise negative dot products. 
Therefore $n \leq d{+}1$. 
\qed
\end{proof}

\begin{proof}[Theorem \ref{thm:bounds}, (iv)]
To show (a), put a point at each pole: $\{{\pm}\b{e}_1,\cdots,{\pm}\b{e}_d\}$. All pairwise angles are exactly $\tfrac{\pi}{2}$. 

To show (b), we show the following equivalent statement: for $r=r_c$, 
there is no configuration of $2d{+}1$ points in $\mathbb{R}^d$ such that all pairwise dot products are non-positive. We show this by induction on $d$. 
Clearly this is true for $d = 2$. Suppose it holds for $d{-}1$, and consider a collection of $n$ points in $\mathbb{R}^d$.
Choose any one of the points and rotate the coordinate system so the point lies on the positive branch of the axis defined by $\b{e}_d$.
All other points have $x_d ≤ 0$, and there is at most one with $x_d = {-}1$. 
Project all points with ${-}1 < x_d ≤ 0$ onto the plane $x_d = 0$, as in the proof of Theorem \ref{thm:bounds}, (iii). They now lie in a $(d{-}1)$-dimensional plane,
so by the induction hypothesis there are at most $2(d{-}1)$ with pairwise non-positive dot products.
As before in the proof of (iii), the projection can only decrease the dot product, so there can be at most $2(d{-}1)$ projected points total.
There are at most two other points (at the poles), so $n ≤ 2(d{-}1){+}2$. 
\qed
\end{proof}

\begin{proof}[Lemma \ref{lem:dot}]
Suppose there are $n > d{+}1$ points parked.
Pick a point $\b{x}$ that has negative dot product with all other points, and rotate the coordinate system to put this point at $\b{e}_d$.
(If there is no such point, we are done). All other points have $x_d ≤ u$, where $u < 0$ is a constant. 

Project the other $n{-}1$ points to the plane $x_d = u$. They now lie in a $(d{-}1)$-dimensional plane.
By Theorem \ref{thm:bounds} (iii), there are at most $d$ points with pairwise negative dot products, so there is at least one pair $\b{a},\b{b}$ such that $P\b{a}\cdot P\b{b} ≥ 0$.
As in the proof of Theorem \ref{thm:bounds} (iii), $\b{a}\cdot\b{b} ≥ P\b{a}\cdot P\b{b} ≥ 0$. Since $\b{a}\cdot\b{b} ≤ 0$, we have $\b{a}\cdot\b{b} = 0$. 
\qed
\end{proof}
%**check this proof. could strengthen, i..e remove equalities, only have inequalities, as in (iii)?

%%%%%%%%%%%%%%%%%%%%%
%%% %%   Algorithm     %%%%%
%%%%%%%%%%%%%%%%%%%%%

\section{Mathematical details of the algorithm}\label{sec:algorithm}

In this section we give details about how to implement each of the steps of the algorithm. We first describe how the coarse stage and fine stage are implemented for a single tile (sections \ref{sec:coarse2}, \ref{sec:fine2}), and then describe the modifications necessary to deal with the boundary conditions induced by the tiling (section \ref{sec:tiles}). 
For this discussion, all distances, angles, areas, etc are measured along geodesics on the surface. 

In the descriptions to follow, we call a new point $\bm\xi = \<\alpha,\beta,\gamma\,\>$ the one which has most recently been chosen on the surface (either parked, or testing to see if it can be parked), and an old point $\bm\zeta = \<\delta,\epsilon,\digamma\!\>$ is any of the other points that have previously been parked. We number the parked points consecutively in the order they park, and will use $i$ for the index of the new point and $\iota$ for the index of any of the old points, so we write $\bm\xi_i$, $\bm\zeta_\iota$ when we want to specify a particular point. 

%\tobeth{can you add a reference for all these formulas? Is there a standard geometry book, for e.g.?}
%\comment{ans: doesn't think so. her students said they couldn't find a good reference. Only her unpublished textbook.} 

\subsection{Coarse stage}\label{sec:coarse2}

 %We will deal with the tiles and boundary conditions later, in section \ref{sec:tiles}.
The steps are as follows.

\paragraph{1. Sample a point $\bm\xi = \<\alpha,\beta,\gamma\,\>$ uniformly from the coarse disc.}

%This can be done in several ways. 
%On the sphere, theoretically, you could set $x = (z_1,z_2,z_3)/|(z_1,z_2,z_3)|$, where $z_i\sim \mathcal{N}(0,1)$ are independent, identically distributed standard normal random variables.
%However in practice, nobody uses such an algorithm because it is extremely slow.
% why do you say that this is a «common» method?? who uses this algorithm?? it must be around 5(?)10(?)20(?) times as slow as the direct uniform distribution.
% computers generate uniform distribution very fast. it takes a lot of work to get normal distribution, from uniform distribution.
% finally, after all that hard work, we just project back to uniform??!!
% this makes no sense whatsoever!! consider carefully that you are submitting to an algorithms journal
%In our code we use a faster algorithm which is to sample uniformly from a disc with radius $\pi$. This method works for all the surfaces under consideration. 

We first choose a point uniformly from a disc enclosing the surface (or tile). 
To sample from the coarse disc with center $\<0,0,1\>$ and area $\varUpsilon$, we proceed as follows. First, we choose two uniform random variables $\varpi\in [-\pi,+\pi]$, $\varOmega\in [0,\varUpsilon]$.
Then, we calculate $\omega$ via
\begin{equation*}\begin{array}{l}
\geoR\textR{72pt}{$\cR\omega = 1{-}{\varOmega\over 2\pi}$}\textR{216pt}{$\sR\omega = \sqrt{{\varOmega\over\pi}(1         {-}{\varOmega\over 4\pi} )}$}\\
\geoB\textB{72pt}{$\cB\omega = 1                        $}\textB{216pt}{$\sB\omega = \sqrt{{\varOmega\over\pi}(1\phantom{{+}{\varOmega\over 4\pi}})}$}\\
\geoG\textG{72pt}{$\cG\omega = 1{+}{\varOmega\over 2\pi}$}\textG{216pt}{$\sG\omega = \sqrt{{\varOmega\over\pi}(1         {+}{\varOmega\over 4\pi} )}$}
\end{array}\end{equation*}

The uniformly distributed point is constructed as
\begin{equation*}\begin{array}{l}
\geoR\textR{300pt}{$\<\alpha,\beta,\gamma\,\> = \<\sR\omega\,\sR\varpi,\sR\omega\,\cR\varpi,\cR\omega\>$}\\
\geoB\textB{300pt}{$\<\alpha,\beta,\gamma\,\> = \<\sB\omega\,\sR\varpi,\sB\omega\,\cR\varpi,\cB\omega\>$}\\
\geoG\textG{300pt}{$\<\alpha,\beta,\gamma\,\> = \<\sG\omega\,\sR\varpi,\sG\omega\,\cR\varpi,\cG\omega\>$}
\end{array}\end{equation*}

\medskip For the \textcolor{redyellow}{sphere}, $\varUpsilon = 4\pi$. For the \textcolor{magentared}{projective plane}, $\varUpsilon = 2\pi$. For other surfaces, $\varUpsilon$ depends on the particular tile (see section \ref{sec:tiles} for details).

The interpretation is as follows: we choose a random area  $0 ≤ \varOmega ≤ \varUpsilon$ uniformly, and then find the corresponding random radius $0 ≤ \omega ≤ \upsilon$. 
The angle $\varpi$ is uniformly distributed. 
Thus this is a uniform distribution over the disc area $\varUpsilon$.
Note that the area $\varOmega$ and radius $\omega$ of a disc are related as follows:
\begin{equation*}\begin{array}{l}
\geoR\textR{90pt}{$\varOmega = \pi(2\sR{\omega\over 2})^2$}\textR{210pt}{${  +  }\varOmega = 2\pi(1{-}\cR\omega)$}\\
\geoB\textB{90pt}{$\varOmega = \pi(2\sB{\omega\over 2})^2$}\textB{210pt}{${\,0\,}\varOmega = 2\pi(1{-}\cB\omega)$}\\
\geoG\textG{90pt}{$\varOmega = \pi(2\sG{\omega\over 2})^2$}\textG{210pt}{${  -  }\varOmega = 2\pi(1{-}\cG\omega)$}
\end{array}\end{equation*}

\paragraph{2. Check distances with all other parked points.}

If  the new point $\bm\xi = \<\alpha,\beta,\gamma\,\>$ has distance $\varsigma < 2\varrho$ away from any old point $\bm\zeta = \<\delta,\epsilon,\digamma\!\>$ that is already parked, then we reject it. Otherwise, we add it to the list of parked points. 

The point space distance between two point vectors is calculated in one of two ways as
\begin{equation}\label{eq:gdist}\begin{array}{l}
\geoR\textR{180pt}{$(2\sR{\varsigma\over 2})^2 = (\alpha{-}\delta)^2{+}(\beta{-}\epsilon)^2{+}(\gamma{-}\digamma\!)^2$}\textR{120pt}{$\cR\varsigma = \gamma\digamma\!{+}\alpha\delta{+}\beta\epsilon$}\\
\geoB\textB{180pt}{$(2\sB{\varsigma\over 2})^2 = (\alpha{-}\delta)^2{+}(\beta{-}\epsilon)^2                          $}\textB{120pt}{$\cB\varsigma = \gamma\digamma\!                               $}\\
\geoG\textG{180pt}{$(2\sG{\varsigma\over 2})^2 = (\alpha{-}\delta)^2{+}(\beta{-}\epsilon)^2{-}(\gamma{-}\digamma\!)^2$}\textG{120pt}{$\cG\varsigma = \gamma\digamma\!{-}\alpha\delta{-}\beta\epsilon$}
\end{array}\end{equation}

\paragraph{3. Calculate boundary arcs.}

We do this in two steps: first, we calculate the \textKY{overlap arcs} (step a). Then, we calculate the \textKB{boundary arcs}, by subtracting the overlap arcs from the exclusion circles (steps b,c). This calculation only needs to be done for the overlaps between each new parked point and the old ones, because the overlaps between the old parked points have already been calculated and do not change.

Both kinds of arcs, overlap arcs and boundary arcs, are stored as lists of vectors, one list for each parked point $i$, where the vectors have the form
\begin{equation}\label{eq:arc}\begin{array}{c}
\text{\textKY{overlap arc} $[\textKG{k,\bm\kappa},\textKR{l,\bm\lambda}] = [\textKG{initial index,point}, \textKR{final index,point}]$}\\
\text{\textKB{boundary arc} $[\textKM{m,\bm\mu},\textKC{n,\bm\nu}] = [\textKM{initial index,point}, \textKC{final index,point}]$}
\end{array}\end{equation}
The initial and final points are the point vectors at the beginning and ends of the arcs, and the indices record which other parked point's exclusion circle intersects the arc at either end. We write $O_i,B_i$ for the set of overlap arcs, boundary arcs for a point $i$ respectively.

 %= [k_{i\iota},\bm\kappa_{i\iota},l_{i\iota},\bm\lambda_{i\iota}] = [\iota,\bm\phi,\iota,\bm\varphi]$. The index 

\begin{enumerate}[(a)]
\item\emph{For the new parked point $\bm\xi_i$, calculate the \textKY{overlap arcs} for each old neighbour $\bm\zeta_\iota$.} 
\end{enumerate}

First, we check if the new point $\bm\xi_i$ has distance $\varsigma < 4\rho$ away from and any old point $\bm\zeta_\iota$, where the geodesic distance is given in \eqref{eq:gdist}. 
If so, then we call such a point a \emph{neighbour} and we calculate the overlap arcs. The associated exclusion circles intersect at $\bm\phi$ $[\psi = {-}1]$ and $\bm\varphi$ $[\psi = {+}1]$.
Here $\bm\phi, \bm\varphi$ are points on the surface, and we always go counter-clockwise around the circle.
The cross product $\bm\xi{\times}\bm\zeta$ determines the orientation of the arc, and the sign $\psi$ selects either the initial endpoint $[\psi = {-}1]$ or final endpoint $[\psi = {+}1]$.

%**$\psi$ = parameter, sign. plus or minus square root. gives roots of quadratic equation. order is important because one is the beginning endpoint, one is the end. when stitching, need to know which is beginning, which is end. {-}1 is the beginning, {+}1 is the end. orientation -- comes from cross product of two centers. always taking center to peripheral .... 

%Since the cross product of two «point vectors» is a «line vector», we need to convert back to a «point vector».

The intersection points are found by solving the quadratic equations for the intersections of the two circles. This gives  
\begin{equation*}\begin{array}{l}
\geoR\textR{300pt}{$\textstyle\bm\phi,\bm\varphi = {\cR 2\rho\over 2(\cR{\varsigma\over 2})_{}^2}(\bm\xi{+}\bm\zeta){+}
\psi{\sqrt{(\sR 2\rho/\sR{\varsigma\over 2})_{}^2{-}1}\over 2(\cR{\varsigma\over 2})_{}^2}\,{\rm conv}_{\rm line}^{\rm point}\,(\bm\xi{\times}\bm\zeta)$}\\
\geoB\textB{300pt}{$\textstyle\bm\phi,\bm\varphi = {\cB 2\rho\over 2(\cB{\varsigma\over 2})_{}^2}(\bm\xi{+}\bm\zeta){+}
\psi{\sqrt{(\sB 2\rho/\sB{\varsigma\over 2})_{}^2{-}1}\over 2(\cB{\varsigma\over 2})_{}^2}\,{\rm conv}_{\rm line}^{\rm point}\,(\bm\xi{\times}\bm\zeta)$}\\
\geoG\textG{300pt}{$\textstyle\bm\phi,\bm\varphi = {\cG 2\rho\over 2(\cG{\varsigma\over 2})_{}^2}(\bm\xi{+}\bm\zeta){+}
\psi{\sqrt{(\sG 2\rho/\sG{\varsigma\over 2})_{}^2{-}1}\over 2(\cG{\varsigma\over 2})_{}^2}\,{\rm conv}_{\rm line}^{\rm point}\,(\bm\xi{\times}\bm\zeta)$}
\end{array}\end{equation*}

The function ${\rm conv}_{\rm line}^{\rm point}$ converts a line vector to a point vector, and is given by % in the appendix, section \ref{app:geometry}. 
\begin{equation*}\begin{array}{l}
\geoR\textR{300pt}{${\rm conv}_{\rm line}^{\rm point}\,\<u,v,w\> = \<u,v,{  +  }w\>$}\\
\geoB\textB{300pt}{${\rm conv}_{\rm line}^{\rm point}\,\<u,v,w\> = \<u,v,{\,0\,}w\>$}\\
\geoG\textG{300pt}{${\rm conv}_{\rm line}^{\rm point}\,\<u,v,w\> = \<u,v,{  -  }w\>$}
\end{array}\end{equation*}

%\medskip For each neighbour $\bm\zeta_\iota$ we calculate the two points where the exclusion circles intersect.
%We call the «overlap arc» the interval where the central exclusion circle and neighbour exclusion %circle overlap.

The overlap arc vector added to $O_i$ is %$[k_{i\iota},\bm\kappa_{i\iota},l_{i\iota},\bm\lambda_{i\iota}] = 
$[\iota,\bm\phi,\iota,\bm\varphi]$, and to $O_\iota$ is 
$[i,\bm\varphi,i,\bm\phi]$. \\

\begin{enumerate}[(a)]\setcounter{enumi}{1}
\item \emph{For the new point $\bm\xi_i$, calculate the \textKB{boundary arcs} $B_i$.}
\end{enumerate}

We now subtract the overlap arcs $O_\iota$ from point $\bm\xi_i$s exclusion circle, to form the boundary arcs $B_i$. 

We initialize the list of \textKB{boundary arcs} as the empty set, $B_i = \{\}$ (entire circle). 
Then, for each neighbour point $\bm\zeta_\iota$, we compare each \textKY{overlap arc} $[\textKG{k,\bm\kappa},\textKR{l,\bm\lambda}]\in O_\iota$,
with each \textKB{boundary arc} $[\textKM{m,\bm\mu},\textKC{n,\bm\nu}]\in B_i$ currently in the list.
Note that we only need to check the new overlap arcs calculated in step (a), for these are the ones associated with point $i$. 
If the list $B_i$ is empty, then the first boundary arc is simply $[\textKM{m,\bm\mu},\textKC{n,\bm\nu}] = [\textKR{l,\bm\lambda},\textKG{k,\bm\kappa}]$:
we reverse the order of the points on the overlap arc, to take its complement.

If the list is not empty, then we must consider how the overlap arcs carve away pieces of the boundary arcs.
There are 6 topological types of intersection, distinguished by determinants and the sign $\psi$ (different from before).
\begin{equation*}\begin{array}{l}
\textK{180pt}{$\textKG{\kappa } = \psi\det\{\phantom{\bm\kappa},\textKR {\bm\lambda},\textKM {\bm\mu},\textKC {\bm\nu}\}$}\geoR\textR{60pt}{$\psi = {-}1\quad(\rho > {\pi\over 4})$}\\
\textK{180pt}{$\textKR{\lambda} = \psi\det\{\textKG {\bm\kappa},\phantom{\bm\lambda},\textKM {\bm\mu},\textKC {\bm\nu}\}$}\geoR\textR{60pt}{$\psi = {+}1\quad(\rho < {\pi\over 4})$}\\
\textK{180pt}{$\textKM{\mu    } = \psi\det\{\textKG {\bm\kappa},\textKR {\bm\lambda},\phantom{\bm\mu},\textKC {\bm\nu}\}$}\geoB\textB{60pt}{$\psi = {+}1$}\\
\textK{180pt}{$\textKC{\nu    } = \psi\det\{\textKG {\bm\kappa},\textKR {\bm\lambda},\textKM {\bm\mu},\phantom{\bm\nu}\}$}\geoG\textG{60pt}{$\psi = {+}1$}
\end{array}\end{equation*}
On the sphere (and projective plane), $\psi$ is undefined at the critical radius $\rho = {\pi\over4}$.

%Here $\psi$ is a different variable for each space: for the \textcolor{redyellow}{sphere} and \textcolor{magentared}{projective plane}, $\psi = {-}1$ for large radius ($2\rho > {\pi\over 2}$), $\psi = {+}1$ for small radius ($2\rho < {\pi\over 2}$), but $\psi$ is undefined at the critical radius ($2\rho = {\pi\over2}$). For other tilings, $\psi = {+}1$ always. 

%** bold = where things intersect. 
%non-bold = represent whether three points are clockwise or anticlockwise. determinant is orientation. signed area of three points. only care about whether positive or negative (ie orientation) used to determine which arcs are inside / overlap / outside which others. 
%** all of this is a method to efficiently calculate who is overlapping who. don't know actual angle, just know relative to other points. 

We modify each \textKB{boundary arc} $[\textKM{m,\bm\mu},\textKC{n,\bm\nu}]$ according to the topological type of overlap add the modified boundary arc to $B_i$.
\begin{equation*}\begin{array}{lcl}
\textKG{(\kappa{>}0)}\wedge\textKR{(\lambda{>}0)}\wedge\textKM{(\mu{<}0)}\wedge\textKC{(\nu{<}0)}&&\text{delete all}\\
\textKG{(\kappa{>}0)}\wedge\textKR{(\lambda{>}0)}\wedge\textKM{(\mu{>}0)}\wedge\textKC{(\nu{>}0)}&[\textKM{m,\bm\mu},\textKC{n,\bm\nu}]&\text{keep all}\\
\textKG{(\kappa{>}0)}\wedge\textKR{(\lambda{<}0)}\wedge\textKM{(\mu{<}0)}\wedge\textKC{(\nu{>}0)}&[\textKM{m,\bm\mu},\textKG{k,\bm\kappa}]\phantom{\,\cup\,[\textKR{l,\bm\lambda},\textKC{n,\bm\nu}]}&\text{keep initial, delete final}\\
\textKG{(\kappa{<}0)}\wedge\textKR{(\lambda{>}0)}\wedge\textKM{(\mu{>}0)}\wedge\textKC{(\nu{<}0)}&\phantom{[m,\bm\mu,k,\bm\kappa]\,\cup\,}[\textKR{l,\bm\lambda},\textKC{n,\bm\nu}]&\text{delete initial, keep final}\\
\textKG{(\kappa{<}0)}\wedge\textKR{(\lambda{<}0)}\wedge\textKM{(\mu{>}0)}\wedge\textKC{(\nu{>}0)}&[\textKM{m,\bm\mu},\textKG{k,\bm\kappa}]\cup[\textKR{l,\bm\lambda},\textKC{n,\bm\nu}]&\text{keep ends, delete middle}\\
\textKG{(\kappa{<}0)}\wedge\textKR{(\lambda{<}0)}\wedge\textKM{(\mu{<}0)}\wedge\textKC{(\nu{<}0)}&[\textKR{l,\bm\lambda},\textKG{k,\bm\kappa}]&\text{delete ends, keep middle}
\end{array}\end{equation*}

\begin{figure*}
\center\includegraphics[width=4.5in,height=3in]{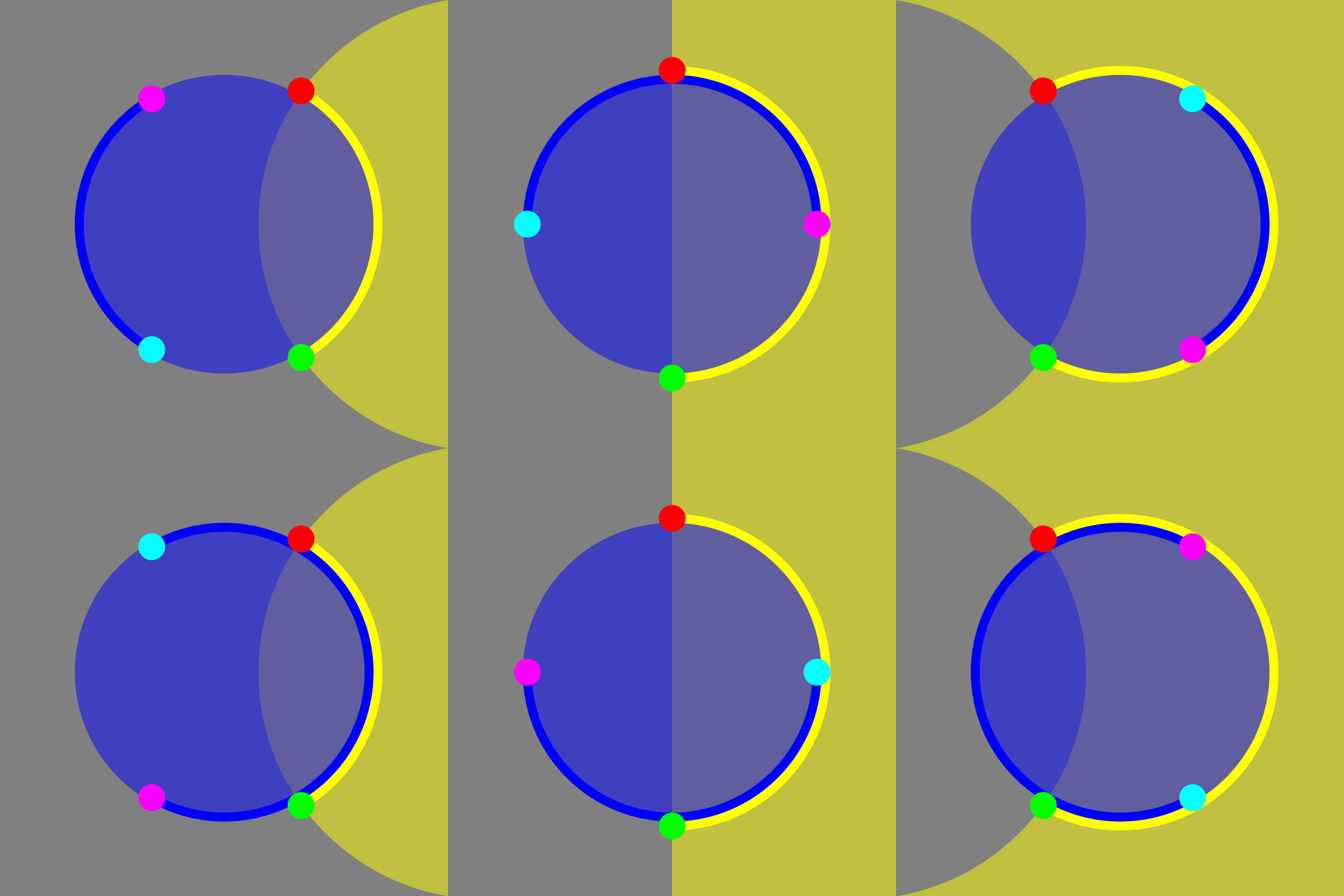}
\caption{Diagram showing the six possible ways that an \textKY{overlap arc} $[\textKG{k,\bm\kappa},\textKR{l,\bm\lambda}]$
can intersect with a \textKB{boundary arc} $[\textKM{m,\bm\mu},\textKC{n,\bm\nu}]$.}\label{fig:sixarcs}
\end{figure*}

If the list $B_i$ becomes empty (after the initialization), then there are no boundary arcs associated with the exclusion circle for point $i$.

\begin{enumerate}[(a)]\setcounter{enumi}{2}
\item \emph{Reverse the above process. For each old point $\bm\zeta_\iota$, use the new point $\bm\xi_i$ to update the list of boundary arcs $B_\iota$.} 
\end{enumerate}

\paragraph{4. Check exit criterion}

We terminate coarse stage, and move to fine stage, once there are a fixed number of rejections $n_{\text{coarse}}$ in a row.
In addition, if there are no more boundary arcs ($B_j$ is empty for all $j$),
then we terminate the sample altogether, and re-start the algorithm with a new sample.

\subsection{Fine stage}\label{sec:fine2}

Implementing the fine stage requires the following steps.

\paragraph{1. Compute the polygonal insertion regions.} 

We  sew the boundary arcs together to form the boundaries of the insertion regions. 
The set of all boundary arcs $\bigcup_{j{=}1}^kB_j$ (where $k$ is the number of  points currently parked) forms an incidence table which tells us, as we traverse a given arc, what is the next arc we should follow. 
By beginning with one arc in the list and following the indices in the incidence table until we return to the original index and original point, we determine the boundaries curved polygon.
One can either always follow the initial arcs, or always follow the final arcs, as long as the choice remains consistent.

In this step, we also test whether the insertion region is an infinite strip.
More details are given in section \ref{sec:tiles}, step 6. 

\paragraph{2. Construct fine discs enclosing the polygons.} 

We enclose each polygon in a fine disc centered at its centroid $\bm\xi_i = \<\alpha,\beta,\gamma\>$, 
whose radius $\delta$ is the distance to the farthest vertex of each polygon, and the area $\Delta$ is
\begin{equation*}\begin{array}{l}
\geoR\textR{90pt}{$\varDelta = \pi(2\sR{\delta\over 2})_{}^2$}\textR{210pt}{${  +  }\varDelta = 2\pi(1{-}\cR\delta)$}\\
\geoB\textB{90pt}{$\varDelta = \pi(2\sB{\delta\over 2})_{}^2$}\textB{210pt}{${\,0\,}\varDelta = 2\pi(1{-}\cB\delta)$}\\
\geoG\textG{90pt}{$\varDelta = \pi(2\sG{\delta\over 2})_{}^2$}\textG{210pt}{${  -  }\varDelta = 2\pi(1{-}\cG\delta)$}
\end{array}\end{equation*}

If the polygon is a trigon, then we could also calculate the circumradius. Unfortunately it does take more cputime, and we cannot generalize it to higher polygons.

The centroid of a polygon on the surface is the sum of the vertex vectors, normalized to the lie on the surface. The normalizations are
\begin{equation}\begin{array}{l}
\geoR\textR{300pt}{$\displaystyle{\rm norm}_{\rm point}^{\vphantom{\rm line}}\,\<x,y,z\> = {\<x,y,z\>\over\sqrt{z^2         {+}x^2{+}y^2} }$}\\
\geoB\textB{300pt}{$\displaystyle{\rm norm}_{\rm point}^{\vphantom{\rm line}}\,\<x,y,z\> = {\<x,y,z\>\over\sqrt{z^2\phantom{{+}x^2{+}y^2}}}$}\\
\geoG\textG{300pt}{$\displaystyle{\rm norm}_{\rm point}^{\vphantom{\rm line}}\,\<x,y,z\> = {\<x,y,z\>\over\sqrt{z^2         {-}x^2{-}y^2} }$}
\end{array}\end{equation}

\paragraph{3. Sample a point uniformly from the insertion regions.}

This is done by sampling uniformly from the total area of the enclosing fine discs. A particular disc is chosen in proportion to its area, and a point is chosen uniformly in this disc.
If the point also lies in the polygon inside the corresponding disc, then it is parked, and the polygons are recomputed. Otherwise, the point is rejected. 

%\comment{\note{remove?} This ensures that points in the polygons are sampled with the correct probability. To see why, let $P_i$, $C_i$ denote the polygon and enclosing disc of the $i$th region, and let $A(\cdot)$ denote the area of an object. The $i$th circle is chosen with probability $A(C_i) / \sum_i A(C_i)$, and the parked point lies in the polygon with probability $A(P_i)/\sum_i A(C_i)$. This equals the desired probability except for a constant of proportionality. 

%\note{remove?} In the coarse mode, the denominator would be $A(S)$, where $S$ is the entire surface, so the acceptance rate is bigger than it would be in coarse mode by a factor of $A(S) / \sum_i A(C_i)$. If the circles are small, this shows that fine mode is much more efficient than coarse mode.

%\note{remove?} Note that this argument holds even when the enclosing circles overlap, provided a point is parked only  if it falls in the polygon corresponding to the chosen circle. (If the insertion regions are small enough, it will typically never happen that one insertion region overlaps with more than one fine disc). }

The algorithm to sample uniformly from a fine disc is the same as in section \ref{sec:coarse}, step 1,
except that now the disc is centered at $\bm\xi = \<\alpha,\beta,\gamma\>$, instead of the origin $\<0,0,1\>$.
%We can overcome this by sampling from a disc centered at the origin, and then performing a geodesic translation to the given center $\bm\xi_i = \<\alpha_i,\beta_i,\gamma_i\>$.
%we no longer do this!! we now sample directly around the new point
%also, i think its important to say $\<0,0,1\>$, otherwise people might assume that the origin is $\<0,0,0,\>$
Below, we give explicit formulas for sampling uniformly from a disc with center $\bm\xi = \<\alpha,\beta,\gamma\>$ and area $\varDelta$.
First, we choose two uniform random variables $\varpi\in [-\pi,+\pi]$, $\delta\in [0,\varDelta]$. Then, we calculate $\delta$ via
\begin{equation*}\begin{array}{l}
\geoR\textR{90pt}{$\cR\delta = 1{-}{\varDelta\over 2\pi}$}\textR{210pt}{$\sR\delta = \sqrt{{\varDelta\over\pi}(1         {-}{\varDelta\over 4\pi} )}$}\\
\geoB\textB{90pt}{$\cB\delta = 1                        $}\textB{210pt}{$\sB\delta = \sqrt{{\varDelta\over\pi}(1\phantom{{+}{\varDelta\over 4\pi}})}$}\\
\geoG\textG{90pt}{$\cG\delta = 1{+}{\varDelta\over 2\pi}$}\textG{210pt}{$\sG\delta = \sqrt{{\varDelta\over\pi}(1         {+}{\varDelta\over 4\pi} )}$}
\end{array}\end{equation*}
\begin{equation*}\begin{array}{l}
\geoR\textR{360pt}{$\textstyle\<x,y,z\> = \cR\omega\<\alpha,\beta,\gamma\>{-}\sR\omega\,\cR\varpi{\<\alpha\!\gamma,\beta\!\gamma,{  -  }(\alpha\!\alpha{+}\beta\!\beta)\>\over\sqrt{\alpha\!\alpha{+}\beta\!\beta}}{+}
\sR\omega\,\sR\varpi{\<{-}\beta,\alpha,0\>\over\sqrt{\alpha\!\alpha{+}\beta\!\beta}}$}\\
\geoB\textB{360pt}{$\textstyle\<x,y,z\> = \cB\omega\<\alpha,\beta,\gamma\>{-}\sB\omega\,\cR\varpi{\<\alpha\!\gamma,\beta\!\gamma,{\,0\,}(\alpha\!\alpha{+}\beta\!\beta)\>\over\sqrt{\alpha\!\alpha{+}\beta\!\beta}}{+}
\sB\omega\,\sR\varpi{\<{-}\beta,\alpha,0\>\over\sqrt{\alpha\!\alpha{+}\beta\!\beta}}$}\\
\geoG\textG{360pt}{$\textstyle\<x,y,z\> = \cG\omega\<\alpha,\beta,\gamma\>{-}\sG\omega\,\cR\varpi{\<\alpha\!\gamma,\beta\!\gamma,{  +  }(\alpha\!\alpha{+}\beta\!\beta)\>\over\sqrt{\alpha\!\alpha{+}\beta\!\beta}}{+}
\sG\omega\,\sR\varpi{\<{-}\beta,\alpha,0\>\over\sqrt{\alpha\!\alpha{+}\beta\!\beta}}$}
\end{array}\end{equation*}

\paragraph{4. Check distances with other parked points.}

As in section \ref{sec:coarse}, step 2.  

Either accept or reject point after this step. 

\paragraph{5. Calculate boundary arcs.}

As in section \ref{sec:coarse}, step 3.

\paragraph{6. Check exit criterion.} 

Fine mode is terminated when there are no more boundary arcs, hence no more space where a point can be parked without overlap.

\subsection{Tiles and boundary conditions}\label{sec:tiles}

Let the ``central tile'' be the one with face center $\<0,0,1\>$ (the origin). 
This has vertices $\bm\upsilon\!\in V$, edge midpoints $\bm\epsilon\in E$, and edge normals $\bm\eta\in H$, where $V$, $E$, $H$ are sets of the appropriate objects.
The tile itself is defined by the set of points $\<x,y,z\>$ which satisfy the inequalities $\bm\eta\cdot\<x,y,z\> ≥ 0$ for all edge normals $\bm\eta\in H$.

We call a \emph{neighbour tile} one which shares an edge or a vertex with the central tile. These have face centers $\bm\digamma\!\in F$ for $F$ the appropriate set. 
The inradius of the tile (the distance from the center to the nearest edge midpoint) is denoted $\epsilon$, and the circumradius (the distance from the center to the farthest vertex) is denoted $\upsilon$. 
Formulas for all of these quantities are given in the appendix, section \ref{app:coords}. 

Figures \ref{fig:2}, \ref{fig:3} show the geometry of the different spaces and tilings, along with some maximal \emph{packings}. Note that these are \emph{not} random parkings, but rather were created separately to help understand the possible arrangements of objects in the different spaces. 
The first column of each figure shows the central tile (yellow) and its nearest neighbours, along with a disc of maximal size $\rho = \epsilon$ parked at the center of the tile. At this radius only one disc fits per tile, and it is tangent to copies of itself in different tiles. Exclusion circles are drawn in red. 
The second column of each figure shows a packing at a value of $\rho$ where exactly 2 discs fit per tile. Exclusion circles for the magenta discs are red, and for the cyan discs are green. Together the exclusion circles completely cover the area of the central tile. 
The third column shows a packing at $\rho = \tfrac{\epsilon}{2}$, where the exclusion circles for a point and its copy are tangent. For $\rho < \tfrac{\epsilon}{2}$, they are disjoint. For $\rho > \tfrac{\epsilon}{2}$, they are overlapping.

We now explain the modifications to the parking algorithm presented in section \ref{sec:algorithm}, that are needed to apply it to a periodic tiling.

\paragraph{1. Coarse mode: sample a point uniformly from the entire surface.}

This is done by first sampling a point uniformly from the «coarse disc»:  a disc containing the tile,  with center $\<0,0,1\>$ and radius $\upsilon$.
Then, we need to check to see if the point lies inside the tile. This  is done by testing whether all inequalities ($\bm\eta\cdot\<x,y,z\> ≥ 0$) hold. 
For the projective plane, this reduces to the single inequality $(z ≥ 0)$. For the sphere, there is no edge boundary condition.
If the point lies outside the tile (at least one inequality does not hold), then we keep resampling until we obtain a point that lies inside the tile.

\paragraph{2. Coarse mode and fine mode: check the new point for overlap with all  old points.}

Each time a point is chosen on the surface, we need to check for overlap not just with all the points parked on the central tile, but also with all translated copies of these points, on all tiles. 
The translation matrices which generate all possible translated tiles are shown in the appendix, section \ref{app:coords}. 

The number of tiles one needs to check depends on the disc radius:  
for small radius, we need only consider the small number of tiles that are immediate neighbours (they share an edge or a vertex), whereas for large radius, we need to consider more tiles. 
There is no analytic formula for how many tiles must be considered for a given radius, so we determine the number by trial and error, adding in tiles until the number of errors we detect (see section \ref{sec:errors}) is sufficiently small. 
The extra tiles are added by considering the edges and vertices of the neighbour tiles that are closest to the origin $\<0,0,1\>$, and adding the tiles which adjoin these. If this is not enough tiles, we add in tiles which touch those edges and vertices which are closest for the current set of tiles. This is repeated until we have enough tiles. 
The more tiles one considers in the list, the fewer the errors the algorithm will make (particularly the kind where holes are an infinite strip), but the less efficient the algorithm will be. The number of tiles we used in our code is given in the appendix section \ref{app:neighbours}.

\paragraph{3. Fine mode: sampling from a fine disc.}

After we sample a new point from a fine disc %with center $\<\alpha_i,\beta_i,\gamma_i\>$ and radius $\delta_i$, 
and determine it lies in the insertion region, we also have to check whether it lies in the central tile. If not, then we find the isometry (some product of translation matrices $\bm\sigma_{\bm\digamma\!}$ or the inverse $\bm\tau_{\bm\digamma\!}$) that maps the point to the central tile, where it is then parked.
The formulas for the translation matrices are given in the appendix sections \ref{app:coords}.

\paragraph{4. Fine mode: storing arcs.}

In fine mode, we need to keep track of slightly more information when we store overlap or boundary arcs, since each parked point implies all its translated copies are also parked so we need to keep track of which particular translation is involved in the arc. 

The format of each stored arc is
\begin{equation}\begin{array}{rl}
\text{\textKY{overlap arc}}  & [\textKG{k,\bm\kappa,\bm\tau},\textKR{l,\bm\lambda,\bm\sigma}] = \text{[\textKG{initial index,point,matrix}, \textKR{final index,point,matrix}]}\\
\text{\textKB{boundary arc}} & [\textKM{m,\bm\mu,\bm\sigma},\textKC{n,\bm\nu,\bm\tau}] = \text{[\textKM{initial index,point,matrix}, \textKC{final index,point,matrix}]}
\end{array}\end{equation}

The initial matrix $\bm\sigma$ is the isometry which brings the new point $\bm\xi_i$ close to the neighbour point $\bm\zeta_\iota$, where close means they are within $4\rho$.
The final matrix $\bm\tau$ is the isometry which brings the $\bm\zeta_\iota$ close to $\bm\xi_i$, and is the inverse of $\bm\sigma$. We use them to match arcs when we sew arcs together to knit the polygon.

\paragraph{5. Fine mode: sewing boundary arcs together to form the polygon.}

First, we initialize the polygon matrices, as either $\bm\varSigma = {\rm identity}$ (to follow initial arcs) or $\bm T = {\rm identity}$ (to follow final arcs).
The arc data gives us the next point and matrix of the polygon. The actual point and its corresponding matrix are found by multiplying the matrices cumulatively:
as we follow initial arcs, the initial matrices transform as $\bm\varSigma\to\bm\sigma\bm\varSigma$, and as we follow final arcs, the final matrices transform as $\bm T\to\bm\tau\bm T$. 

To check whether the arcs match up, we make sure that $[($final index of previous arc$) = ($initial index of next arc$)]$ and $[($final matrix of previous arc$) = ($initial matrix of next arc$)]$. There are at least two reasons why arcs might not match up locally: (i) we didn't use enough tiles in our neighbour list, so are not able to find the entire correct polygon,  (ii) roundoff error. 
If arcs do not match up at some point, then we throw out the sample. This will induce an error in our statistics and to make this error small, we choose the number of neighbour tiles so that an error of type (i) occurs less than about 1-2 times per $10^4$ samples. To mitigate errors of type (ii), we pre-computed all the translation matrices in Mathematica, and imported the analytic formulas into our code. 

 If we return to the original index and original point and original matrix (identity), then the polygon is finished. 
If, on the other hand, we continue following arcs and matrices and they do not return to the original after a given large number of steps, then it is possible that we have an infinite strip, so we abandon the sample and start again. 

One minor modification is required for the projective plane, where there is one positive tile $({\rm isometry} = {+}{\rm identity})$ and one negative tile $({\rm isometry} = {-}{\rm identity})$. Since the negative isometry reverses orientation, whenever we pick a disc from the negative tile,
we need to reverse the arc direction so it becomes 
%$[{\rm initial\;index},{\rm initial\;endpoint},{\rm initial\;matrix},{\rm final\;index},{\rm final\;endpoint},{\rm final\;matrix}] = 
$[l,\bm\lambda,\bm\tau,k,\bm\kappa,\bm\sigma]$.

%%%%%%%%%%%%%%%%%%%%%
%%% %%   Discussion     %%%%%
%%%%%%%%%%%%%%%%%%%%%

\section{Conclusion}\label{sec:conclusion}

We have introduced an algorithm to simulate complete random parkings on spaces of constant curvature, and we applied this to the sphere, plane, hyperboloid, and projective plane. Each of the plane and hyperboloid may be covered by translations of tiles in different ways, and we considered two possible tilings each, allowing us to carefully compare the effect of different boundary conditions. We investigated the distribution of the number of discs parked as a function of disc radius and discovered several interesting properties. At large disc radius, certain spaces had critical values of disc radius where there were always the same number of discs parked, most notably on the sphere where at a critical radius there were four points parked. We proved this statement rigorously, and also showed that parkings on the surfaces of higher-dimensional spheres would have the same deterministic behaviour at the same critical radius. On the plane with periodic boundary conditions,  there is a near-critical radius where 99.3\% of the parkings contain 4 discs, which is related to a peculiar property that the average number of discs parked is not a monotonic function of disc radius. We traced these observations to the nonlocality induced by the boundary conditions.  
For small discs, the density, i.e. the average fraction of the area covered by the discs, appears to approach the same constant for all spaces, though it approaches from above for a sphere and from below for the hyperboloids, with implications for parkings on more general curved spaces. 

Our algorithm has considered identical discs but it could in principle be adapted to discs with unequal radii. A necessary modification is that all the boundary arcs would need to be computed each time a new disc is tentatively parked, because the insertion regions depend on the size of the new disc. The algorithm would be slower but perhaps one could find a way to re-use computations to partly mitigate the loss in efficiency.

A question of interest in applications is whether there critical radii on other spaces, where greater than four points are parked. If so, one could use these spaces as templates for adsorbing a given, constant number of small objects, something that is hard to do with current techniques. For example, one may wish to consider ellipsoids, or other surfaces. Unfortunately, we have not yet found a way to easily adapt our algorithm to spaces with non-constant curvature, because we lack an efficient way to compute distances on these spaces, but this would be an interesting question for future research.

Our algorithm and discussion of critical radii suggests possible theoretical direction to understanding random parkings. The number of points parked must lie between maximum and minimum values which are the solutions to optimization problems. 
The maximum number of discs that can be parked is the maximum number that can be \emph{packed} in a given space, so is equivalent to a well-known optimization problem \cite{CONWAY:1999bv}. The minimum number is the number needed to completely cover the space with discs of radius $2\rho$ (the exclusion circles), subject to the condition that the centers of the discs are never within $2\rho$ of each other (the actual discs can't overlap). This is almost equivalent to the \emph{covering problem} \cite{Fowler:1996wp,CONWAY:1999bv}, but with an extra non-overlap constraint. It was conjectured in \cite{Schade:2013ee} that all optimal solutions to the covering problem satisfy the non-overlap constraint, though this has not yet been proven. 
Without the constraint, the covering problem provides a lower bound to the minimum number in the random parking problem. 
Solutions to these optimization problems combined with statistical analysis of how close we can expect to be to these bounds, such as an estimate of the variance of number parked, could provide insight into the small-$\rho$ behaviour of random parkings and also insight into whether there are critical radii or near-critical radii on other spaces.

%-- other spaces\\
%-- unequal disc radii\\
%-- explicit formula\\
%-- max/min\\
%**effect of geometrical issue... why it's not important, how to (potentially) get around it.
%**reason this worked is we have explicit formulas for distances involved, and geodesics. won't have this on arbitrary surfaces. 

%%%%%%%%%%%%%%%%%%%%%
%%% %%   Acknowledgments     %%%%%
%%%%%%%%%%%%%%%%%%%%%

\clearpage
\begin{figure*}
\center\includegraphics[width=4.5in,height=6in]{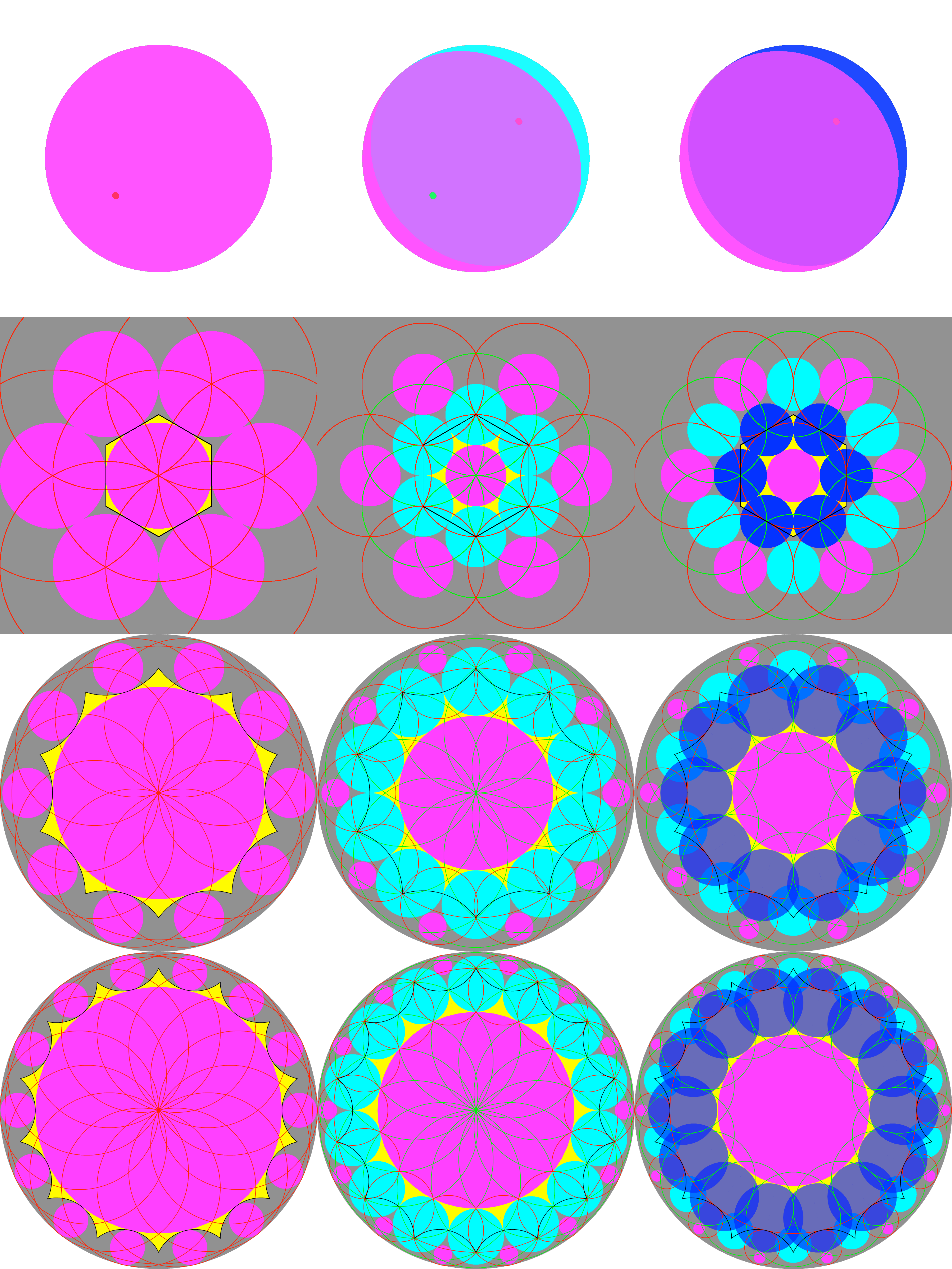}
\caption{Illustrations of maximal \emph{packings}. Discs are solid colors (magenta, cyan, blue), and exclusion circles are thin circles surrounding them. 
First column: maximum possible radius $\rho = \eps$, where exactly one disc fits per tile. 
Second column: Radius such that there is space for exactly 2 discs per tile. 
Third column: Radius $\rho = \tfrac{\eps}{2}$ where exclusion circles are tangent to their translated copies on each tile.
From top to bottom: sphere [2,1], plane [6,3],  hyperboloid [10,5], and hyperboloid [14,7] (not implemented in the random parking algorithm).}
%\caption{neighbors and ${\rm max{\cdot}radius}$ • tiles: $\{\textcolor{redyellow}{[2,1]},\textcolor{bluemagenta}{[6,3]},\textcolor{greencyan}{[10,5]},[14,7]\}$}
\label{fig:2}
\end{figure*}

\clearpage
\begin{figure*}
\center\includegraphics[width=4.5in,height=6in]{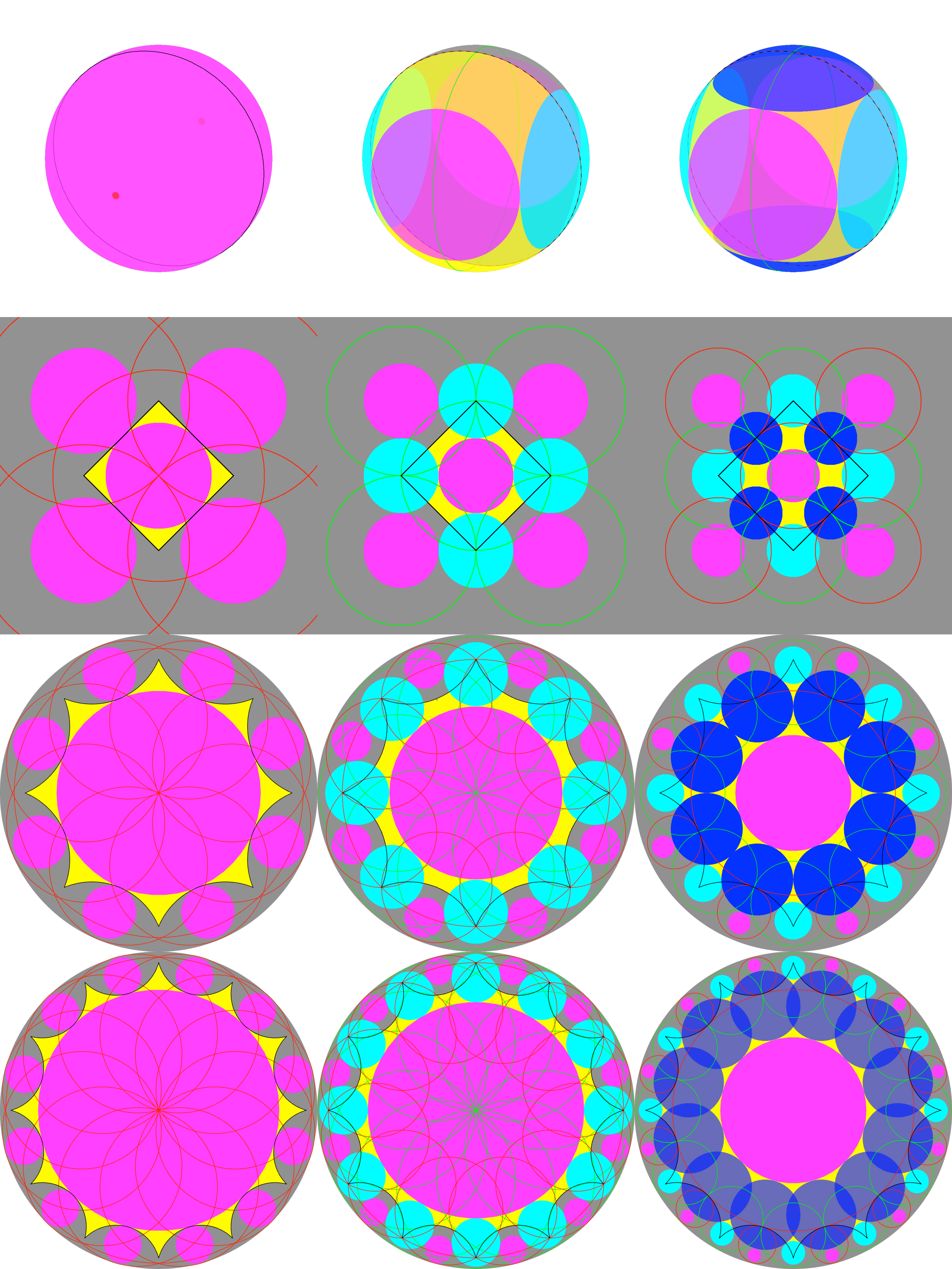}
\caption{Same as figure \ref{fig:2}, but with different spaces. 
From top to bottom: projective plane $[p,2]$, plane [4,4], hyperboloid [8,8], and hyperboloid [12,12] (not implemented in the random parking algorithm).}
%\caption{neighbors and ${\rm max{\cdot}radius}$ • tiles: $\{\textcolor{magentared}{[p,2]},\textcolor{cyanblue}{[4,4]},\textcolor{yellowgreen}{[8,8]},[12,12]\}$}
\label{fig:3}
\end{figure*}

\clearpage

\begin{acknowledgements} 
We would like to thank Michael Brenner for helpful comments. 
ERC acknowledges NSF MSPRF grant DMS-1204686.
MHC acknowledges the Harvard Materials Research Science and Engineering Center Grant DMR-1420570, the Division of Mathematical Sciences Grant DMS-1411694, and the US Department of Energy grant DE-SC0012296.
\end{acknowledgements}

%%%%%%%%%%%%%%%%%%%%%
%%% %%   Bibliography     %%%%%
%%%%%%%%%%%%%%%%%%%%%

\bibliographystyle{spmpsci} 
\bibliography{rsabib,ColloidBib}

% end of figures at end section

% end of commenting out figure captions section

%%%%%%%%%%%%%%%%%%%%%
%%% %%   Appendix     %%%%%
%%%%%%%%%%%%%%%%%%%%%

\clearpage
\newpage
\section{Appendix}

\subsection{Geometry of the surfaces}\label{app:geometry}

The geometric structures of the spaces we consider is given by thinking of them as collections of either 
«point vectors»  or «line vectors»; there is a duality between these spaces. 
«Point vectors» $\<x,y,z\>$ live in «point space», «line vectors» $\<u,v,w\>$ live in «line space». %If a vector does not live on the surface, then we need to normalize it.
Sometimes we need to convert between «point vectors» and «line vectors».
\begin{equation*}\begin{array}{l}
\geoK\textK{162pt}{point·space equation}\textK{162pt}{line·space equation}\\
\geoR\textR{162pt}{$zz = 1{-}xx{-}yy$}\textR{162pt}{$uu{+}vv = 1{-}ww$}\\
\geoB\textB{162pt}{$zz = 1          $}\textB{162pt}{$uu{+}vv = 1     $}\\
\geoG\textG{162pt}{$zz = 1{+}xx{+}yy$}\textG{162pt}{$uu{+}vv = 1{+}ww$}
\end{array}\end{equation*}
\begin{equation*}\begin{array}{l}
\geoK\textK{162pt}{point·space normalize}\textK{162pt}{line·space normalize}\\
\geoR\textR{162pt}{$\displaystyle{\rm norm}_{\rm point}^{\vphantom{\rm line}}\,\<x,y,z\> = {\<x,y,z\>\over\sqrt{zz         {+}xx{+}yy }}$}
     \textR{162pt}{$\displaystyle{\rm norm}_{\rm line}^{\vphantom{\rm point}}\,\<u,v,w\> = {\<u,v,w\>\over\sqrt{uu{+}vv         {+}ww }}$}\\
\geoB\textB{162pt}{$\displaystyle{\rm norm}_{\rm point}^{\vphantom{\rm line}}\,\<x,y,z\> = {\<x,y,z\>\over\sqrt{zz\phantom{{+}xx{+}yy}}}$}
     \textB{162pt}{$\displaystyle{\rm norm}_{\rm line}^{\vphantom{\rm point}}\,\<u,v,w\> = {\<u,v,w\>\over\sqrt{uu{+}vv\phantom{{+}ww}}}$}\\
\geoG\textG{162pt}{$\displaystyle{\rm norm}_{\rm point}^{\vphantom{\rm line}}\,\<x,y,z\> = {\<x,y,z\>\over\sqrt{zz         {-}xx{-}yy }}$}
     \textG{162pt}{$\displaystyle{\rm norm}_{\rm line}^{\vphantom{\rm point}}\,\<u,v,w\> = {\<u,v,w\>\over\sqrt{uu{+}vv         {-}ww }}$}
\end{array}\end{equation*}
\begin{equation*}\begin{array}{l}
\geoK\textK{162pt}{convert point $\to$ line}\textK{162pt}{convert line $\to$ point}\\
\geoR\textR{162pt}{${\rm conv}_{\rm point}^{\rm line}\,\<x,y,z\> = \<{  +  }x,{  +  }y,z\>$}\textR{162pt}{${\rm conv}_{\rm line}^{\rm point}\,\<u,v,w\> = \<u,v,{  +  }w\>$}\\
\geoB\textB{162pt}{${\rm conv}_{\rm point}^{\rm line}\,\<x,y,z\> = \<{\,0\,}x,{\,0\,}y,z\>$}\textB{162pt}{${\rm conv}_{\rm line}^{\rm point}\,\<u,v,w\> = \<u,v,{\,0\,}w\>$}\\
\geoG\textG{162pt}{${\rm conv}_{\rm point}^{\rm line}\,\<x,y,z\> = \<{  -  }x,{  -  }y,z\>$}\textG{162pt}{${\rm conv}_{\rm line}^{\rm point}\,\<u,v,w\> = \<u,v,{  -  }w\>$}
\end{array}\end{equation*}
\begin{equation*}\begin{array}{l}
\geoK\textK{162pt}{point $\<\alpha,\beta,\gamma\>$, point $\<\delta,\epsilon,\digamma\!\>$}\textK{162pt}{line $\<\alpha,\beta,\gamma\>$, line $\<\delta,\epsilon,\digamma\!\>$}\\
\geoK\textK{162pt}{share line $\<\alpha,\beta,\gamma\>\times\<\delta,\epsilon,\digamma\!\>$}\textK{162pt}{share point $\<\alpha,\beta,\gamma\>\times\<\delta,\epsilon,\digamma\!\>$}\\
\end{array}\end{equation*}
\begin{equation*}\begin{array}{l}
\geoK\textK{162pt}{point·space distance (point,point)}\textK{162pt}{line·space distance (line,line)}\\
\geoR\textR{162pt}{$\cR{\rm dist} = \gamma\digamma\!{+}\alpha\delta{+}\beta\epsilon$}\textR{162pt}{$\cR{\rm dist} = \alpha\delta{+}\beta\epsilon{+}\gamma\digamma\!$}\\
\geoB\textB{162pt}{$\cB{\rm dist} = \gamma\digamma\!                               $}\textB{162pt}{$\cR{\rm dist} = \alpha\delta{+}\beta\epsilon                   $}\\
\geoG\textG{162pt}{$\cG{\rm dist} = \gamma\digamma\!{-}\alpha\delta{-}\beta\epsilon$}\textG{162pt}{$\cR{\rm dist} = \alpha\delta{+}\beta\epsilon{-}\gamma\digamma\!$}
\end{array}\end{equation*}
\begin{equation*}\begin{array}{l}
\geoK\textK{162pt}{point·space distance (point,line)}\textK{162pt}{line·space distance (line,point)}\\
\geoR\textR{162pt}{$\sR{\rm dist} = \gamma\digamma\!{+}\alpha\delta{+}\beta\epsilon$}\textR{162pt}{$\sR{\rm dist} = \alpha\delta{+}\beta\epsilon{+}\gamma\digamma\!$}\\
\geoB\textB{162pt}{$\sB{\rm dist} = \gamma\digamma\!{+}\alpha\delta{+}\beta\epsilon$}\textB{162pt}{$\sR{\rm dist} = \alpha\delta{+}\beta\epsilon{+}\gamma\digamma\!$}\\
\geoG\textG{162pt}{$\sG{\rm dist} = \gamma\digamma\!{+}\alpha\delta{+}\beta\epsilon$}\textG{162pt}{$\sR{\rm dist} = \alpha\delta{+}\beta\epsilon{+}\gamma\digamma\!$}
\end{array}\end{equation*}

\clearpage
\subsection{Tilings}\label{app:tiles}

\subsubsection{Possible tilings}

The table below gives a subset of the possible tilings for different spaces. Recall a $[p,q]$-tiling consists of $p$-polygons, meeting $q$ at each vertex. The patterns below generalize to infinite families of tilings. 
\begin{equation*}\begin{array}{l}
\textMR{36pt}{$[\phantom{0}2,\phantom{0}2]$}\textMR{84pt}{proj plane}\textK {36pt}{$[\phantom{0}0,\phantom{0}0]$}\textK {84pt}{undefined  }\textRY{36pt}{$[\phantom{0}2,\phantom{0}1]$}\textRY{84pt}{sphere     }\\
\textMR{36pt}{$[\phantom{0}4,\phantom{0}2]$}\textMR{84pt}{proj plane}\textCB{36pt}{$[\phantom{0}4,\phantom{0}4]$}\textCB{84pt}{plane      }\textBM{36pt}{$[\phantom{0}6,\phantom{0}3]$}\textBM{84pt}{plane      }\\
\textMR{36pt}{$[\phantom{0}6,\phantom{0}2]$}\textMR{84pt}{proj plane}\textYG{36pt}{$[\phantom{0}8,\phantom{0}8]$}\textYG{84pt}{hyperboloid}\textGC{36pt}{$[          10,\phantom{0}5]$}\textGC{84pt}{hyperboloid}\\
\textMR{36pt}{$[\phantom{0}8,\phantom{0}2]$}\textMR{84pt}{proj plane}\textK {36pt}{$[          12,          12]$}\textK {84pt}{hyperboloid}\textK {36pt}{$[          14,\phantom{0}7]$}\textK {84pt}{hyperboloid}\\
\textMR{36pt}{$[          10,\phantom{0}2]$}\textMR{84pt}{proj plane}\textK {36pt}{$[          16,          16]$}\textK {84pt}{hyperboloid}\textK {36pt}{$[          18,\phantom{0}9]$}\textK {84pt}{hyperboloid}\\
\textMR{36pt}{$[          12,\phantom{0}2]$}\textMR{84pt}{proj plane}\textK {36pt}{$[          20,          20]$}\textK {84pt}{hyperboloid}\textK {36pt}{$[          22,          11]$}\textK {84pt}{hyperboloid}
\end{array}\end{equation*}

\subsubsection{Tiling representations and formulae}\label{app:coords}

The central tile (a $[p,q]$-tile) has face center $\<0,0,1\>$ (the origin), vertices $\bm\upsilon$, edge midpoints $\bm\epsilon$, and neighbour tiles with face centers $\bm\digamma\!$. These objects are all «point vectors» which live in «point space».
They have distances $\upsilon$, $\epsilon$, $\digamma\! = 2\epsilon$ (respectively) from the («point space») origin $\<0,0,1\>$.
The inradius of the tile is $\epsilon$ and the circumradius is $\upsilon$. 

The coordinates of these are given by the following, where 
$\theta$ is any odd multiple of ${\pi\over p}$, and 
$\vartheta$ is any even multiple of ${\pi\over p}$:
\begin{equation*}\begin{array}{l}
\geoR\textR{324pt}{$\bm\upsilon\! = \<\sR\upsilon\,\sR\theta,\sR\upsilon\,\cR\theta,\cR\upsilon\>$}\\
\geoB\textB{324pt}{$\bm\upsilon\! = \<\sB\upsilon\,\sR\theta,\sB\upsilon\,\cR\theta,\cB\upsilon\>$}\\
\geoG\textG{324pt}{$\bm\upsilon\! = \<\sG\upsilon\,\sR\theta,\sG\upsilon\,\cR\theta,\cG\upsilon\>$}
\end{array}\end{equation*}
\begin{equation*}\begin{array}{l}
\geoR\textR{324pt}{$\bm\epsilon = \<\sR\epsilon\,\sR\vartheta,\sR\epsilon\,\cR\vartheta,\cR\epsilon\>$}\\
\geoB\textB{324pt}{$\bm\epsilon = \<\sB\epsilon\,\sR\vartheta,\sB\epsilon\,\cR\vartheta,\cB\epsilon\>$}\\
\geoG\textG{324pt}{$\bm\epsilon = \<\sG\epsilon\,\sR\vartheta,\sG\epsilon\,\cR\vartheta,\cG\epsilon\>$}
\end{array}\end{equation*}
\begin{equation*}\begin{array}{l}
\geoR\textR{324pt}{$\bm\digamma\! = \<\sR\digamma\sin\vartheta,\sR\digamma\cos\vartheta,\cR\digamma\!\>$}\\
\geoB\textB{324pt}{$\bm\digamma\! = \<\sB\digamma\sin\vartheta,\sB\digamma\cos\vartheta,\cB\digamma\!\>$}\\
\geoG\textG{324pt}{$\bm\digamma\! = \<\sG\digamma\sin\vartheta,\sG\digamma\cos\vartheta,\cG\digamma\!\>$}
\end{array}\end{equation*}

The central tile has edge normals $\bm\eta\!$.
These are «line vectors» which live in «line space». They have distance $\eta = \epsilon$ from the («point space») origin $\<0,0,1\>$. $\bm\eta$ «line vectors» are the line·normalized cross·product of two adjacent $\bm\upsilon$ «point vectors». Conversely/dually $\bm\upsilon$ «point vectors» are the point·normalized cross·product of two adjacent $\bm\eta$ «line vectors».
The edge normal vectors are computed as shown in the table below, where $\vartheta$ runs over all  integer multiples of ${2\pi\over p}$.
\begin{equation*}\begin{array}{l}
\geoR\textR{324pt}{$\bm\eta\! = \<{-}\cR\eta\,\cR\vartheta,{-}\cR\eta\,\sR\vartheta,\sR\eta\>$}\\
\geoB\textB{324pt}{$\bm\eta\! = \<{-}\cB\eta\,\cR\vartheta,{-}\cB\eta\,\sR\vartheta,\sB\eta\>$}\\
\geoG\textG{324pt}{$\bm\eta\! = \<{-}\cG\eta\,\cR\vartheta,{-}\cG\eta\,\sR\vartheta,\sG\eta\>$}
\end{array}\end{equation*}

To calculate the inradius $\epsilon$ and circumradius $\upsilon$, it is helpful to think about triangulating the tile polygons.
Let the triangles be $\bm\chi$ with angles $\{\chi_{\bm\digamma}^{},\chi_{\bm\epsilon}^{},\chi_{\bm\upsilon}^{}\} = \{{\pi\over p},{\pi\over 2},{\pi\over q}\}$
at the tile face center $\<0,0,1\>$, vertices $\bm\upsilon\!$, edge midpoints $\bm\epsilon$, and neighbour tile face centers $\bm\digamma\!$.
The triangle edge lengths can be found by using basic trigonometry. 

\medskip Each triangle $\bm\chi$ has area $\chi$, and each tile has area $X = 2p\chi$. For the plane, $\chi$ can scale freely. For the other tilings, $\chi$ is determined by the angles, as follows:
\begin{equation*}\begin{array}{l}
\geoR\textR{324pt}{$\pi         {+}\chi  = \chi_{\bm\digamma}^{}{+}\chi_{\bm\epsilon}^{}{+}\chi_{\bm\upsilon}^{} = {\pi\over p}{+}{\pi\over 2}{+}{\pi\over q}$}\\
\geoB\textB{324pt}{$\pi\phantom{{+}\chi} = \chi_{\bm\digamma}^{}{+}\chi_{\bm\epsilon}^{}{+}\chi_{\bm\upsilon}^{} = {\pi\over p}{+}{\pi\over 2}{+}{\pi\over q}$}\\
\geoG\textG{324pt}{$\pi         {-}\chi  = \chi_{\bm\digamma}^{}{+}\chi_{\bm\epsilon}^{}{+}\chi_{\bm\upsilon}^{} = {\pi\over p}{+}{\pi\over 2}{+}{\pi\over q}$}
\end{array}\end{equation*}

%** triangles are how you figure out what these different distances are. law of sines and cosines belong with triangulations. use to figure out area, hence density.
 
\medskip The circumradius  $\upsilon\!$ of a tile can be determined by the law of sines, once we know the inradius $\epsilon$.
This in turn is determined by the law of cosines, except for the plane, where we can choose it arbitrarily. The formulae are as follows:
\begin{equation*}\begin{array}{l}
\geoR\textR{90pt}{$\displaystyle{\cR{\pi\over q}\over\sR{\pi\over p}} = \cR\epsilon$}\textR{234pt}{$\sR{\pi\over q} = \displaystyle{\sR\epsilon\over\sR\upsilon}$}\\
\geoB\textB{90pt}{$\displaystyle{\cR{\pi\over q}\over\sR{\pi\over p}} = \cB\epsilon$}\textB{234pt}{$\sR{\pi\over q} = \displaystyle{\sB\epsilon\over\sB\upsilon}$}\\
\geoG\textG{90pt}{$\displaystyle{\cR{\pi\over q}\over\sR{\pi\over p}} = \cG\epsilon$}\textG{234pt}{$\sR{\pi\over q} = \displaystyle{\sG\epsilon\over\sG\upsilon}$}
\end{array}\end{equation*}

For the cases we have implemented, we have
\begin{equation*}\begin{array}{l}
\geoR\textRY{36pt}{$[\phantom{0}2,1]$}\textRY{90pt}{$\upsilon = {\pi\over 2}                $}\textRY{90pt}{$\epsilon = \pi                                             $}\textRY{90pt}{$\digamma\! = 2\pi                        $}\\
\geoR\textMR{36pt}{$[\phantom{0}p,2]$}\textMR{90pt}{$\upsilon = {\pi\over 2}                $}\textMR{90pt}{$\epsilon = {\pi\over 2}                                    $}\textMR{90pt}{$\digamma\! = \pi                         $}\\
\geoB\textBM{36pt}{$[\phantom{0}6,3]$}\textBM{90pt}{$\upsilon = {2\over 3}\!\sqrt 3         $}\textBM{90pt}{$\epsilon = 1                                               $}\textBM{90pt}{$\digamma\! = 2                           $}\\
\geoB\textCB{36pt}{$[\phantom{0}4,4]$}\textCB{90pt}{$\upsilon = \sqrt 2                     $}\textCB{90pt}{$\epsilon = 1                                               $}\textCB{90pt}{$\digamma\! = 2                           $}\\
\geoG\textGC{36pt}{$[          10,5]$}\textGC{90pt}{$\upsilon = \cosh^{{-}1}\!(2{+}1\sqrt 5)$}\textGC{90pt}{$\epsilon = \cosh^{{-}1}\!({3\over 2}{+}{1\over 2}\!\sqrt 5)$}\textGC{90pt}{$\digamma\! = \cosh^{{-}1}\!(6{+}3\sqrt 5)$}\\
\geoG\textYG{36pt}{$[\phantom{0}8,8]$}\textYG{90pt}{$\upsilon = \cosh^{{-}1}\!(3{+}2\sqrt 2)$}\textYG{90pt}{$\epsilon = \cosh^{{-}1}\!(1{+}1\sqrt 2)                    $}\textYG{90pt}{$\digamma\! = \cosh^{{-}1}\!(5{+}4\sqrt 2)$}
\end{array}\end{equation*}

\medskip Note that for the plane, $\epsilon$ \& $\upsilon$ can scale freely, however their ratio is always fixed; we choose $\epsilon = 1$.

\medskip 
For a given point in the central tile, we will need to find the corresponding points in the translated tiles. This is done by translating the point along geodesics on the surface, which is practice is done by multiplying the point by a matrix. 
The symmetry group is generated by translations $\bm\tau_{\bm\digamma\!}$ (from the origin $\<0,0,1\>$ to the neighbour tile face centers $\bm\digamma\!$).
A translation $\bm\tau_{\<\alpha,\beta,\gamma\>}$ from the origin $\<0,0,1\>$ to the point $\<\alpha,\beta,\gamma\>$ has inverse $\bm\sigma_{\<\alpha,\beta,\gamma\>} = \bm\tau_{\<{-}\alpha,{-}\beta,\gamma\>}$, where these are computed as
\begin{equation*}\begin{array}{l}
\geoR\textR{144pt}{$\bm\sigma_{\<\alpha,\beta,\gamma\>} = \begin{pmatrix}
1{-}{\alpha\!\alpha\over 1{+}\gamma} & 0{-}{\beta\!\alpha\over 1{+}\gamma} & {-}\alpha\\
0{-}{\alpha\!\beta\over 1{+}\gamma} & 1{-}{\beta\!\beta\over 1{+}\gamma} & {-}\beta\\
\phantom{0}{+}\alpha\phantom{\beta\,} & \phantom{0}{+}\beta\phantom{\alpha\,} & \phantom{-}\!\!\gamma\end{pmatrix}$}
     \textR{144pt}{$\bm\tau_{\<\alpha,\beta,\gamma\>} = \begin{pmatrix}
1{-}{\alpha\!\alpha\over 1{+}\gamma} & 0{-}{\beta\!\alpha\over 1{+}\gamma} & {+}\alpha\\
0{-}{\alpha\!\beta\over 1{+}\gamma} & 1{-}{\beta\!\beta\over 1{+}\gamma} & {+}\beta\\
\phantom{0}{-}\alpha\phantom{\beta\,} & \phantom{0}{-}\beta\phantom{\alpha\,} & \phantom{+}\!\!\gamma\end{pmatrix}$}\\
\geoB\textB{144pt}{$\bm\sigma_{\<\alpha,\beta,\gamma\>} = \begin{pmatrix}
1\phantom{{+}{\alpha\!\alpha\over 1{+}\gamma}} & 0\phantom{{+}{\beta\!\alpha\over 1{+}\gamma}} & {-}\alpha\\
0\phantom{{+}{\alpha\!\beta\over 1{+}\gamma}} & 1\phantom{{+}{vv\over 1{+}\gamma}} & {-}\beta\\
\phantom{0}{\,0\,}\alpha\phantom{\beta\,} & \phantom{0}{\,0\,}\beta\phantom{\alpha\,} & \phantom{-}\!\!\gamma\end{pmatrix}$}
     \textB{144pt}{$\bm\tau_{\<\alpha,\beta,\gamma\>} = \begin{pmatrix}
1\phantom{{+}{\alpha\!\alpha\over 1{+}\gamma}} & 0\phantom{{+}{\beta\!\alpha\over 1{+}\gamma}} & {+}\alpha\\
0\phantom{{+}{\alpha\!\beta\over 1{+}\gamma}} & 1\phantom{{+}{vv\over 1{+}\gamma}} & {+}\beta\\
\phantom{0}{\,0\,}\alpha\phantom{\beta\,} & \phantom{0}{\,0\,}\beta\phantom{\alpha\,} & \phantom{+}\!\!\gamma\end{pmatrix}$}\\
\geoG\textG{144pt}{$\bm\sigma_{\<\alpha,\beta,\gamma\>} = \begin{pmatrix}
1{+}{\alpha\!\alpha\over 1{+}\gamma} & 0{+}{\beta\!\alpha\over 1{+}\gamma} & {-}\alpha\\
0{+}{\alpha\!\beta\over 1{+}\gamma} & 1{+}{\beta\!\beta\over 1{+}\gamma} & {-}\beta\\
\phantom{0}{-}\alpha\phantom{\beta\,} & \phantom{0}{-}\beta\phantom{\alpha\,} & \phantom{-}\!\!\gamma\end{pmatrix}$}
     \textG{144pt}{$\bm\tau_{\<\alpha,\beta,\gamma\>} = \begin{pmatrix}
1{+}{\alpha\!\alpha\over 1{+}\gamma} & 0{+}{\beta\!\alpha\over 1{+}\gamma} & {+}\alpha\\
0{+}{\alpha\!\beta\over 1{+}\gamma} & 1{+}{\beta\!\beta\over 1{+}\gamma} & {+}\beta\\
\phantom{0}{+}\alpha\phantom{\beta\,} & \phantom{0}{+}\beta\phantom{\alpha\,} & \phantom{+}\!\!\gamma\end{pmatrix}$}
\end{array}\end{equation*}

These three families have several nice properties. At each vertex $\bm\upsilon$, there are no `duplicate' vertices which map to it under the translations or products of translations $\bm\sigma_{\bm\digamma\!}$ (or inverses $\bm\tau_{\bm\digamma\!}$).
Exactly 2 tiles share each edge, which are opposite edges. Thus the $p$-polygons must have an even number of sides. Since $p$ is even, this means that polygons can be translated between neighbors.
Thus there is no conflict/collision among points parked near edges or vertices.
Also, the translations about each vertex $\bm\upsilon$ are compatible: if you perform $q$ translations, then you will end up in the original tile, with the original orientation.

In what follows, for the sphere, when we say `translations', we really mean $\{{\rm identity}\}$; in other words there is no need to translate points. For the projective plane, when we say `translations', we really mean $\{{\pm}{\rm identity}\}$: each parked point on a unit sphere implies a second parked point diametrically opposite. Otherwise, these have the same properties as the other tilings.

\subsubsection{Tile neighbours}\label{app:neighbours}

The list below shows the number of translated tiles we used, for small, medium, and large radius (the cases are separated by dots). These numbers were found by trial and error to be sufficient that the number of geometrical errors was small enough. We did not keep track of the exact cutoff radius. 
\begin{equation*}\begin{array}{l}
\geoR\textRY{36pt}{$[\phantom{0}2,1]$}\textRY{72pt}{sphere     }\textRY{180pt}{1 tile\phantom{s}\quad $\{{+}{\rm identity}\}$}\\
\geoR\textMR{36pt}{$[\phantom{0}p,2]$}\textMR{72pt}{proj plane }\textMR{180pt}{2 tiles\quad $\{\pm{\rm identity}\}$}\\
\geoB\textBM{36pt}{$[\phantom{0}6,3]$}\textBM{72pt}{plane      }\textBM{180pt}{7 • 13 • 19 tiles}\\
\geoB\textCB{36pt}{$[\phantom{0}4,4]$}\textCB{72pt}{plane      }\textCB{180pt}{9 • 21 tiles}\\
\geoG\textGC{36pt}{$[          10,5]$}\textGC{72pt}{hyperboloid}\textGC{180pt}{31 • 121 • 481 tiles}\\
\geoG\textYG{36pt}{$[\phantom{0}8,8]$}\textYG{72pt}{hyperboloid}\textYG{180pt}{49 • 337 • 1681 tiles}
\end{array}\end{equation*}

Note that especially for small discs $2\rho < \epsilon$, 
we need not consider all discs in neighbour tiles. When the neighbour tile shares a vertex, we consider only disc centers located less than $\varsigma$ away from the vertex («point vector»). When the neighbour tile shares an edge, we consider only disc centers located less than $\varsigma$ away from the edge («line vector»).

\clearpage

For each tiling, we list the neighbouring vertex vectors $\bm\upsilon$
\begin{equation*}\begin{array}{l}
\textRY{36pt}{$[\phantom{0}2,1]$}\textRY{300pt}{none}\\
\end{array}\end{equation*}
\begin{equation*}\begin{array}{l}
\textMR{36pt}{$[\phantom{0}p,2]$}\textMR{300pt}{along equator}\\
\end{array}\end{equation*}
\begin{equation*}\begin{array}{l}
\textBM{36pt}{$[\phantom{0}6,3]$}\textBM{300pt}{$\<{+}1,{+}\!{1\over 3}\sqrt 3,1\>$}\\
\textBM{36pt}{$[\phantom{0}6,3]$}\textBM{300pt}{$\<\phantom{+}0,{+}\!{2\over 3}\sqrt 3,1\>$}\\
\textBM{36pt}{$[\phantom{0}6,3]$}\textBM{300pt}{$\<{-}1,{+}\!{1\over 3}\sqrt 3,1\>$}\\
\textBM{36pt}{$[\phantom{0}6,3]$}\textBM{300pt}{$\<{-}1,{-}\!{1\over 3}\sqrt 3,1\>$}\\
\textBM{36pt}{$[\phantom{0}6,3]$}\textBM{300pt}{$\<\phantom{+}0,{-}\!{2\over 3}\sqrt 3,1\>$}\\
\textBM{36pt}{$[\phantom{0}6,3]$}\textBM{300pt}{$\<{+}1,{-}\!{1\over 3}\sqrt 3,1\>$}\\
\end{array}\end{equation*}
\begin{equation*}\begin{array}{l}
\textCB{36pt}{$[\phantom{0}4,4]$}\textCB{300pt}{$\<{+}1,{+}1,1\>$}\\
\textCB{36pt}{$[\phantom{0}4,4]$}\textCB{300pt}{$\<{-}1,{+}1,1\>$}\\
\textCB{36pt}{$[\phantom{0}4,4]$}\textCB{300pt}{$\<{-}1,{-}1,1\>$}\\
\textCB{36pt}{$[\phantom{0}4,4]$}\textCB{300pt}{$\<{+}1,{-}1,1\>$}\\
\end{array}\end{equation*}
\begin{equation*}\begin{array}{l}
\textGC{36pt}{$[          10,5]$}\textGC{300pt}{$\<{+}\!{1\over 2}\!\sqrt{30{+}14\sqrt 5},{+}\!{1\over 2}\!\sqrt{02{+}02\sqrt 5},2{+}\sqrt 5\>$}\\
\textGC{36pt}{$[          10,5]$}\textGC{300pt}{$\<{+}\!{1\over 2}\!\sqrt{10{+}06\sqrt 5},{+}\!{1\over 2}\!\sqrt{22{+}10\sqrt 5},2{+}\sqrt 5\>$}\\
\textGC{36pt}{$[          10,5]$}\textGC{300pt}{$\<0,{+}\!\sqrt{8{+}4\sqrt 5},2{+}\sqrt 5\>$}\\
\textGC{36pt}{$[          10,5]$}\textGC{300pt}{$\<{-}\!{1\over 2}\!\sqrt{10{+}06\sqrt 5},{+}\!{1\over 2}\!\sqrt{22{+}10\sqrt 5},2{+}\sqrt 5\>$}\\
\textGC{36pt}{$[          10,5]$}\textGC{300pt}{$\<{-}\!{1\over 2}\!\sqrt{30{+}14\sqrt 5},{+}\!{1\over 2}\!\sqrt{02{+}02\sqrt 5},2{+}\sqrt 5\>$}\\
\textGC{36pt}{$[          10,5]$}\textGC{300pt}{$\<{-}\!{1\over 2}\!\sqrt{30{+}14\sqrt 5},{-}\!{1\over 2}\!\sqrt{02{+}02\sqrt 5},2{+}\sqrt 5\>$}\\
\textGC{36pt}{$[          10,5]$}\textGC{300pt}{$\<{-}\!{1\over 2}\!\sqrt{10{+}06\sqrt 5},{-}\!{1\over 2}\!\sqrt{22{+}10\sqrt 5},2{+}\sqrt 5\>$}\\
\textGC{36pt}{$[          10,5]$}\textGC{300pt}{$\<0,{-}\!\sqrt{8{+}4\sqrt 5},2{+}\sqrt 5\>$}\\
\textGC{36pt}{$[          10,5]$}\textGC{300pt}{$\<{+}\!{1\over 2}\!\sqrt{10{+}06\sqrt 5},{-}\!{1\over 2}\!\sqrt{22{+}10\sqrt 5},2{+}\sqrt 5\>$}\\
\textGC{36pt}{$[          10,5]$}\textGC{300pt}{$\<{+}\!{1\over 2}\!\sqrt{30{+}14\sqrt 5},{-}\!{1\over 2}\!\sqrt{02{+}02\sqrt 5},2{+}\sqrt 5\>$}\\
\end{array}\end{equation*}
\begin{equation*}\begin{array}{l}
\textYG{36pt}{$[\phantom{0}8,8]$}\textYG{300pt}{$\<{+}\sqrt{14{+}10\sqrt 2},{+}\sqrt{02{+}02\sqrt 2},3{+}2\sqrt 2\>$}\\
\textYG{36pt}{$[\phantom{0}8,8]$}\textYG{300pt}{$\<{+}\sqrt{02{+}02\sqrt 2},{+}\sqrt{14{+}10\sqrt 2},3{+}2\sqrt 2\>$}\\
\textYG{36pt}{$[\phantom{0}8,8]$}\textYG{300pt}{$\<{-}\sqrt{02{+}02\sqrt 2},{+}\sqrt{14{+}10\sqrt 2},3{+}2\sqrt 2\>$}\\
\textYG{36pt}{$[\phantom{0}8,8]$}\textYG{300pt}{$\<{-}\sqrt{14{+}10\sqrt 2},{+}\sqrt{02{+}02\sqrt 2},3{+}2\sqrt 2\>$}\\
\textYG{36pt}{$[\phantom{0}8,8]$}\textYG{300pt}{$\<{-}\sqrt{14{+}10\sqrt 2},{-}\sqrt{02{+}02\sqrt 2},3{+}2\sqrt 2\>$}\\
\textYG{36pt}{$[\phantom{0}8,8]$}\textYG{300pt}{$\<{-}\sqrt{02{+}02\sqrt 2},{-}\sqrt{14{+}10\sqrt 2},3{+}2\sqrt 2\>$}\\
\textYG{36pt}{$[\phantom{0}8,8]$}\textYG{300pt}{$\<{+}\sqrt{02{+}02\sqrt 2},{-}\sqrt{14{+}10\sqrt 2},3{+}2\sqrt 2\>$}\\
\textYG{36pt}{$[\phantom{0}8,8]$}\textYG{300pt}{$\<{+}\sqrt{14{+}10\sqrt 2},{-}\sqrt{02{+}02\sqrt 2},3{+}2\sqrt 2\>$}
\end{array}\end{equation*}

\clearpage

For each tiling, we list the neighbouring edge midpoint vectors $\bm\epsilon$
\begin{equation*}\begin{array}{l}
\textRY{36pt}{$[\phantom{0}2,1]$}\textRY{300pt}{none}\\
\end{array}\end{equation*}
\begin{equation*}\begin{array}{l}
\textMR{36pt}{$[\phantom{0}p,2]$}\textMR{300pt}{along equator}\\
\end{array}\end{equation*}
\begin{equation*}\begin{array}{l}
\textBM{36pt}{$[\phantom{0}6,3]$}\textBM{300pt}{$\<{+}1,0,1\>$}\\
\textBM{36pt}{$[\phantom{0}6,3]$}\textBM{300pt}{$\<{+}\!{1\over 2},{+}\!{1\over 2}\!\sqrt 3,1\>$}\\
\textBM{36pt}{$[\phantom{0}6,3]$}\textBM{300pt}{$\<{-}\!{1\over 2},{+}\!{1\over 2}\!\sqrt 3,1\>$}\\
\textBM{36pt}{$[\phantom{0}6,3]$}\textBM{300pt}{$\<{-}1,0,1\>$}\\
\textBM{36pt}{$[\phantom{0}6,3]$}\textBM{300pt}{$\<{-}\!{1\over 2},{-}\!{1\over 2}\!\sqrt 3,1\>$}\\
\textBM{36pt}{$[\phantom{0}6,3]$}\textBM{300pt}{$\<{+}\!{1\over 2},{-}\!{1\over 2}\!\sqrt 3,1\>$}\\
\end{array}\end{equation*}
\begin{equation*}\begin{array}{l}
\textCB{36pt}{$[\phantom{0}4,4]$}\textCB{300pt}{$\<{+}1,0,1\>$}\\
\textCB{36pt}{$[\phantom{0}4,4]$}\textCB{300pt}{$\<0,{+}1,1\>$}\\
\textCB{36pt}{$[\phantom{0}4,4]$}\textCB{300pt}{$\<{-}1,0,1\>$}\\
\textCB{36pt}{$[\phantom{0}4,4]$}\textCB{300pt}{$\<0,{-}1,1\>$}\\
\end{array}\end{equation*}
\begin{equation*}\begin{array}{l}
\textGC{36pt}{$[          10,5]$}\textGC{300pt}{$\<{+}\!{1\over 2}\!\sqrt{10{+}6\sqrt 5},0,{1\over 2}(3{+}\sqrt 5)\>$}\\
\textGC{36pt}{$[          10,5]$}\textGC{300pt}{$\<{+}\!{1\over 4}\!\sqrt{30{+}14\sqrt 5},{+}\!{1\over 4}\!\sqrt{10{+}10\sqrt 5},{1\over 2}(3{+}\sqrt 5)\>$}\\
\textGC{36pt}{$[          10,5]$}\textGC{300pt}{$\<{+}\!{1\over 2}\!\sqrt{00{+}01\sqrt 5},{+}\!{1\over 2}\!\sqrt{10{+}05\sqrt 5},{1\over 2}(3{+}\sqrt 5)\>$}\\
\textGC{36pt}{$[          10,5]$}\textGC{300pt}{$\<{-}\!{1\over 2}\!\sqrt{00{+}01\sqrt 5},{+}\!{1\over 2}\!\sqrt{10{+}05\sqrt 5},{1\over 2}(3{+}\sqrt 5)\>$}\\
\textGC{36pt}{$[          10,5]$}\textGC{300pt}{$\<{-}\!{1\over 4}\!\sqrt{30{+}14\sqrt 5},{+}\!{1\over 4}\!\sqrt{10{+}10\sqrt 5},{1\over 2}(3{+}\sqrt 5)\>$}\\
\textGC{36pt}{$[          10,5]$}\textGC{300pt}{$\<{-}\!{1\over 2}\!\sqrt{10{+}6\sqrt 5},0,{1\over 2}(3{+}\sqrt 5)\>$}\\
\textGC{36pt}{$[          10,5]$}\textGC{300pt}{$\<{-}\!{1\over 4}\!\sqrt{30{+}14\sqrt 5},{-}\!{1\over 4}\!\sqrt{10{+}10\sqrt 5},{1\over 2}(3{+}\sqrt 5)\>$}\\
\textGC{36pt}{$[          10,5]$}\textGC{300pt}{$\<{-}\!{1\over 2}\!\sqrt{00{+}01\sqrt 5},{-}\!{1\over 2}\!\sqrt{10{+}05\sqrt 5},{1\over 2}(3{+}\sqrt 5)\>$}\\
\textGC{36pt}{$[          10,5]$}\textGC{300pt}{$\<{+}\!{1\over 2}\!\sqrt{00{+}01\sqrt 5},{-}\!{1\over 2}\!\sqrt{10{+}05\sqrt 5},{1\over 2}(3{+}\sqrt 5)\>$}\\
\textGC{36pt}{$[          10,5]$}\textGC{300pt}{$\<{+}\!{1\over 4}\!\sqrt{30{+}14\sqrt 5},{-}\!{1\over 4}\!\sqrt{10{+}10\sqrt 5},{1\over 2}(3{+}\sqrt 5)\>$}\\
\end{array}\end{equation*}
\begin{equation*}\begin{array}{l}
\textYG{36pt}{$[\phantom{0}8,8]$}\textYG{300pt}{$\<{+}\!\sqrt{2{+}2\sqrt 2},0,1{+}\sqrt 2\>$}\\
\textYG{36pt}{$[\phantom{0}8,8]$}\textYG{300pt}{$\<{+}\!\sqrt{1{+}1\sqrt 2},{+}\!\sqrt{1{+}1\sqrt 2},1{+}\sqrt 2\>$}\\
\textYG{36pt}{$[\phantom{0}8,8]$}\textYG{300pt}{$\<0,{+}\!\sqrt{2{+}2\sqrt 2},1{+}\sqrt 2\>$}\\
\textYG{36pt}{$[\phantom{0}8,8]$}\textYG{300pt}{$\<{-}\!\sqrt{1{+}1\sqrt 2},{+}\!\sqrt{1{+}1\sqrt 2},1{+}\sqrt 2\>$}\\
\textYG{36pt}{$[\phantom{0}8,8]$}\textYG{300pt}{$\<{-}\!\sqrt{2{+}2\sqrt 2},0,1{+}\sqrt 2\>$}\\
\textYG{36pt}{$[\phantom{0}8,8]$}\textYG{300pt}{$\<{-}\!\sqrt{1{+}1\sqrt 2},{-}\!\sqrt{1{+}1\sqrt 2},1{+}\sqrt 2\>$}\\
\textYG{36pt}{$[\phantom{0}8,8]$}\textYG{300pt}{$\<0,{-}\!\sqrt{2{+}2\sqrt 2},1{+}\sqrt 2\>$}\\
\textYG{36pt}{$[\phantom{0}8,8]$}\textYG{300pt}{$\<{+}\!\sqrt{1{+}1\sqrt 2},{-}\!\sqrt{1{+}1\sqrt 2},1{+}\sqrt 2\>$}
\end{array}\end{equation*}

\clearpage

For each tiling, we list the neighbouring face center vectors $\bm\digamma\!$
\begin{equation*}\begin{array}{l}
\textRY{36pt}{$[\phantom{0}2,1]$}\textRY{300pt}{none}\\
%\textRY{36pt}{$[\phantom{0}2,1]$}\textRY{300pt}{$\<0,0,1\>$}\\
\end{array}\end{equation*}
\begin{equation*}\begin{array}{l}
%\textMR{36pt}{$[\phantom{0}p,2]$}\textMR{300pt}{$\<0,0,{+}1\>$}\\
\textMR{36pt}{$[\phantom{0}p,2]$}\textMR{300pt}{$\<0,0,{-}1\>$}\\
\end{array}\end{equation*}
\begin{equation*}\begin{array}{l}
%\textBM{36pt}{$[\phantom{0}6,3]$}\textBM{300pt}{$\<0,0,1\>$}\\
\textBM{36pt}{$[\phantom{0}6,3]$}\textBM{300pt}{$\<{+}2,0,1\>$}\\
\textBM{36pt}{$[\phantom{0}6,3]$}\textBM{300pt}{$\<{+}1,{+}\sqrt 3,1\>$}\\
\textBM{36pt}{$[\phantom{0}6,3]$}\textBM{300pt}{$\<{-}1,{+}\sqrt 3,1\>$}\\
\textBM{36pt}{$[\phantom{0}6,3]$}\textBM{300pt}{$\<{-}2,0,1\>$}\\
\textBM{36pt}{$[\phantom{0}6,3]$}\textBM{300pt}{$\<{-}1,{-}\sqrt 3,1\>$}\\
\textBM{36pt}{$[\phantom{0}6,3]$}\textBM{300pt}{$\<{+}1,{-}\sqrt 3,1\>$}\\
\end{array}\end{equation*}
\begin{equation*}\begin{array}{l}
%\textCB{36pt}{$[\phantom{0}4,4]$}\textCB{300pt}{$\<0,0,1\>$}\\
\textCB{36pt}{$[\phantom{0}4,4]$}\textCB{300pt}{$\<{+}2,0,1\>$}\\
\textCB{36pt}{$[\phantom{0}4,4]$}\textCB{300pt}{$\<0,{+}2,1\>$}\\
\textCB{36pt}{$[\phantom{0}4,4]$}\textCB{300pt}{$\<{-}2,0,1\>$}\\
\textCB{36pt}{$[\phantom{0}4,4]$}\textCB{300pt}{$\<0,{-}2,1\>$}\\
\end{array}\end{equation*}
\begin{equation*}\begin{array}{l}
%\textGC{36pt}{$[          10,5]$}\textGC{300pt}{$\<0,0,1\>$}\\
\textGC{36pt}{$[          10,5]$}\textGC{300pt}{$\<{+}\!{1\over 2}\!\sqrt{80{+}36\sqrt 5},0,6{+}3\sqrt 5\>$}\\
\textGC{36pt}{$[          10,5]$}\textGC{300pt}{$\<{+}\!{1\over 2}\!\sqrt{210{+}94\sqrt 5},{+}\!{1\over 2}\!\sqrt{110{+}14\sqrt 5},6{+}3\sqrt 5\>$}\\
\textGC{36pt}{$[          10,5]$}\textGC{300pt}{$\<{+}\!{1\over 2}\!\sqrt{030{+}14\sqrt 5},{+}\!{1\over 2}\!\sqrt{290{+}94\sqrt 5},6{+}3\sqrt 5\>$}\\
\textGC{36pt}{$[          10,5]$}\textGC{300pt}{$\<{-}\!{1\over 2}\!\sqrt{030{+}14\sqrt 5},{+}\!{1\over 2}\!\sqrt{290{+}94\sqrt 5},6{+}3\sqrt 5\>$}\\
\textGC{36pt}{$[          10,5]$}\textGC{300pt}{$\<{-}\!{1\over 2}\!\sqrt{210{+}94\sqrt 5},{+}\!{1\over 2}\!\sqrt{110{+}14\sqrt 5},6{+}3\sqrt 5\>$}\\
\textGC{36pt}{$[          10,5]$}\textGC{300pt}{$\<{-}\!{1\over 2}\!\sqrt{80{+}36\sqrt 5},0,6{+}3\sqrt 5\>$}\\
\textGC{36pt}{$[          10,5]$}\textGC{300pt}{$\<{-}\!{1\over 2}\!\sqrt{210{+}94\sqrt 5},{-}\!{1\over 2}\!\sqrt{110{+}14\sqrt 5},6{+}3\sqrt 5\>$}\\
\textGC{36pt}{$[          10,5]$}\textGC{300pt}{$\<{-}\!{1\over 2}\!\sqrt{030{+}14\sqrt 5},{-}\!{1\over 2}\!\sqrt{290{+}94\sqrt 5},6{+}3\sqrt 5\>$}\\
\textGC{36pt}{$[          10,5]$}\textGC{300pt}{$\<{+}\!{1\over 2}\!\sqrt{030{+}14\sqrt 5},{-}\!{1\over 2}\!\sqrt{290{+}94\sqrt 5},6{+}3\sqrt 5\>$}\\
\textGC{36pt}{$[          10,5]$}\textGC{300pt}{$\<{+}\!{1\over 2}\!\sqrt{210{+}94\sqrt 5},{-}\!{1\over 2}\!\sqrt{110{+}14\sqrt 5},6{+}3\sqrt 5\>$}\\
\end{array}\end{equation*}
\begin{equation*}\begin{array}{l}
%\textYG{36pt}{$[\phantom{0}8,8]$}\textYG{300pt}{$\<0,0,1\>$}\\
\textYG{36pt}{$[\phantom{0}8,8]$}\textYG{300pt}{$\<{+}\!\sqrt{56{+}40\sqrt 2},0,5{+}4\sqrt 2\>$}\\
\textYG{36pt}{$[\phantom{0}8,8]$}\textYG{300pt}{$\<{+}\!\sqrt{28{+}20\sqrt 2},{+}\!\sqrt{28{+}20\sqrt 2},5{+}4\sqrt 2\>$}\\
\textYG{36pt}{$[\phantom{0}8,8]$}\textYG{300pt}{$\<0,{+}\!\sqrt{56{+}40\sqrt 2},5{+}4\sqrt 2\>$}\\
\textYG{36pt}{$[\phantom{0}8,8]$}\textYG{300pt}{$\<{-}\!\sqrt{28{+}20\sqrt 2},{+}\!\sqrt{28{+}20\sqrt 2},5{+}4\sqrt 2\>$}\\
\textYG{36pt}{$[\phantom{0}8,8]$}\textYG{300pt}{$\<{-}\!\sqrt{56{+}40\sqrt 2},0,5{+}4\sqrt 2\>$}\\
\textYG{36pt}{$[\phantom{0}8,8]$}\textYG{300pt}{$\<{-}\!\sqrt{28{+}20\sqrt 2},{-}\!\sqrt{28{+}20\sqrt 2},5{+}4\sqrt 2\>$}\\
\textYG{36pt}{$[\phantom{0}8,8]$}\textYG{300pt}{$\<0,{-}\!\sqrt{56{+}40\sqrt 2},5{+}4\sqrt 2\>$}\\
\textYG{36pt}{$[\phantom{0}8,8]$}\textYG{300pt}{$\<{+}\!\sqrt{28{+}20\sqrt 2},{-}\!\sqrt{28{+}20\sqrt 2},5{+}4\sqrt 2\>$}
\end{array}\end{equation*}

\clearpage

For each tiling, we list the neighbouring edge normal vectors $\bm\eta$ 
\begin{equation*}\begin{array}{l}
\textRY{36pt}{$[\phantom{0}2,1]$}\textRY{300pt}{none}\\
\end{array}\end{equation*}
\begin{equation*}\begin{array}{l}
\textMR{36pt}{$[\phantom{0}p,2]$}\textMR{300pt}{$\<0,0,1\>$}\\
\end{array}\end{equation*}
\begin{equation*}\begin{array}{l}
\textBM{36pt}{$[\phantom{0}6,3]$}\textBM{300pt}{$\<{-}1,0,1\>$}\\
\textBM{36pt}{$[\phantom{0}6,3]$}\textBM{300pt}{$\<{-}\!{1\over 2},{-}\!{1\over 2}\!\sqrt 3,1\>$}\\
\textBM{36pt}{$[\phantom{0}6,3]$}\textBM{300pt}{$\<{+}\!{1\over 2},{-}\!{1\over 2}\!\sqrt 3,1\>$}\\
\textBM{36pt}{$[\phantom{0}6,3]$}\textBM{300pt}{$\<{+}1,0,1\>$}\\
\textBM{36pt}{$[\phantom{0}6,3]$}\textBM{300pt}{$\<{+}\!{1\over 2},{+}\!{1\over 2}\!\sqrt 3,1\>$}\\
\textBM{36pt}{$[\phantom{0}6,3]$}\textBM{300pt}{$\<{-}\!{1\over 2},{+}\!{1\over 2}\!\sqrt 3,1\>$}\\
\end{array}\end{equation*}
\begin{equation*}\begin{array}{l}
\textCB{36pt}{$[\phantom{0}4,4]$}\textCB{300pt}{$\<{-}1,0,1\>$}\\
\textCB{36pt}{$[\phantom{0}4,4]$}\textCB{300pt}{$\<0,{-}1,1\>$}\\
\textCB{36pt}{$[\phantom{0}4,4]$}\textCB{300pt}{$\<{+}1,0,1\>$}\\
\textCB{36pt}{$[\phantom{0}4,4]$}\textCB{300pt}{$\<0,{+}1,1\>$}\\
\end{array}\end{equation*}
\begin{equation*}\begin{array}{l}
\textGC{36pt}{$[          10,5]$}\textGC{300pt}{$\<{-}{1\over 2}(3{+}\sqrt 5),0,{1\over 2}\!\sqrt{10{+}6\sqrt 5}\>$}\\
\textGC{36pt}{$[          10,5]$}\textGC{300pt}{$\<{-}{1\over 2}(2{+}\sqrt 5),{-}\!{1\over 2}\!\sqrt{05{+}02\sqrt 5},{1\over 2}\!\sqrt{10{+}6\sqrt 5}\>$}\\
\textGC{36pt}{$[          10,5]$}\textGC{300pt}{$\<{-}{1\over 4}(1{+}\sqrt 5),{-}\!{1\over 4}\!\sqrt{50{+}22\sqrt 5},{1\over 2}\!\sqrt{10{+}6\sqrt 5}\>$}\\
\textGC{36pt}{$[          10,5]$}\textGC{300pt}{$\<{+}{1\over 4}(1{+}\sqrt 5),{-}\!{1\over 4}\!\sqrt{50{+}22\sqrt 5},{1\over 2}\!\sqrt{10{+}6\sqrt 5}\>$}\\
\textGC{36pt}{$[          10,5]$}\textGC{300pt}{$\<{+}{1\over 2}(2{+}\sqrt 5),{-}\!{1\over 2}\!\sqrt{05{+}02\sqrt 5},{1\over 2}\!\sqrt{10{+}6\sqrt 5}\>$}\\
\textGC{36pt}{$[          10,5]$}\textGC{300pt}{$\<{+}{1\over 2}(3{+}\sqrt 5),0,{1\over 2}\!\sqrt{10{+}6\sqrt 5}\>$}\\
\textGC{36pt}{$[          10,5]$}\textGC{300pt}{$\<{+}{1\over 2}(2{+}\sqrt 5),{+}\!{1\over 2}\!\sqrt{05{+}02\sqrt 5},{1\over 2}\!\sqrt{10{+}6\sqrt 5}\>$}\\
\textGC{36pt}{$[          10,5]$}\textGC{300pt}{$\<{+}{1\over 4}(1{+}\sqrt 5),{+}\!{1\over 4}\!\sqrt{50{+}22\sqrt 5},{1\over 2}\!\sqrt{10{+}6\sqrt 5}\>$}\\
\textGC{36pt}{$[          10,5]$}\textGC{300pt}{$\<{-}{1\over 4}(1{+}\sqrt 5),{+}\!{1\over 4}\!\sqrt{50{+}22\sqrt 5},{1\over 2}\!\sqrt{10{+}6\sqrt 5}\>$}\\
\textGC{36pt}{$[          10,5]$}\textGC{300pt}{$\<{-}{1\over 2}(2{+}\sqrt 5),{+}\!{1\over 2}\!\sqrt{05{+}02\sqrt 5},{1\over 2}\!\sqrt{10{+}6\sqrt 5}\>$}\\
\end{array}\end{equation*}
\begin{equation*}\begin{array}{l}
\textYG{36pt}{$[\phantom{0}8,8]$}\textYG{300pt}{$\<{-}(1{+}\sqrt 2),0,\sqrt{2{+}2\sqrt 2}\>$}\\
\textYG{36pt}{$[\phantom{0}8,8]$}\textYG{300pt}{$\<{-}\!{1\over 2}(2{+}\sqrt 2),{-}\!{1\over 2}(2{+}\sqrt 2),\sqrt{2{+}2\sqrt 2}\>$}\\
\textYG{36pt}{$[\phantom{0}8,8]$}\textYG{300pt}{$\<0,{-}(1{+}\sqrt 2),\sqrt{2{+}2\sqrt 2}\>$}\\
\textYG{36pt}{$[\phantom{0}8,8]$}\textYG{300pt}{$\<{+}\!{1\over 2}(2{+}\sqrt 2),{-}\!{1\over 2}(2{+}\sqrt 2),\sqrt{2{+}2\sqrt 2}\>$}\\
\textYG{36pt}{$[\phantom{0}8,8]$}\textYG{300pt}{$\<{+}(1{+}\sqrt 2),0,\sqrt{2{+}2\sqrt 2}\>$}\\
\textYG{36pt}{$[\phantom{0}8,8]$}\textYG{300pt}{$\<{+}\!{1\over 2}(2{+}\sqrt 2),{+}\!{1\over 2}(2{+}\sqrt 2),\sqrt{2{+}2\sqrt 2}\>$}\\
\textYG{36pt}{$[\phantom{0}8,8]$}\textYG{300pt}{$\<0,{+}(1{+}\sqrt 2),\sqrt{2{+}2\sqrt 2}\>$}\\
\textYG{36pt}{$[\phantom{0}8,8]$}\textYG{300pt}{$\<{-}\!{1\over 2}(2{+}\sqrt 2),{+}\!{1\over 2}(2{+}\sqrt 2),\sqrt{2{+}2\sqrt 2}\>$}
\end{array}\end{equation*}

\clearpage

For each tiling, we list the neighbouring translation matrices $\bm\sigma_{\bm\digamma\!}$
\begin{equation*}\begin{array}{l}
\textRY{36pt}{$[\phantom{0}2,1]$}\textRY{300pt}{none}\\
%\textRY{36pt}{$[\phantom{0}2,1]$}\textRY{300pt}{$\scriptsize\begin{pmatrix}1 & 0 & 0\\0 & 1 & 0\\0 & 0 & 1\end{pmatrix}$}\\
\end{array}\end{equation*}
\begin{equation*}\begin{array}{l}
%\textMR{36pt}{$[\phantom{0}p,2]$}\textMR{300pt}{$\scriptsize\begin{pmatrix}1 & 0 & 0\\0 & 1 & 0\\0 & 0 & 1\end{pmatrix}$}\\
\textMR{36pt}{$[\phantom{0}p,2]$}\textMR{300pt}{$\scriptsize\begin{pmatrix}{-}1 & 0 & 0\\0 & {-}1 & 0\\0 & 0 & {-}1\end{pmatrix}$}\\
\end{array}\end{equation*}
\begin{equation*}\begin{array}{l}
%\textBM{36pt}{$[\phantom{0}6,3]$}\textBM{300pt}{$\scriptsize\begin{pmatrix}1 & 0 & 0\\0 & 1 & 0\\0 & 0 & 1\end{pmatrix}$}\\
\textBM{36pt}{$[\phantom{0}6,3]$}\textBM{300pt}{$\scriptsize\begin{pmatrix}1 & 0 & {+}2\\0 & 1 & 0\\0 & 0 & 1\end{pmatrix}$}\\
\textBM{36pt}{$[\phantom{0}6,3]$}\textBM{300pt}{$\scriptsize\begin{pmatrix}1 & 0 & {+}1\\0 & 1 & {+}\sqrt 3\\0 & 0 & 1\end{pmatrix}$}\\
\textBM{36pt}{$[\phantom{0}6,3]$}\textBM{300pt}{$\scriptsize\begin{pmatrix}1 & 0 & {-}1\\0 & 1 & {+}\sqrt 3\\0 & 0 & 1\end{pmatrix}$}\\
\textBM{36pt}{$[\phantom{0}6,3]$}\textBM{300pt}{$\scriptsize\begin{pmatrix}1 & 0 & {-}2\\0 & 1 & 0\\0 & 0 & 1\end{pmatrix}$}\\
\textBM{36pt}{$[\phantom{0}6,3]$}\textBM{300pt}{$\scriptsize\begin{pmatrix}1 & 0 & {-}1\\0 & 1 & {-}\sqrt 3\\0 & 0 & 1\end{pmatrix}$}\\
\textBM{36pt}{$[\phantom{0}6,3]$}\textBM{300pt}{$\scriptsize\begin{pmatrix}1 & 0 & {+}1\\0 & 1 & {-}\sqrt 3\\0 & 0 & 1\end{pmatrix}$}\\
\end{array}\end{equation*}
\begin{equation*}\begin{array}{l}
%\textCB{36pt}{$[\phantom{0}4,4]$}\textCB{300pt}{$\scriptsize\begin{pmatrix}1 & 0 & 0\\0 & 1 & 0\\0 & 0 & 1\end{pmatrix}$}\\
\textCB{36pt}{$[\phantom{0}4,4]$}\textCB{300pt}{$\scriptsize\begin{pmatrix}1 & 0 & {+}2\\0 & 1 & 0\\0 & 0 & 1\end{pmatrix}$}\\
\textCB{36pt}{$[\phantom{0}4,4]$}\textCB{300pt}{$\scriptsize\begin{pmatrix}1 & 0 & 0\\0 & 1 & {+}2\\0 & 0 & 1\end{pmatrix}$}\\
\textCB{36pt}{$[\phantom{0}4,4]$}\textCB{300pt}{$\scriptsize\begin{pmatrix}1 & 0 & {-}2\\0 & 1 & 0\\0 & 0 & 1\end{pmatrix}$}\\
\textCB{36pt}{$[\phantom{0}4,4]$}\textCB{300pt}{$\scriptsize\begin{pmatrix}1 & 0 & 0\\0 & 1 & {-}2\\0 & 0 & 1\end{pmatrix}$}\\
\end{array}\end{equation*}
\begin{equation*}\begin{array}{l}
%\textGC{36pt}{$[          10,5]$}\textGC{300pt}{$\scriptsize\begin{pmatrix}1 & 0 & 0\\ 0 & 1 & 0\\ 0 & 0 & 1\end{pmatrix}$}\\
\textGC{36pt}{$[          10,5]$}\textGC{300pt}{$\scriptsize\begin{pmatrix}6{+}3\sqrt 5 & 0 & {+}\!\sqrt{80{+}36\sqrt 5}\\0 & 1 & 0\\{+}\!\sqrt{80{+}36\sqrt 5} & 0 & 6{+}3\sqrt 5\end{pmatrix}$}\\
\textGC{36pt}{$[          10,5]$}\textGC{300pt}{$\scriptsize\begin{pmatrix}{1\over 4}(19{+}7\sqrt 5) & {+}\!{1\over 4}\!\sqrt{250{+}110\sqrt 5} & {+}\!{1\over 2}\!\sqrt{210{+}\phantom{0}94\sqrt 5}\\{+}\!{1\over 4}\!\sqrt{250{+}110\sqrt 5} & {1\over 4}(9{+}5\sqrt 5) & {+}\!{1\over 2}\!\sqrt{110{+}\phantom{0}50\sqrt 5}\\{+}\!{1\over 2}\!\sqrt{210{+}\phantom{0}94\sqrt 5} & {+}\!{1\over 2}\!\sqrt{110{+}\phantom{0}50\sqrt 5} & 6{+}3\sqrt 5\end{pmatrix}$}\\
\textGC{36pt}{$[          10,5]$}\textGC{300pt}{$\scriptsize\begin{pmatrix}{1\over 2}(2{+}\sqrt 5) & {+}\!{1\over 2}\!\sqrt{\phantom{0}25{+}\phantom{0}10\sqrt 5} & {+}\!{1\over 2}\!\sqrt{\phantom{0}30{+}\phantom{0}14\sqrt 5}\\{+}\!{1\over 2}\!\sqrt{\phantom{0}25{+}\phantom{0}10\sqrt 5} & {1\over 2}(12{+}5\sqrt 5) & {+}\!{1\over 2}\!\sqrt{290{+}130\sqrt 5}\\{+}\!{1\over 2}\!\sqrt{\phantom{0}30{+}\phantom{0}14\sqrt 5} & {+}\!{1\over 2}\!\sqrt{290{+}130\sqrt 5} & 6{+}3\sqrt 5\end{pmatrix}$}\\
\textGC{36pt}{$[          10,5]$}\textGC{300pt}{$\scriptsize\begin{pmatrix}{1\over 2}(2{+}\sqrt 5) & {-}\!{1\over 2}\!\sqrt{\phantom{0}25{+}\phantom{0}10\sqrt 5} & {-}\!{1\over 2}\!\sqrt{\phantom{0}30{+}\phantom{0}14\sqrt 5}\\{-}\!{1\over 2}\!\sqrt{\phantom{0}25{+}\phantom{0}10\sqrt 5} & {1\over 2}(12{+}5\sqrt 5) & {+}\!{1\over 2}\!\sqrt{290{+}130\sqrt 5}\\{-}\!{1\over 2}\!\sqrt{\phantom{0}30{+}\phantom{0}14\sqrt 5} & {+}\!{1\over 2}\!\sqrt{290{+}130\sqrt 5} & 6{+}3\sqrt 5\end{pmatrix}$}\\
\textGC{36pt}{$[          10,5]$}\textGC{300pt}{$\scriptsize\begin{pmatrix}{1\over 4}(19{+}7\sqrt 5) & {-}\!{1\over 4}\!\sqrt{250{+}110\sqrt 5} & {-}\!{1\over 2}\!\sqrt{210{+}\phantom{0}94\sqrt 5}\\{-}\!{1\over 4}\!\sqrt{250{+}110\sqrt 5} & {1\over 4}(9{+}5\sqrt 5) & {+}\!{1\over 2}\!\sqrt{110{+}\phantom{0}50\sqrt 5}\\{-}\!{1\over 2}\!\sqrt{210{+}\phantom{0}94\sqrt 5} & {+}\!{1\over 2}\!\sqrt{110{+}\phantom{0}50\sqrt 5} & 6{+}3\sqrt 5\end{pmatrix}$}\\
\textGC{36pt}{$[          10,5]$}\textGC{300pt}{$\scriptsize\begin{pmatrix}6{+}3\sqrt 5 & 0 & {-}\!\sqrt{80{+}36\sqrt 5}\\0 & 1 & 0\\{-}\!\sqrt{80{+}36\sqrt 5} & 0 & 6{+}3\sqrt 5\end{pmatrix}$}\\
\textGC{36pt}{$[          10,5]$}\textGC{300pt}{$\scriptsize\begin{pmatrix}{1\over 4}(19{+}7\sqrt 5) & {+}\!{1\over 4}\!\sqrt{250{+}110\sqrt 5} & {-}\!{1\over 2}\!\sqrt{210{+}\phantom{0}94\sqrt 5}\\{+}\!{1\over 4}\!\sqrt{250{+}110\sqrt 5} & {1\over 4}(9{+}5\sqrt 5) & {-}\!{1\over 2}\!\sqrt{110{+}\phantom{0}50\sqrt 5}\\{-}\!{1\over 2}\!\sqrt{210{+}\phantom{0}94\sqrt 5} & {-}\!{1\over 2}\!\sqrt{110{+}\phantom{0}50\sqrt 5} & 6{+}3\sqrt 5\end{pmatrix}$}\\
\textGC{36pt}{$[          10,5]$}\textGC{300pt}{$\scriptsize\begin{pmatrix}{1\over 2}(2{+}\sqrt 5) & {+}\!{1\over 2}\!\sqrt{\phantom{0}25{+}\phantom{0}10\sqrt 5} & {-}\!{1\over 2}\!\sqrt{\phantom{0}30{+}\phantom{0}14\sqrt 5}\\{+}\!{1\over 2}\!\sqrt{\phantom{0}25{+}\phantom{0}10\sqrt 5} & {1\over 2}(12{+}5\sqrt 5) & {-}\!{1\over 2}\!\sqrt{290{+}130\sqrt 5}\\{-}\!{1\over 2}\!\sqrt{\phantom{0}30{+}\phantom{0}14\sqrt 5} & {-}\!{1\over 2}\!\sqrt{290{+}130\sqrt 5} & 6{+}3\sqrt 5\end{pmatrix}$}\\
\textGC{36pt}{$[          10,5]$}\textGC{300pt}{$\scriptsize\begin{pmatrix}{1\over 2}(2{+}\sqrt 5) & {-}\!{1\over 2}\!\sqrt{\phantom{0}25{+}\phantom{0}10\sqrt 5} & {+}\!{1\over 2}\!\sqrt{\phantom{0}30{+}\phantom{0}14\sqrt 5}\\{-}\!{1\over 2}\!\sqrt{\phantom{0}25{+}\phantom{0}10\sqrt 5} & {1\over 2}(12{+}5\sqrt 5) & {-}\!{1\over 2}\!\sqrt{290{+}130\sqrt 5}\\{+}\!{1\over 2}\!\sqrt{\phantom{0}30{+}\phantom{0}14\sqrt 5} & {-}\!{1\over 2}\!\sqrt{290{+}130\sqrt 5} & 6{+}3\sqrt 5\end{pmatrix}$}\\
\textGC{36pt}{$[          10,5]$}\textGC{300pt}{$\scriptsize\begin{pmatrix}{1\over 4}(19{+}7\sqrt 5) & {-}\!{1\over 4}\!\sqrt{250{+}110\sqrt 5} & {+}\!{1\over 2}\!\sqrt{210{+}\phantom{0}94\sqrt 5}\\{-}\!{1\over 4}\!\sqrt{250{+}110\sqrt 5} & {1\over 4}(9{+}5\sqrt 5) & {-}\!{1\over 2}\!\sqrt{110{+}\phantom{0}50\sqrt 5}\\{+}\!{1\over 2}\!\sqrt{210{+}\phantom{0}94\sqrt 5} & {-}\!{1\over 2}\!\sqrt{110{+}\phantom{0}50\sqrt 5} & 6{+}3\sqrt 5\end{pmatrix}$}\\
\end{array}\end{equation*}
\begin{equation*}\begin{array}{l}
%\textYG{36pt}{$[\phantom{0}8,8]$}\textYG{300pt}{$\scriptsize\begin{pmatrix}1 & 0 & 0\\ 0 & 1 & 0\\ 0 & 0 & 1\end{pmatrix}$}\\
\textYG{36pt}{$[\phantom{0}8,8]$}\textYG{300pt}{$\scriptsize\begin{pmatrix}5{+}4\sqrt 2 & 0 & {+}\!\sqrt{56{+}40\sqrt 2}\\0 & 1 & 0\\{+}\!\sqrt{56{+}40\sqrt 2} & 0 & 5{+}4\sqrt 2\end{pmatrix}$}\\
\textYG{36pt}{$[\phantom{0}8,8]$}\textYG{300pt}{$\scriptsize\begin{pmatrix}{+}3{+}2\sqrt 2 & {+}2{+}2\sqrt 2 & {+}\!\sqrt{28{+}20\sqrt 2}\\{+}2{+}2\sqrt 2 & {+}3{+}2\sqrt 2 & {+}\!\sqrt{28{+}20\sqrt 2}\\{+}\!\sqrt{28{+}20\sqrt 2} & {+}\!\sqrt{28{+}20\sqrt 2} & 5{+}4\sqrt 2\end{pmatrix}$}\\
\textYG{36pt}{$[\phantom{0}8,8]$}\textYG{300pt}{$\scriptsize\begin{pmatrix}1 & 0 & 0\\0 & 5{+}4\sqrt 2 & {+}\sqrt{56{+}40\sqrt 2}\\0 & {+}\!\sqrt{56{+}40\sqrt 2} & 5{+}4\sqrt 2\end{pmatrix}$}\\
\textYG{36pt}{$[\phantom{0}8,8]$}\textYG{300pt}{$\scriptsize\begin{pmatrix}{+}3{+}2\sqrt 2 & {-}2{-}2\sqrt 2 & {-}\!\sqrt{28{+}20\sqrt 2}\\{-}2{-}2\sqrt 2 & {+}3{+}2\sqrt 2 & {+}\!\sqrt{28{+}20\sqrt 2}\\{-}\!\sqrt{28{+}20\sqrt 2} & {+}\!\sqrt{28{+}20\sqrt 2} & 5{+}4\sqrt 2\end{pmatrix}$}\\
\textYG{36pt}{$[\phantom{0}8,8]$}\textYG{300pt}{$\scriptsize\begin{pmatrix}5{+}4\sqrt 2 & 0 & {-}\!\sqrt{56{+}40\sqrt 2}\\0 & 1 & 0\\{-}\!\sqrt{56{+}40\sqrt 2} & 0 & 5{+}4\sqrt 2\end{pmatrix}$}\\
\textYG{36pt}{$[\phantom{0}8,8]$}\textYG{300pt}{$\scriptsize\begin{pmatrix}{+}3{+}2\sqrt 2 & {+}2{+}2\sqrt 2 & {-}\!\sqrt{28{+}20\sqrt 2}\\{+}2{+}2\sqrt 2 & {+}3{+}2\sqrt 2 & {-}\!\sqrt{28{+}20\sqrt 2}\\{-}\!\sqrt{28{+}20\sqrt 2} & {-}\!\sqrt{28{+}20\sqrt 2} & 5{+}4\sqrt 2\end{pmatrix}$}\\
\textYG{36pt}{$[\phantom{0}8,8]$}\textYG{300pt}{$\scriptsize\begin{pmatrix}1 & 0 & 0\\0 & 5{+}4\sqrt 2 & {-}\sqrt{56{+}40\sqrt 2}\\0 & {-}\!\sqrt{56{+}40\sqrt 2} & 5{+}4\sqrt 2\end{pmatrix}$}\\
\textYG{36pt}{$[\phantom{0}8,8]$}\textYG{300pt}{$\scriptsize\begin{pmatrix}{+}3{+}2\sqrt 2 & {-}2{-}2\sqrt 2 & {+}\!\sqrt{28{+}20\sqrt 2}\\{-}2{-}2\sqrt 2 & {+}3{+}2\sqrt 2 & {-}\!\sqrt{28{+}20\sqrt 2}\\{+}\!\sqrt{28{+}20\sqrt 2} & {-}\!\sqrt{28{+}20\sqrt 2} & 5{+}4\sqrt 2\end{pmatrix}$}
\end{array}\end{equation*}

\end{document}